\documentclass[a4paper,11pt]{article}
\usepackage{a4wide}
\usepackage{amsmath,amssymb}
\usepackage{graphics,graphicx}
\usepackage{color}
\usepackage{cite}
\usepackage{hyperref}

\numberwithin{equation}{section}

\newcommand{\beq}{\begin{equation}}
\newcommand{\eeq}{\end{equation}}
\newcommand{\beqa}{\begin{eqnarray}}
\newcommand{\eeqa}{\end{eqnarray}}
\newcommand{\dd}{{\rm d}}

\newcommand{\Z}{{\mathbb Z}}

\newcommand{\Q}{{\mathbb Q}}

\newcommand{\R}{{\mathbb R}}

\newcommand{\C}{{\mathbb C}}

\newcommand{\PP}{{\mathbb P}}
\newcommand{\e}{\,{\rm e}}

\def\to{\rightarrow}

\newcommand{\nn}{\nonumber}

\newcommand{\wt}{\widetilde}
\newcommand{\wh}{\widehat}

\begin{document}
\thispagestyle{empty}
\begin{flushright}
\end{flushright}
\vspace{1cm}
\begin{center}
{\LARGE\bf A pair of Calabi-Yau manifolds from a two parameter
 non-Abelian gauged linear sigma model}
\end{center}
\vspace{8mm}
\begin{center}
{\large Kentaro Hori\footnote{{\tt kentaro.hori@ipmu.jp}}${}^{\ast}$,
  Johanna Knapp\footnote{{\tt
      knapp@hep.itp.tuwien.ac.at}}${}^{\dagger}$}
\end{center}
\vspace{3mm}
\begin{center}
{\em ${}^{\ast}$ Kavli Institute for the Physics and Mathematics of the 
Universe (WPI)\\ The University of Tokyo,
Kashiwa, Chiba 277-8583, Japan}\\ {\em ${}^{\dagger}$ Institute for
  Theoretical Physics, TU Wien\\ Wiedner Hauptstrasse 8-10, 1040
  Vienna, Austria}
\end{center}
\vspace{15mm}
\begin{abstract}
\noindent 
We construct and study a two parameter
gauged linear sigma model with gauge group $(U(1)^2\times O(2))/\Z_2$
that has a dual model with gauge group $(U(1)^2\times SO(4))/\Z_2$.
The model has two geometric phases, three hybrid phases and one phase whose
character is unknown. 
One of the geometric phases is strongly coupled and the other
is weakly coupled, where strong versus weak is exchanged under the duality.
They correspond to two Calabi-Yau manifolds with
$(h^{1,1},h^{2,1})=(2,24)$
which are birationally inequivalent but are expected to be
derived equivalent.
A region of the discriminant locus in the space of
Fayet-Iliopoulos-theta parameters
supports a mixed Coulomb-confining branch which
is mapped to a mixed Coulomb-Higgs branch in the dual model.

\end{abstract}
\newpage
\setcounter{tocdepth}{1}
\tableofcontents
\setcounter{footnote}{0}
\section{Introduction}

Gauged linear sigma models (GLSMs) \cite{Witten:1993yc}
have been useful tools in the construction and analysis of
two-dimensional (2,2) superconformal field theories that can be used
for supersymmetric string compactifications.
The model has two classes of coupling constants that descend to
exactly marginal parameters of the superconformal field theory ---
the superpotential couplings and the Fayet-Iliopoulos (FI) - theta parameters.
The space of FI parameters is decomposed into chambers
called the ``phases'' according to the pattern of gauge symmetry breaking,
and the low energy theory in each phase has its own character.
For example, many models have geometric phases where the gauge symmetry
is completely Higgsed and the low energy theory is
a non-linear sigma model with Calabi-Yau target space.

In the early days, a class of
models with Abelian gauge groups has been studied extensively,
partly because they have geometric phases corresponding to
complete intersection Calabi-Yaus in toric varieties, for which
a body of mathematical results are available \cite{Morrison:1994fr}.
Also, mirror symmetry is well understood when the gauge group is Abelian
\cite{Hori:2000kt,Morrison:1995yh}.
More recently, GLSMs with non-Abelian gauge groups have started to be
considered.
Unlike in Abelian models,
non-Abelian theories may have ``strongly coupled phases'', where
continuous subgroups of the gauge group
 remain unbroken, and yet massless charged matter exists.
In such a phase, the classical analysis is not reliable
and it is in general difficult to understand the nature of
the low energy theory.
However, it is sometimes possible to obtain relevant results in
the strongly coupled gauge sector \cite{Hori:2006dk,Hori:2011pd}
with which we can understand the low energy behaviour of the models.
We may end up with a non-linear sigma model with a
Calabi-Yau target space, in a way quite different
from the classical Higgs mechanism.

In \cite{Hori:2006dk,Hori:2011pd,Hori:2013gga,Gerhardus:2015sla},
non-Abelian GLSMs with such strongly coupled and yet geometric phases
have been constructed and studied. The Calabi-Yau manifolds that
appear are some kind of determinantal varieties in (weighted)
projective spaces.  These models also have the standard weakly coupled
geometric phases where different Calabi-Yau manifolds appear.  When
two Calabi-Yau manifolds, say $X$ and $Y$, appear in two different
regimes of a common FI-theta parameter space, a number of interesting
conclusions can be drawn.  For example, the mirrors of $X$ and $Y$
must be in the same complex deformation family and in particular the
Gromov-Witten theories of $X$ and $Y$ must be governed by the same
Picard-Fuchs system.  Also, the topological B-models of $X$ and $Y$
must be equivalent and in particular the derived categories of
coherent sheaves on $X$ and $Y$ must be equivalent: $\mathrm{D}^b_{\it
  Coh}(X)\cong \mathrm{D}^b_{\it Coh}(Y)$.  The consequences apply to
all GLSMs, including Abelian ones, but the distinguished feature of the
models in
\cite{Hori:2006dk,Hori:2011pd,Hori:2013gga,Gerhardus:2015sla} with
weakly and strongly coupled phases is that $X$ and $Y$ are
birationally inequivalent.  In fact, these works were partly motivated
by such mathematical results and in return have impact on mathematics
as well\footnote{The works \cite{Hori:2006dk}, \cite{Hori:2011pd} and
  \cite{Gerhardus:2015sla} are motivated by such pairs of Calabi-Yau
  threefolds found by R\o dland \cite{rodland98}, Hosono-Takagi
  \cite{Hosono:2011np,Hosono:2012hc} and Miura
  \cite{Miura,GalkinTalk}, respectively.  The models in
  \cite{Hori:2013gga} realize the Pfaffian Calabi-Yau threefolds
  listed in \cite{Kanazawa:2012xya} along with another
  ``new'' determinantal Calabi-Yau in the strongly coupled phases,
  and have hybrid models in the weakly coupled phases.  (Such a
  manifold-hybrid pair had also been found in \cite{Caldararu:2007tc}
  in an Abelian GLSM which has a non-Abelian dual \cite{Hori:2011pd}.)
  Recently, another pair of Calabi-Yau threefolds of Picard number one
  was found by Ito et al \cite{IMOU,KuznetsovG2}, which begs for a
  physics understanding. In the other direction, the works
  \cite{Hori:2006dk,Hori:2011pd} motivated the proofs
  \cite{BorisovCaldararu,KuznetsovPfGr,ADS,RennemoThesis} of the derived
  equivalence. The duality \cite{Hori:2011pd} motivated to
  establish its categorical counterpart \cite{Rennemo:2016oiu}.  Also,
  the work \cite{Hori:2013gga} presents predictions on derived
  equivalences.}.

The GLSMs in
\cite{Hori:2006dk,Hori:2011pd,Hori:2013gga,Gerhardus:2015sla} are all
``one parameter models'' in the sense that they have a single FI-theta
parameter and the resulting Calabi-Yau manifolds have Picard number
one.  A natural task then is to generalize them to ``multiparameter
models''.  In fact, there are natural targets --- we take the gauge
group to be of the form 
\beq 
G={U(1)^L\times H\over \Gamma} 
\eeq 
where $H$ is a symplectic or (special) orthogonal group and $\Gamma$ is a
discrete subgroup of $U(1)^L\times H$, and the matter consisting of a
number of $H$-singlets and $H$-fundamentals with various charges under
$U(1)^L$.  Indeed, all the models in
\cite{Hori:2006dk,Hori:2011pd,Hori:2013gga,Gerhardus:2015sla} are of
this type with $L=1$.  It is possible that such a generalization will
yield a systematic construction of a large number of Calabi-Yau
manifolds, in the same way as one parameter Calabi-Yau hypersurfaces
in weighted projective spaces are generalized to complete intersection
Calabi-Yaus in toric varieties.

\begin{figure}[htb]
\centerline{\includegraphics{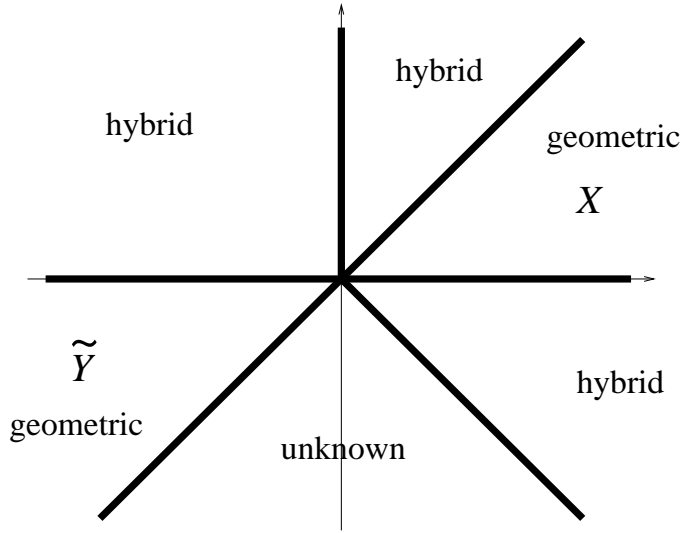}}
%\centerline{\includegraphics{introphases.pdf}}
\caption{The phases of the model}
\label{fig:introphases}
\end{figure}

In this paper, we make a modest first step toward generalization
--- to construct
 one explicit example of this form and study it in as much detail as possible.
If possible, we would like to find a model with
a weakly coupled geometric phase as well as a strongly coupled geometric phase.
After some trials, we found a simple model with such properties.
It is a two parameter model with gauge group
\beq
G={U(1)\times U(1)\times O(2)\over \{(\pm 1, \pm 1, \pm {\bf 1}_2)\}}.
\eeq
It has a dual model with gauge group
\beq
\wt{G}={U(1)\times U(1)\times SO(4)\over \{(\pm 1, \pm 1, \pm {\bf 1}_4)\}}.
\eeq
It can be regarded as a two parameter generalization of
the model \cite{Hori:2011pd} for Hosono-Takagi's Calabi-Yau pair
\cite{Hosono:2011np,Hosono:2012hc}.
The model has six phases as depicted in Figure~\ref{fig:introphases}.
There are two geometric phases corresponding to 
Calabi-Yau threefolds $X$ and $\wt{Y}$, three hybrid phases and one phase
whose character is unknown.
In the original model (gauge group $G$),
the geometric phase of $X$ and the hybrid phases are weakly coupled.
The geometric phase of $\wt{Y}$ is strongly coupled in the original model
but weakly coupled in the dual (gauge group $\wt{G}$).
The remaining phase is strongly coupled in both the original
and dual models, and that is why we refer to it as ``unknown''.

The Calabi-Yau manifold $X$ is a free $\Z_2$ quotient of
a complete intersection of hypersurfaces in an
eight dimensional toric variety.
In particular, it is not simply connected,
$|\pi_1(X)|\in 2\Z$. It has Hodge numbers $(h^{1,1},h^{2,1})=(2,24)$.
On the other hand, the Calabi-Yau manifold $\wt{Y}$ is a $\Z_2$ cover of
a symmetric determinantal variety in a four dimensional toric variety.
It can also be realized as a free $\wt{G}_{\C}$ quotient of
an open part of an affine variety.
It is simply connected, $\pi_1(\wt{Y})=\{1\}$. We have not
computed its Hodge numbers yet, but we obtained $h^{1,1}-h^{2,1}=-22$,
consistent with $(h^{1,1},h^{2,1})=(2,24)$.
Note that
\beq
\pi_1(X)\not\cong\pi_1(\wt{Y}).
\eeq
Since the fundamental group is a birational invariant, $X$ and $\wt{Y}$
cannot be birationally equivalent. However, since they appear in two regimes of
a common FI-theta parameter space, they must have equivalent derived
categories,
\beq
\mathrm{D}^b_{\it Coh}(X)\cong \mathrm{D}^b_{\it Coh}(\wt{Y}).
\label{DEXY}
\eeq Thus, $X$ and $\wt{Y}$ must be another example of a birationally
inequivalent but derived equivalent pair of Calabi-Yau manifolds.

As in any GLSM, the FI-theta parameter space has a discriminant locus
that supports a non-compact flat direction in the scalar component
of the vector multiplet, such as the Coulomb branch or 
a mixed Coulomb-Higgs branch. When projected to the FI parameter space,
it descends to the phase boundaries in the asymptotic directions.
One interesting feature of the present model is that there is a region of
the discriminant locus that supports a branch
where the effective theory includes a strongly coupled gauge sector,
in addition to a free Maxwell theory. 
The horizontal phase boundary
between the geometric phase of $\wt{Y}$ and the hybrid phase above
(see Figure~\ref{fig:introphases})
lifts to a region of a discriminant component that supports such a 
 ``mixed Coulomb-confining branch'' in the original model,
while it supports a mixed Coulomb-Higgs branch in the dual.
The other regime of the same component, descending to
the boundary between the geometric phase of $X$ and the hybrid phase below,
supports a standard mixed Coulomb-Higgs branch in the original
but a mixed Coulomb-confining branch in the dual.

Mirror symmetry for non-abelian GLSMs is still an open
problem. Nevertheless it is sometimes possible to construct a mirror
for the Calabi-Yaus that arise as phases of non-abelian GLSMs. In our
example we are in the lucky situation that the Calabi-Yau in one of
the phases, $X$, is a free quotient of a
complete intersection in a toric variety. Therefore it is possible to
construct its mirror $X^{\vee}$ by
standard methods. In fact, the calculation of the mirror is completely
analogous to the mirror construction by Hosono and Takagi of their
Calabi-Yau
\cite{Hosono:2011np}. This allows us to determine the Picard-Fuchs
operators and the Gromov-Witten invariants associated to $X$. Once we
have the Picard-Fuchs operators we are also able to extract some
information about the other phases.

We would like to note that multiparameter GLSMs with non-Abelian gauge
groups were studied also in a nice work \cite{Jockers:2012zr}.  All
the phases in these models are weakly coupled.  In a subclass of
models called ``linear PAX'', the phases are all geometric as well and
correspond to Calabi-Yau manifolds which are mutually birationally
equivalent. The authors of\cite{Jockers:2012zr} worked out three
two-parameter examples of this class with $(h^{1,1},h^{2,1})=(2,52),
(2,34), (2,52)$.

The rest of the paper is organized as follows. In Section
\ref{sec-glsm} we recall some basic properties of the GLSM. After
summarizing the defining data, we give an overview over the different
types of phases that can occur in non-Abelian GLSMs. We also discuss
how to extract the information about the discriminants from the
GLSM. In Section \ref{sec-model} we introduce our model and its
dual. The remaining sections contain a detailed analysis of this
GLSM. In Section \ref{sec-phasedetails} we discuss the phases,
focusing in particular on the geometric ones. We compute the
topological characteristics of $X$ and $\widetilde{Y}$, where for the
latter we make use of the duality. We furthermore give a brief
analysis of the other phases. In Section \ref{sec-kahler} we identify
the discriminant locus of the FI-theta parameters by determining the
Coulomb and mixed Coulomb-Higgs branches of the GLSM and its dual. The
complement of the discriminant in the FI-theta parameter space
determines the ``K\"ahler moduli space'' $\mathfrak{M}_K$, i.e. the
space of exactly marginal twisted chiral parameters of the infra-red
superconformal field theories (SCFTs). In this section we also discuss
the mixed Coulomb-confining branch mentioned above. In Section
\ref{sec:genericity} we discuss the regularity condition for the GLSM
superpotential such that the Higgs branch is compact.  We find one
condition, Condition (C), that works in all phases.  It determines the
``complex moduli space'' $\mathfrak{M}_C$, i.e. the space of exactly
marginal chiral parameters of the infra-red SCFTs. We derive some
important consequences of Condition (C) that are used in
Section~\ref{sec-phasedetails} for the analysis of each phase.  In
Section \ref{sec-mirror}, we compute the mirror of $X$ and then
determine the Picard-Fuchs operators and the
Gromov-Witten invariants of $X$.
%This is possible because $X$ is a free quotient
%of a complete intersection in a toric variety, and so well-known
%mirror symmetry methods can be applied.
With the Picard-Fuchs operators at hand, we are
also able to extract the Gromov-Witten invariants of $\widetilde{Y}$
in the strongly coupled phase, up to normalization. We end with some
outlook on future directions of research in Section
\ref{sec-outlook}. Further details on the mirror symmetry calculations
can be found in the appendix.

%%%%%%%%%%%%%%%%%%%%%%%%%%%%%%%%%%%%%%%%%%%%%%%%%%%%%%%%%%%%%%%%%%%%%%%%%%%%%%
\section{Basics of GLSMs}
\label{sec-glsm}
In this section we recall some basic properties of GLSMs
  which we will need for the discussion of our model.
See 
\cite{Witten:1993yc,Morrison:1994fr,Hori:2006dk,Hori:2011pd}
for more details.
\subsection{The data}
To specify a GLSM, we choose a gauge group $G$, a matter
representation $V$, a superpotential $W$, and a twisted superpotential
$\wt{W}$.  We assume that $G$ is a compact Lie group and $\rho_V:
G\rightarrow GL(V)$ is a faithful complex representation.  The
superpotential $W$ is a $G$-invariant polynomial of the scalar
component $\phi$ of the matter chiral multiplet which takes values in
$V$.  The twisted superpotential $\wt{W}$ is a $G$-invariant
polynomial of the scalar component $\sigma$ of the vector multiplet
which takes values in the complexified Lie algebra $\mathfrak{g}_{\C}$
of $G$.  We also choose $G$-invariant (hermitian) inner products
on $V$ and $i\mathfrak{g}$. The hermitian
inner product on $V$ determines the moment map $\mu: V\rightarrow
i\mathfrak{g}^{\ast}$ and the inner product on $i\mathfrak{g}$ is
parametrized by the gauge coupling constants $e$.

We are interested in models with vector and axial $U(1)$
R-symmetries with charge integrality. 
A vector $U(1)$ R-symmetry exists when we can assign R-charges on $\phi$,
given by $R\in {\rm End}(V)^G$, under which the superpotential $W(\phi)$
has R-charge $2$.
It has charge integrality when $e^{i\pi R}=\rho_V(J)$ for some $J\in G$.
The axial R-charge of $\sigma$ must be $2$ and
it is a symmetry of the classical system when $\wt{W}(\sigma)$ is linear.
It is anomaly free under the Calabi-Yau condition 
$\rho_V:\:G\rightarrow SL(V)$.
In the Abelian case this reduces to the
well-known condition that the gauge charges sum up to zero.
Under these conditions, the theory is expected to flow in the infra-red limit
to an SCFT of central charge
$\wh{c}={\rm tr}_V(1-R)-\dim G$, with spectral flows between
Ramond and Neveu-Schwarz sectors.
In the following we will only consider such GLSMs.

The linear twisted superpotential is written as
\begin{equation}
\widetilde{W}(\sigma)=-\langle t,\sigma\rangle,%+\widetilde{W}_{\Theta}(\sigma),
\end{equation}
for a $t\in\mathfrak{g}^{\ast G}_{\mathbb{C}}$ which
can be decomposed as $t=\zeta-i\theta$ into the Fayet-Iliopoulos (FI)
parameters $\zeta \in i\mathfrak{g}^{\ast G}$ and
the $\theta$-angles $\theta \in i\mathfrak{g}^{\ast G}$.
The theta angles are subject to appropriate periodicities.
A discrete theta angle and/or discrete torsion must be specified when
$\mathrm{Tors}(\pi_1(G))$ and/or $\pi_0(G)$ are non-trivial.

%where the last term encodes discrete contribution to the
%$\theta$-angle which depends on the choice of gauge group and
%regularity conditions \cite{Hori:2013gga}\footnote{This definition of
%$\widetilde{W}_{\Theta}(\sigma)$ differs from \cite{Hori:2013gga} in
%that respect that we do not include the continuous theta angle
%contribution into the definition.}.
\subsection{Phases}

The classical vacua of the theory are determined by the zeroes of the
potential
\begin{align}
\label{classical-potential}
U=\frac{1}{8e^2}|[\sigma,\bar{\sigma}]|^2
+\frac{1}{2}\left(|\sigma\phi|^2+|\bar{\sigma}\phi|^2\right)
+\frac{e^2}{2}(\mu(\phi)-\zeta)^2+|dW(\phi)|^2.
\end{align}
The first term constrains $\sigma$ to take values in
a Cartan subalgebra
$\mathfrak{t}_{\mathbb{C}}\subset\mathfrak{g}_{\mathbb{C}}$.
The last two terms, depending only on $\phi$,
are called the D-term and the F-term respectively, and yield
the D-term equations 
\begin{align}
\label{d-term}
\mu(\phi)-\zeta=0,
\end{align}
and the
F-term equations
\begin{align}
dW(\phi)=0.
\end{align}

Depending on the value of $\zeta$, some components of $\phi$
are forced to be non-zero by the D-term equations.
This breaks some part of the gauge group, and
hence the components of $\sigma$ corresponding to the broken
generators are forced to vanish.
The pattern of gauge symmetry breaking by the
solutions of the D-term and F-term equations divides the
FI-parameter space into chambers, known as phases of the GLSM.
The nature of the low energy physics in the different phases
can be quite different.

In the interior of a phase,
typically, and always when the gauge group is Abelian,
the gauge symmetry is broken to a finite subgroup,
and all components of $\sigma$ are forced to vanish.
The continuous part of the gauge group is completely Higgsed
and we can study the physics reliably by the classical analysis.
In such a case, the space of solutions to the D-term equations alone modulo
the gauge group action,
which is the symplectic quotient $\mu^{-1}(\zeta)/G$,
is a smooth manifold or an orbifold.
It can also be described as the complex quotient
\begin{align}
\mu^{-1}(\zeta)/G\simeq(V-F_{\zeta})/G_{\mathbb{C}},
\end{align}
where $F_{\zeta}\subset V$ is the locus of $\phi\in V$ 
whose $G_{\mathbb{C}}$-orbit does not hit $\mu^{-1}(\zeta)$, called the
deleted set. 
The superpotential $W$ induces a holomorphic function $W_{\zeta}$
on $\mu^{-1}(\zeta)/G$, and the space of classical vacua
is the critical locus of this function,
\begin{align}
dW^{-1}(0)\cap\mu^{-1}(\zeta)/G={\rm Crit}(W_{\zeta}).
\label{HiggsB}
\end{align} 
If $W_{\zeta}$ is a Bott-Morse function on $\mu^{-1}(\zeta)/G$,
all modes transverse to ${\rm Crit}(W_{\zeta})$ are massive and can be
integrated out. The theory reduces at low energies
to the non-linear sigma model with target ${\rm Crit}(W_{\zeta})$.
This target space is a Calabi-Yau manifold (or a Calabi-Yau orbifold)
whose K\"ahler and B-field classes are determined by $\zeta$ and $\theta$.
Such a phase is referred to as a geometric (or orbifold) phase.
If $W_{\zeta}$ has a single isolated critical point,
the low energy theory is the Landau-Ginzburg model or an orbifold thereof.
Such a phase is referred to as a Landau-Ginzburg phase.
If neither of the above holds for $(\mu^{-1}(\zeta)/G,W_{\zeta})$,
the phase is referred to as a hybrid phase. 
%Characters of hybrid phases have been studied in \cite{Aspinwall:2009qy}.
%In particular, the notion of good and bad hybrids are introduced.

On the interface between different phases,
some of the solutions $\phi$ to the D-term and F-term equations
leave continuous subgroups of the gauge group unbroken.
Accordingly, $\sigma$ can take arbitrary values in the Cartan
subalgebra of the unbroken gauge group. That is, we have a non-compact
flat direction in the effective target space,
called the Coulomb branch.
To be precise, when some of the gauge group is broken by
$\phi$, we shall call it mixed Coulomb-Higgs branch.

\subsection{Strongly coupled phases}
\label{intro:strong}

When the gauge group is non-Abelian, we may have a phase
in which some of the solutions $\phi$
to the D-term and F-term equations
leave continuous subgroups of the gauge group unbroken,
and yet $\sigma$ for the unbroken gauge group
cannot take large values. 
Then, the classical analysis is invalid to understand the
nature of the low energy theory.
Such a strongly coupled phase is very difficult
to study in general.

In \cite{Hori:2006dk,Hori:2011pd}, some useful results were obtained
to deal with such a strongly coupled phase.  In particular a
two-dimensional analog of Seiberg duality has been identified.  
(See \cite{Aharony:2016jki}
for a recent discussion on an important point on such results.)
When
it is applied to a GLSM, strongly coupled phases in the model are
sometimes mapped to weakly coupled phases in the dual, where the gauge
symmetry is broken to finite subgroups and the classical analysis can
be reliably used to understand the low energy physics.

%%%%%%
\subsection{Discriminants}
\label{sec-coulombmixed}

We are primarily interested in regular models which have a discrete spectrum
when formulated on $\R\times S^1$. This is ensured when the
effective target space is compact. But this is not always the case ---
when the parameters of the theory are fine tuned to a special locus called
the discriminant, a non-compact flat direction
emerges in the effective target space.
There are two classes of parameters, 
the chiral parameters in $W$
and the twisted chiral parameters in $\wt{W}$, i.e., the FI-theta parameters.
On the discriminant locus of the chiral parameters,
the Higgs branch (\ref{HiggsB}) becomes non-compact.\footnote{This criterion
of the (ir)regularity of the superpotential couplings was
pointed out in \cite{Aspinwall:2015zia,AspPleTalk}, and will
significantly simplify our analysis.}
It is exactly determined by the classical analysis.
On the discriminant locus of the FI-theta parameters 
a Coulomb branch or a mixed Coulomb-Higgs branch emerges.
As discussed above, Coulomb or mixed branches appear at
the phase boundaries. However, as we will discuss below,
we should take into account the effect of the theta angles
\cite{Coleman:1976uz} and quantum corrections to find the exact location.

Let us first determine the precise locus which supports a pure Coulomb branch
where the gauge symmetry is broken to a maximal torus $T$ and $\sigma$
takes large values in $\mathfrak{t}_{\C}=\mathrm{Lie}(T)_{\C}$.  On
such a branch, all the $T$-charged matter fields are heavy and should be
integrated out, along with the W-bosons. This generates an effective
twisted superpotential for such a $\sigma$:
\begin{align}
\label{coulombweff}
\widetilde{W}_{\it eff}(\sigma)=\widetilde{W}(\sigma)
+\pi i\sum_{\alpha>0}\langle \alpha,\sigma\rangle
-\sum_{Q}\langle Q,\sigma\rangle\left(\log \langle Q,\sigma\rangle-1\right).
\end{align}  
The sums of the second and the third terms are over the positive roots of
$G$ and the weights of the representation $\rho_V$, respectively.
This determines the effective FI-theta parameters
$t_{\it eff}(\sigma)=-\dd\widetilde{W}_{\it eff}(\sigma)$
which enter into the effective potential \cite{Witten:1993yc,Coleman:1976uz} as
\begin{align}
U_{\it eff}=\min_{n\in \mathrm{P}}
\frac{e_{\it eff}^2}{2}\left|t_{\it eff}(\sigma)+2\pi i n\right|^2,
\end{align}  
where $\mathrm{P}$ is the weight lattice of $T$.
The vacuum equation is therefore
\beq
t_{\it eff}(\sigma)\equiv 0\quad\mbox{modulo $2\pi i\mathrm{P}$.}
\label{VECoulomb}
\eeq 
By the Calabi-Yau condition, if $\sigma_*$ is such a vacuum,
then its arbitrary complex multiple is also a vacuum.
That is, the non-compact Coulomb branch exists if and only if the FI-theta
parameter $t$ allows such a solution. In fact, equation
(\ref{VECoulomb})
 provides a parametric representation
of the discriminant locus which supports the pure Coulomb branch.

Let us next determine the locus that supports a mixed Coulomb-Higgs
branch where the gauge symmetry
is broken to a non-maximal torus $T_L\subset G$ and
$\sigma$ takes large values $\sigma_L$ in
$\mathfrak{t}_{L\C}=\mathrm{Lie}(T_L)_{\C}$.
On such a branch, one has to divide the matter fields
$\phi=(\dot{\phi},\hat{\phi})$ into those which
receive mass by $\sigma_L$ (hatted) and those which do not (dotted).
Similarly, we divide $\sigma$ into $(\dot{\sigma},\sigma_L,\hat{\sigma})$,
where $\hat{\sigma}$ receive mass by $\sigma_L$ and
$(\dot{\sigma},\sigma_L)$ do not, i.e., are in
$\mathfrak{c}_{L\C}=\mathrm{Lie}(C_L)_{\C}$
where $C_L\subset G$ is the centralizer of $T_L$.
Integrating out the hatted fields, we obtain the effective theory
of the matter fields $\dot{\phi}$ and the gauge group $C_L$
with the effective twisted superpotential
$\wt{W}_{\it eff,C_L}(\dot{\sigma},\sigma_L)$ given by the same formula as
(\ref{coulombweff}) except that we only
sum over the hatted roots and weights.
The classical potential of the effective theory is as follows:
\begin{align}
U=
{1\over 8(e_{\it eff}^{\mathfrak{c}_L})^2}|{[}\dot{\sigma},\bar{\dot{\sigma}}{]}|^2
+\frac{1}{2}\left(|\dot{\sigma}\dot{\phi}|^2
+|\dot{\bar{\sigma}}\dot{\phi}|^2\right)
+\frac{(e^{\mathfrak{c}_L}_{\it eff})^2}{2}\left(\mu^{\mathfrak{c}_L}(\dot{\phi})
-\zeta_{\it eff}^{\mathfrak{c}_L}(\dot{\sigma},\sigma_L)\right)^2
+|d\dot{W}(\dot{\phi})|^2.
\end{align}
Here $\mu^{\mathfrak{c}_L}(\dot{\phi})$ is the restriction of
$\mu(\dot{\phi})$ to $i\mathfrak{c}_L$,
$\zeta_{\it eff}^{\mathfrak{c}_L}$ is the real part of
$t_{\it eff}^{\mathfrak{c}_L}=-\dd \wt{W}_{\it eff,C_L}$, 
and $\dot{W}$ is the restriction of $W$ to $\dot{\phi}$.
We have the mixed branch when there is a solution to the vacuum
equations which breaks $C_L$ to $T_L$, that is, $\dot{\sigma}=0$ is forced by
$\dot{\phi}$ where $(\dot{\phi},\sigma_L)$ satisfy
\beqa
&&t_{\it eff}^{\mathfrak{t}_L}(0,\sigma_L)\equiv 0\quad
\mbox{modulo $2\pi i {\rm P}_L$},
\label{VEmixedC}\\
&&\mu^{\mathfrak{c}_L}(\dot{\phi})
=\zeta_{\it eff}^{\mathfrak{c}_L}(0,\sigma_L),\quad 
d\dot{W}(\dot{\phi})=0,
\label{VEmixedH}
\eeqa
with $\mathrm{P}_L$ being the weight lattice of $T_L$.
Again, the equation (\ref{VEmixedC}) provides a parametric representation
of the discriminant locus which supports the mixed Coulomb-Higgs
branch, under the condition
that (\ref{VEmixedH}) has a solution.

It is also possible to have a mixed branch where the effective 
theory at large $\sigma_L$ is strongly coupled in the same sense
as in Section~\ref{intro:strong} --- the effective gauge group $C_L$
has an unbroken subgroup bigger than $T_L$ and yet $\dot{\sigma}$ cannot
have large values.
The results of \cite{Hori:2006dk,Hori:2011pd} can again be of help
in such a situation. In this paper, we shall indeed find in that way
a discriminant locus supporting such a branch which may be called
a ``mixed Coulomb-confining branch''.

The complement of the discriminant descends to the space of
exactly marginal parameters of the infra-red SCFTs,
the ``K\"ahler moduli space'' $\mathfrak{M}_K$ for the FI-theta parameters
and the ``complex moduli space'' $\mathfrak{M}_C$ for the chiral parameters.
It is known that the moduli space of 2d (2,2) SCFTs is a direct product
$\mathfrak{M}_K\times \mathfrak{M}_C$
(see \cite{Gomis:2016sab} for a fine point and references therein)
possibly up to a discrete identification.
This means that the discriminant locus
for the superpotential couplings should not depend on the phases,
even though the analysis itself depends on them.
Unlike in Abelian models where a general argument exists for
the independence \cite{Aspinwall:2015zia,AspPleTalk}, in non-Abelian models
with strongly coupled phases, a quite non-trivial work has to be done
in each model, to the best of our knowledge and ability.
(See \cite{BorisovCaldararu}, \cite{Hori:2011pd,Hori:2013gga}
and also Section~\ref{sec:genericity}.)

%%%%%%%%%%%%%%%%%%%%%%%%%%%%%%%%%%%%%%%%%%%%%%%%%%%%%%%%%%%%%%%%%%%%%%%%%%%%%%
\section{The Model}
\label{sec-model}

In this section, we introduce the GLSM we study in this paper,
and describe the classical phase structure. We also
describe the dual model.

%%%%%%%%%%%%%%%%%%%%%%%%%%%%%
\subsection{GLSM data and classical phases}
\label{sec-phases}
The gauge group of the model is 
\begin{align}
\label{gaugegroup}
G=\frac{U(1)_1\times U(1)_2\times O(2)}{ \{(\pm 1,\pm 1,\pm {\bf 1}_2)\}}.
\end{align} 
By $O(2)$ we mean $O(2)_+$ as defined in \cite{Hori:2011pd}, but we
will omit the subscript here. The chiral matter consists of six $O(2)$
singlets $p^I$ and five $O(2)$ doublets $x_I$ which are charged under
$U(1)_1\times U(1)_2$ as in the following table:
\begin{align}
\begin{array}{c|rrrrr||c}
&p^{1...4}&p^5&p^6&x_{1...4}&x_5&{\rm FI}\\
\hline
U(1)_1&-2&-1&1&1&0&\zeta_1\\
U(1)_2&0&-1&-1&0&1&\zeta_2\\
O(2)&{\bf 1}&{\bf 1}&{\bf 1}&\Box&\Box&{\rm -} 
\end{array}
\end{align}
We see that the group $G$ as in (\ref{gaugegroup})
acts effectively on these variables.
 We denote by $\zeta_1,\zeta_2$ the FI-parameters associated to the
 two $U(1)$s. The superpotential is 
\begin{align}
\label{superpotential}
W=\sum_{I,J=1}^5S^{IJ}(p)(x_Ix_J).%=\sum_{I,J=1}^5S^{IJ}(p)u_Iv_J.
\end{align}
The entries of the symmetric matrix $S^{IJ}(p)$ are determined by
gauge invariance:
\begin{align}
S^{ij}(p)=&\sum_{k=1}^4S^{ij}_kp^k,\\
S^{5j}(p)=&S^{5j}_5p^5+\sum_{k=1}^4S^{5j}_kp^kp^6,\\
S^{55}(p)=&S^{55}_5p^5p^6+\sum_{k=1}^4S^{55}_kp^k(p^6)^2.
\end{align}

The model has a vector $U(1)$ R-symmetry 
with the R-charges $2$ for $p^{1... 5}$ and $0$ for $p^6,x_{1... 5}$,
or equivalently, $0$ for $p^{1...6}$ and $1$ for $x_{1... 5}$.
The model also has an axial $U(1)$ R-symmetry since each gauge
transformation has determinant $1$.
Both satisfy charge integrality.
Therefore, the model is expected to flow to a family of
superconformal field theories of central charge
$\wh{c}=5(1-2)+11-3$ (or $6-3)=3$ 
that can be used as string backgrounds with spacetime supersymmetry.
The number of K\"ahler parameters of the family
is $2$ since there are two FI-theta parameters, while the number of
complex parameters is $24$ (see Section~\ref{subsec:countcplx} for the count).
In fact, these are the full numbers of exactly marginal K\"ahler and complex
parameters, since we shall see that there are two geometric phases
and the Hodge numbers of one of the
Calabi-Yau manifolds are computed to be $(h^{1,1},{h^{2,1}})=(2,24)$.

It is sometimes convenient to use
different parametrizations of the fields and the group elements.\footnote{We
choose the convention $i,j,k,\ldots \in
  \{1,\ldots,4\}$, $I,J,K,\ldots\in \{1,\ldots,5\}$. Furthermore we
  use the short-hand notation $|u_{1,2,3}|^2=|u_1|^2+|u_2|^2+|u_3|^2$
  or $u_{1,2,3}=0$ for $u_1=u_2=u_3=0$, etc.}
Let us put $u_I:=x_I^1+ix_I^2$ and $v_I:=x_I^1-ix_I^2$. They
have charge $1$ and $-1$, respectively, under the identity component
$SO(2)\cong U(1)$ of $O(2)$ and are exchanged under
${\rm diag}(1,-1)\in O(2)$. The charge table for these variables is
\begin{align}
\label{basis1}
\begin{array}{c|rrrrrrr||c}
&p^{1...4}&p^5&p^6&u_{1...4}&u_5&v_{1...4}&v_5&{\rm FI}\\
\hline
U(1)_1&-2&-1&1&1&0&1&0&\zeta_1\\
U(1)_2&0&-1&-1&0&1&0&1&\zeta_2\\
SO(2)&0&0&0&1&1&-1&-1&{\rm -}
\end{array}
\end{align}
The identity component $G_0$ of $G$ is isomorphic to $U(1)^3$ via
\beq
\begin{array}{ccc}
\displaystyle
\frac{U(1)_1\times U(1)_2\times SO(2)}
{\{(\pm 1,\pm 1,\pm {\bf 1}_2)\}}&\cong& 
U(1)_0\times U(1)_3\times U(1)_4\\
{[}z_1,z_2,h{]} &\longmapsto&(z_1^2, z_1z_2,z_1h).
\end{array}
\label{o2toruspar}
\eeq
The $U(1)_0\times U(1)_3\times U(1)_4$ charges of the fields are
\begin{align}
\label{basis2}
\begin{array}{c|rrrrrrr||c}
&p^{1...4}&p^5&p^6&u_{1...4}&u_5&v_{1...4}&v_5&{\rm FI}\\
\hline
U(1)_0&-1&0&1&0&-1&1&0&\frac{\zeta_1-\zeta_2}{2}\\
U(1)_3&0&-1&-1&0&1&0&1&\zeta_2\\
U(1)_4&0&0&0&1&1&-1&-1&-
\end{array}
\end{align}

\begin{figure}
\begin{center}
\input{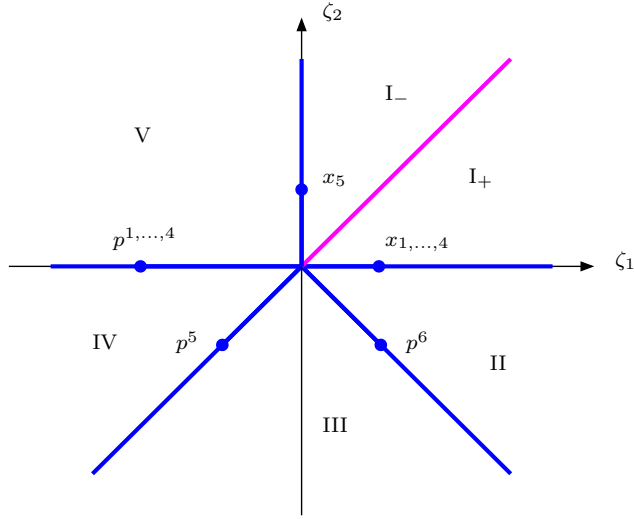}
\end{center}\caption{Classical phases. The non-Abelian D-term leads to an 
extra phase boundary.}\label{fig-twoparphases}
\end{figure}

To determine the classical vacua we have to solve the D-term and
F-term equations. The D-term equations can be read off for instance from
(\ref{basis1})
\begin{align}
|u_{1\ldots 4}|^2+|v_{1\ldots 4}|^2+|p^6|^2-2|p^{1\ldots 4}|^2-|p^5|^2
=&\zeta_1,\label{2paro2d1}\\
|u_5|^2+|v_5|^2-|p^5|^2-|p^6|^2=&\zeta_2,\label{2paro2d2}\\
|u_{1\ldots 4}|^2+|u_5|^2-|v_{1\ldots 4}|^2-|v_5|^2=&0\label{2paro2d3}.
\end{align}
The last one is the $O(2)$ D-term equation and does not come with an
FI parameter. The F-term equations are
\begin{align}
S_k^{ij}u_iv_j+S^{5j}_kp^6(u_5v_j+v_5u_j)+S^{55}_k(p^6)^2u_5v_5=&0,
\quad k=1,...,4,\label{2paro2f1}\\
S^{5j}_5(u_5v_j+v_5u_j)+S^{55}_5p^6u_5v_5=&0,\label{2paro2f2}\\
S^{5j}_kp^k(u_5v_j+v_5u_j)+(S^{55}_5p^5+2S^{55}_kp^kp^6)u_5v_5=&0,
\label{2paro2f3}\\
S^{IJ}(p)u_J=S^{IJ}(p)v_J=&0,\quad I=1,...,5\label{2paro2f4}.
\end{align}
The phase structure is determined by the pattern of gauge symmetry breaking
by the classical vacua.
In the Abelian theory with $W=0$, the phase boundary
is spanned by the charge vectors of the matter fields, and
coincides with the secondary fan of the associated toric variety. 
When the superpotential $W$ is turned on, the F-term equations may lift some
of these phase boundaries. The
D-term equations associated to the non-Abelian factors of the gauge group may
also alter the structure of the phase diagram. We indeed find a phase
boundary associated to the non-Abelian D-term in our example. The
phase diagram is depicted in figure \ref{fig-twoparphases}.
Let us confirm that there is indeed an additional phase boundary at
$\zeta_1=\zeta_2=:\zeta>0$. In this case there are solutions to the
D-term and F-term equations where all fields vanish except
$|u_{1...4}|=|v_5|=\sqrt{\zeta}$ ({\it
  resp}. $|u_5|=|v_{1...4}|=\sqrt{\zeta}$) for which a $U(1)$
subgroup of elements with $z_1=z_2^{-1}=h^{-1}$ ({\it resp}.
$z_1=z_2^{-1}=h$) is unbroken.

The unbroken gauge groups at the classical vacua are all finite
in phases I${}_+$, I${}_-$, II and V.
Thus, these are weakly coupled phases where
a simple classical analysis is enough to identify the low energy physics.
We shall see that I${}_+$ is a geometric phase
while I${}_-$, II and V are hybrid LG/sigma model phases.
In phase IV, the unbroken gauge group is $O(2)\subset G$
at each classical vacuum, while in phase III, it
ranges from the identity to $O(2)$ depending on the vacuum.
These are strongly coupled phases where the classical analysis
is invalid.  To see what we get, we may employ the result of
\cite{Hori:2011pd}.
We shall see that IV is a geometric phase while we are unable to find 
the nature of the low energy physics of phase III.
Once we have
determined the Picard-Fuchs operator, we will provide further evidence
by computing the monodromy matrices around the limiting points
--- only phases I${}_+$ and IV have maximally unipotent
monodromy.

%%%%%%%%%%%%%%%%%%%%%%%%%%%%%%%%%%%%%%%%%%%%%%%%%%%%%%%%%%%%%%%%%%%%%%%%%
\subsection{The dual model}\label{subsec:dualmodel}

One of the results of \cite{Hori:2011pd} 
which can be useful is the 2d Seiberg duality.
Employing that, we obtain the dual of our GLSM. It has gauge group
\beq
\wt{G}={U(1)_1\times U(1)_2\times SO(4)\over
\{(\pm 1,\pm 1,\pm{\bf 1}_4)\}},
\eeq
the matter content
\beq
\begin{array}{c|rrrrrrrr||c}
&p^{1\ldots 4}&p^5&p^6&\wt{x}^{1\ldots 4}&\wt{x}^5&s_{ij}&s_{i5}&s_{55}&
{\rm FI}\\
\hline
U(1)_1&-2~&-1~&1~&-1~&0~&2~&1~&0~&\wt{\zeta}_1\\
U(1)_2&0~&-1~&-1~&0~&-1~&0~&1~&2~&\wt{\zeta}_2\\
SO(4)&{\bf 1}~&{\bf 1}~&{\bf 1}~&\Box~&\Box~&{\bf 1}~&{\bf 1}~&{\bf 1}~&-
\end{array}
\eeq
and the superpotential
\beq
W=\sum_{IJ}s_{IJ}(\wt{x}^I\wt{x}^J)+\sum_{IJ}S^{IJ}(p)s_{IJ},
\eeq
where the variables $(s_{IJ})_{I,J=1}^5$ form a $5\times 5$ symmetric matrix.
The dual model also have vector and axial $U(1)$ R-symmetries with
charge integrality. The vector R-charges of the fields are
$2$ for $p^{1\ldots 5}$, $1$ for $\wt{x}^I$ and $0$ for
$p^6$ and $s_{IJ}$, or equivalently, $2$ for $s_{IJ}$ and $0$ for all others.

We obtain exactly the same phase structure as in the original model,
i.e., that of Figure~\ref{fig-twoparphases}. Note that the ``new'' phase
boundary $\wt{\zeta}_1=\wt{\zeta}_2>0$
between I${}_+$ and I${}_-$ is just the ray spanned by the charge vector of
$s_{i5}$. What is interesting here is that the ray
$\wt{\zeta}_1=0$ and $\wt{\zeta}_2<0$ spanned by the charge vector of
$\wt{x}^5$ is {\it not} a phase boundary.
This is because there is no vacuum where only $\wt{x}^5$ is nonzero;
the F-term equation implies $(\wt{x}^5\wt{x}^5)=0$ but this is
not compatible with the $SO(4)$ D-term equations if $\wt{x}^5\ne 0$.

In phase IV, the dual model is weakly coupled and we will indeed see,
most decisively in this way, that the theory reduces to a Calabi-Yau
sigma model. On the other hand, in phase III, the dual model also
has classical vacua with continuous unbroken gauge symmetry,
and we are unable to find the nature of the low energy theory.

%%%%%%%%%%%%%%%%%%%%%%%%%%%%%%%%%%%%%%%%%%%%%%%%%%%%%%%%%%%%%%%%%%%%%%%%
\section{The Phases}
\label{sec-phasedetails}
In this section, we provide a description of each phase,
with a detailed discussion of phases I${}_+$ and IV.
In particular, we describe the topology of the Calabi-Yau manifolds
that appear in these two phases.
Throughout this section, we assume Condition (C)
described in Section~\ref{sec:genericity}, which is
a genericity condition on the coefficients $S^{ij}_k,\ldots$
of the superpotential (\ref{superpotential}).
A lot of details associated with Condition (C) will be explained
in Section~\ref{sec:genericity}. Here we just state the results.

\subsection{Phase I${}_+$}
\label{sec-i+}

In phase I${}_+$ where $\zeta_1>\zeta_2>0$, the D-term equations forbid
\beq
F_{\rm I_+}=\{u_5=v_5=0\}\cup
\{p^6=u_{1\ldots 4}=0\}\cup
\{p^6=v_{1\ldots 4}=0\}\cup
\{u_{1\ldots 5}=0\}\cup
\{v_{1\ldots 5}=0\}.
\label{FIplus}
\eeq
The identity component $G_0$ of the gauge group acts freely on the
space of solutions to the D-term equations, and the quotient defines
a smooth non-compact toric variety
$V_{\rm I_+}=(V-F_{\rm I_+})/G_{0\C}$. Under Condition (C),
the superpotential $W=\sum_{I=1}^5p^IS_I(u,v,p^6)$
is a Bott-Morse function on $V_{\rm I_+}$ with the critical point set
\beq
p^I=S_I(u,v,p^6)=0,\qquad
I=1,\ldots, 5.
\eeq
Under the same condition (C),
the critical locus $\wt{X}$ is a simply connected smooth Calabi-Yau manifold
on which the component group $\Z_2=G/G_0$ acts freely.
Hence, the theory reduces at low energies to the non-linear sigma model
whose target space is the free quotient $X=\wt{X}/\Z_2$,
which is a smooth Calabi-Yau manifold with
$|\pi_1(X)|=2|\pi_1(\wt{X})|$.

In what follows, we describe the topology of the Calabi-Yau manifolds $\wt{X}$
and $X$. Let $\PP_{\rm I_+}$ be the smooth compact toric variety of dimension 8
obtained by setting $p^1=\cdots =p^5=0$ in $V_{\rm I_+}$, or
more directly as the quotient
\beq
\PP_{\rm I_+}=(V_{B,{\rm I_+}}- F_{\rm I_+})/G_{0\C}
\eeq
where $V_{B,{\rm I_+}}$ is the space of $(u,v,p^6)$.
The weights of the variables under $G_0=U(1)_0\times U(1)_3\times U(1)_4$
can be found in (\ref{basis2}).
Then, $\wt{X}$ is the complete intersection of the five hypersurfaces
$S_1(u,v,p^6)=\cdots =S_5(u,v,p^6)=0$ in $\PP_{\rm I_+}$.
The group $\Z_2=G/G_0$ acts on $\wt{X}\subset \PP_{\rm I_+}$
via the exchange $u\leftrightarrow v$ of the variables and
the involution $(g_0,g_3,g_4)\mapsto (g_0,g_3, g_0g_4^{-1})$
on the group $G_0$.

Let us first describe the topology of the toric variety $\PP=\PP_{\rm I_+}$.
We denote by $H_a$ the divisor class of the line bundle
associated with the charge $1$ representation of $U(1)_a$
($a=0,3,4$). It is simpler to work with the combination
\beq
x:=H_4,\quad y:=H_0-H_4,\quad z:=H_3,
\eeq
on which the $\Z_2$ acts as $x\leftrightarrow y$, $z\to z$.
The classes of the homogeneous coordinates are
$p^6: H_0-H_3=x+y-z$, $u_{1\ldots 4}: H_4=x$,
$u_5: -H_0+H_3+H_4=z-y$,
$v_{1\ldots 4}: H_0-H_4=y$,
$v_5: H_3-H_4=z-x$.
Since the deleted set is (\ref{FIplus}),
we see that the relations among these classes are
\beqa
&&(z-y)(z-x)=0,\nn\\
&&(x+y-z)x^4=(x+y-z)y^4=0,\\
&&x^4(z-y)=y^4(z-x)=0.\nn
\eeqa
Since there is exactly one point with $u_{1\ldots 4}=v_{1\ldots 4}=0$,
\beq
\int_{\PP}x^4y^4=1.
\eeq
Finally the Chern class of $\PP$ is
\beq
c(\PP)=(1+x+y-z)(1+x)^4(1+z-y)(1+y)^4(1+z-x).
\eeq
It is a simple exercise to see that non-zero Hodge numbers of $\PP$
are
$$h^{0,0},\ldots, h^{8,8}=1,3,5,7,9,7,5,3,1,$$
and hence
$\chi(\PP)=41$, which can be checked with $\int_{\PP}c_8(\PP)=41$.

Next we analyze the topology of $\wt{X}\subset \PP$
defined by the zero of a section of
${\mathcal O}(H_0)^{\oplus 4}\oplus {\mathcal O}(H_3)$.
The intersection numbers in $\wt{X}$ can be found from
$\int_{\wt{X}}\tilde\eta=\int_{\PP}(x+y)^4z\tilde\eta$ for a class
$\tilde\eta\in {\rm H}^6(\PP)$:
\beqa
\label{3partriple}
&&x^3=y^3=5,\quad
x^2y=xy^2=10,\quad
x^2z=y^2z=11,\quad
xyz=14,\nn\\
&&
xz^2=yz^2=15,\quad
z^3=16.\nn
\eeqa
The Chern class is
$c(\wt{X})=c(\PP)/((1+x+y)^4(1+z))$, i.e., $c_1(\wt{X})=0$ and
\beq
c_2(\wt{X})=-x^2-y^2+(x+y)z+4xy,
~\,
c_3(\wt{X})=-3x^2y-3xy^2-2xyz.
\eeq
In particular,
\beq
c_2(\wt{X})\cdot x=c_2(\wt{X})\cdot y=50,\quad
c_2(\wt{X})\cdot z=64,\quad
\chi(\wt{X})=-88.
\eeq

Finally we can compute the topology of $X=\wt{X}/\Z_2$.
Note that $\Z_2$ acts on the classes $x,y,z$ on $\wt{X}$
as $x\leftrightarrow y$, $z\mapsto z$.
 The generating divisor classes of $X$ are
$x+y=H_0$ and $z=H_3$, and the intersection numbers can be found from
$\int_{X}\eta={1\over 2}\int_{\wt{X}}\pi^*\eta$ where $\pi:\wt{X}\to X$
is the projection map:
\beq
H_0^3=35,\quad
H_0^2H_3=25,\quad
H_0H_3^2=15,\quad
H_3^3=8,
\eeq
\beq
c_2(X)\cdot H_0=50,\quad
c_2(X)\cdot H_3=32,
\eeq
and
\beq
\chi(X)=-{88\over 2}=-44.
\eeq
Since $h^{1,1}(X)=2$, the above Euler number tells us that
$h^{2,1}(X)=24$.

In \cite{Candelas:2015amz} two Calabi-Yaus with
$(h^{1,1},h^{2,1})=(2,24)$, constructed as free quotients of complete
intersections in toric ambient spaces, are listed. The geometry of
these examples looks different to ours. However, due to redundancies
in the description of complete intersection Calabi-Yaus one still may
have the same Calabi-Yau. It would be interesting to know the other
topological characteristics of the Calabi-Yaus of
\cite{Candelas:2015amz} in order to see whether there is a match.

\subsection{Phase IV}

In phase IV, where $\zeta_1<\zeta_2<0$,
the D-term equations forbid
$(p^{1\ldots 4},u_5,v_5)=0$, $p^{1\ldots 5}=0$ and $p^{5,6}=0$.
This alone would allow the possibility to have $p^{1\ldots 4}=0$ but
$(u_5,v_5)\ne 0$ and $p^5\ne 0$, but that is eliminated by
the F-term equations. Indeed, $S(p)u=S(p)v=0$ imply in this case
$S^{i5}_5p^5u_5=S^{i5}p^5v_5=0$, and we know from Condition (C) in
Section~\ref{sec:genericity}
that $S^{i5}_5\ne 0$. Thus, this enforces $u_5=v_5=0$, in contradiction to
$(u_5,v_5)\ne 0$. Therefore, the deleted set in this phase is
\beq
F_{\rm IV}=\{p^{1\ldots 4}=0\}\cup\{p^{5,6}=0\}.
\label{defFIV}
\eeq
Under Condition (C), the vacuum equations also imply $u=v=0$.
The unbroken gauge group is $O(2)\subset G$ at every classical vacuum
--- we are in a strongly coupled phase.
The manifold of classical vacua is
\beq
\PP_{\rm IV}=(V_{B,{\rm IV}}-F_{\rm IV})/\C^*_0\times \C^*_3,
\label{PPIV}
\eeq
where $V_{B,{\rm IV}}$ is the space of $(p^1,\ldots, p^6)$ and
$\C^*_0\times \C^*_3$ is the complexification of the group
$U(1)_0\times U(1)_3$, with the weights given in (\ref{basis2}). 
It is a smooth compact toric variety of dimension $4$.

We can proceed as in the analysis \cite{Hori:2011pd}
of the strongly coupled phase of Hosono-Takagi model.
First, we may try to work in the Born-Oppenheimer approximation,
where we first ``solve'' the $O(2)$ sector for a fixed value of $p$
and then consider the fluctuation in the $p$ field.
Under Condition (C), the $5\times 5$ mass matrix
$S(p)$ is at least $3$ as long as $p$ is away from the deleted set
$F_{\rm IV}$. Let $Y$ and $C$ be the loci of $[p]\in \PP_{\rm IV}$
where ${\rm rank}\,S(p)\leq 4$ and ${\rm rank}\,S(p)=3$, respectively.
$Y\subset \PP_{\rm IV}$ is a hypersurface (a three-fold)
with an A${}_1$ singularity along $C$ (a curve).
The result of the $O(2)$ gauge theory in \cite{Hori:2011pd} implies that
the low energy theory must be the non-linear sigma model whose
the target space is a double cover
of $Y$ which is ramified along $C$ so that the A${}_1$ singularity
is unfolded as $\C^2\to \C^2/\Z_2$.
From this analysis, however,
it is not clear whether such a double cover exists globally.
A natural construction of the cover is provided by the dual model,
as we described in Section~\ref{subsec:dualmodel}.

In the dual model,
phase IV corresponds to $\wt{\zeta}_1<\wt{\zeta}_2<0$. In this phase,
as will be shown later in Section~\ref{subsub:detail},
the vacuum equations require $p$ to be away from the deleted set
$F_{\rm IV}$ in (\ref{defFIV}), breaking the gauge group
$\wt{G}$ to the subgroup
$(\{(\pm 1, \pm 1)\}\times SO(4))/\{(\pm 1,\pm 1,\pm {\bf 1}_4)\} 
\cong SO(4)$. The equations also imply $s_{IJ}=0$ for all $I,J$,
under Condition (C) as proven in
Section~\ref{subsub:iiandiii}.
The remaining F-term equations are
\beq
S^{IJ}(p)+(\wt{x}^I\wt{x}^J)=0\qquad\forall I,J=1,\ldots, 5.
\label{wtYeq}
\eeq
Since $S(p)$ for $p\not\in F_{\rm IV}$ has at least rank $3$ under (C),
$\wt{x}^I_a$ obeying these equations must always have rank $3$ or more,
completely breaking the residual gauge group $SO(4)$.
Thus, we are in a weakly coupled phase. The vacuum manifold is
\beq
\wt{Y}\,\,=\,\,\left\{\,\,\,(p,\wt{x})\,\,\, \Bigl| \,\,\,
p\not\in F_{\rm IV},~\mbox{$SO(4)$ stability,}~(\ref{wtYeq}).\right\}
\mbox{\Large $/$}
\wt{G}_{\C}.
\eeq
All modes transverse to this are massive, and the theory reduces at low energies
to the non-linear sigma model with this target space.
Since the equations (\ref{wtYeq}) imply that $S(p)$ has rank $4$ or less,
$[(p,\wt{x})]\mapsto [p]$ defines a map $\pi:\wt{Y}\to Y$.
It is surjective --- two to one over $Y- C$ and one to one over $C$ ---
and behaves as the quotient map $\C^2\to \C^2/\Z_2$
in the directions transverse to $C$.
Thus, $\wt{Y}$ realizes the wanted double cover
of $Y$ that unfolds the A${}_1$ singularity along $C$.

\subsubsection{Topology of $\wt{Y}$}
\label{sec-iv}

Let us study the topology of the Calabi-Yau threefold $\wt{Y}$ by
employing the method discussed in Appendix D of \cite{Hori:2013gga}.
As stated above, it is a $\Z_2$ cover of a hypersurface
$Y=\{\det S(p)=0\}$ of $\PP_{\rm IV}$ unfolding the A${}_1$ singularity
along the curve $C=\{{\rm rank}\,S(p)=3\}\subset Y$.
Let us describe the topology of the four dimensional toric variety
$\PP=\PP_{\rm IV}$ (\ref{PPIV}).
We denote by $H_a$ the divisor class of
the line bundle associated with the charge $-1$ representation
of $\C^*_a$ ($a=0,3$). In view of the deleted set (\ref{defFIV}),
we find that these classes are related as
\beq
H_0^4=0,\qquad H_3(H_3-H_0)=0.
\label{relPIV}
\eeq
Since there is one point with $p^1=p^2=p^3=p^5=0$,
\beq
\int_{\PP}H_0^3H_3=1.
\label{volPIV}
\eeq
The non-zero Hodge numbers are $h^{0,0},\ldots, h^{4,4}=1,2,2,2,1$.
It is simply connected: $\pi_1(\PP)=\{1\}$.

We first resolve the A${}_1$ singularity of $Y$ along $C$ by
inserting $\PP({\rm Ker}S(p))$, which is a $\PP^1$ over $C$ and one point
elsewhere on $Y$. Since $S(p)$ can be regarded as
the bundle map ${\mathcal E}\to {\mathcal E}^*(H_0)$ for
${\mathcal E}={\mathcal O}^{\oplus 4}\oplus {\mathcal O}(H_0-H_3)$,
the resolution  is given by
\beq
Z=\left\{\,\,\,[(p,x)]\in \PP({\mathcal E})\,\,\Bigr|\,\,
S(p)x=0\,\,\,\right\}.
\eeq
The forgetful map $Z\to Y$ is one to one over $Y- C$ and
is a $\PP^1$ bundle $D$ over $C$.
We introduce $\wt{Z}$ as the fiber product
\beq
\begin{array}{ccccc}
\wt{Z}&\longrightarrow&Z&\subset&\PP({\mathcal E})\\
\downarrow&&\downarrow&&\downarrow\\
\wt{Y}&\stackrel{\pi}{\longrightarrow}&Y&\subset &\PP
\end{array}
\label{diaG}
\eeq
We shall denote the pre-images of $C\subset Y$ and $D\subset Z$
in $\wt{Y}$ and $\wt{Z}$ by the same symbols.
Since $\wt{Z}- D$ is an unramified double cover over $Z- D$
by the upper-horizontal arrow of (\ref{diaG}) and
is mapped isomorphically onto $\wt{Y}- C$ by the left-vertical arrow,
we find the following relations among the Euler numbers
\beq
\chi(\wt{Z}- D)
=\left\{\begin{array}{l}
2\chi(Z- D)=2\chi(Z)-2\chi(D)\\[0.2cm]
\chi(\wt{Y}- C)=\chi(\wt{Y})-\chi(C).
\end{array}\right.
\eeq
Note also that $\chi(D)=2\chi(C)$ as $D$ is a $\PP^1$ bundle on $C$.
Combining these, we find
\beq
\chi(\wt{Y})=2\chi(Z)-3\chi(C).
\label{chitilY}
\eeq

The ambient space $\PP({\mathcal E})$
is realized as the vacuum manifold in a certain phase of the GLSM
with gauge group $G_{\rm aux}=U(1)_0\times U(1)_3\times U(1)_5$,
matter content
\beq
\begin{array}{c|rrrrr||c}
&\,\,p^{1\ldots 4}&\,\,p^5&\,\,p^6&\,\,x_{1\ldots 4}&\,\,x_5&{\rm FI}\\
\hline
U(1)_0&-1&0&1&0&-1&\zeta^0\\
U(1)_3&0&-1&-1&0&1&\zeta^3\\
U(1)_5&0&0&0&-1&-1&\zeta^5
\end{array}
\eeq
and vanishing superpotential. The relevant phase is $\zeta^3<0$,
$\zeta^5<\zeta^0<0$
where the deleted set is
\beq
F_{\rm aux}=\{\,p^{1\ldots 4}=0\,\}\cup \{\,p^{5,6}=0\,\}\cup
\{\,x_{1\ldots 5}=0\,\}.
\label{defFaux}
\eeq
We denote by $h_a$ the divisor class of the line bundle associated with
the charge $-1$ representation of $U(1)_a$ ($a=0,3,5$).
In view of the deleted set $F_{\rm aux}$, we see that the relations
of these classes are
\beq
h_0^4=0,\quad h_3(h_0-h_3)=0,\quad h_5^4(h_5+h_0-h_3)=0.
\label{relaux}
\eeq
Since the zeroes of $p^1,p^2,p^3,p^5,x_{1\ldots 4}$ intersect
transversally at one point,
\beq
\int_{\PP({\mathcal E})}h_0^3h_3h_5^4=1.
\eeq
Finally the Chern class is
\beq
c(\PP({\mathcal E}))=(1+h_0)^4(1+h_3)(1-h_0+h_3)(1+h_5)^4(1+h_5+h_0-h_3).
\label{abovec}
\eeq
This determines the topology of $\PP({\mathcal E})$.
It follows from (\ref{relaux}) that the non-zero
Hodge numbers are $h^{0,0},\ldots, h^{8,8}=1,3,5,7,8,7,5,3,1$ and in particular
the Euler number is $40$. Also, (\ref{abovec}) yields
$\int_{\PP({\mathcal E})}c_8(\PP({\mathcal E}))=40$ as well.

To determine the class of the divisor $D\subset Z$,
we consider the diagram of vector bundles on $\PP({\mathcal E})$,
\beq
\begin{array}{rcl}
0~\longrightarrow~ {\mathcal O}(-h_5) ~\longrightarrow
\!\!\!\!\!\!&\varpi^*{\mathcal E}&\!\!\!\!\!
\longrightarrow ~\,\,{\mathcal F}\,\,~\longrightarrow~ 0\\
&\downarrow&\!\!\!\!\!\!\!\!\!\!\!\!\! S(p)\\
0~\longrightarrow ~\,\,{\mathcal F}^*(h_0)\,\, ~\longrightarrow
\!&\varpi^*{\mathcal E}^*(h_0)&\!\!
\longrightarrow~ {\mathcal O}(h_0+h_5)~\longrightarrow~ 0
\end{array}
\eeq
The upper line is the tautological exact sequence on the fiber of
$\varpi:\PP({\mathcal E})\to \PP$, and the lower line is its dual tensored
with ${\mathcal O}(h_0)$.
Note that $\det{\mathcal F}={\mathcal O}(h_0-h_3+h_5)$.
The submanifold $Z$ is the locus where the map from ${\mathcal O}(-h_5)$
to $\varpi^*{\mathcal E}^*(h_0)$ vanishes.
By the symmetry of $S(p)$, it is the same as the locus where
the map from $\varpi^*{\mathcal E}$ to ${\mathcal O}(h_0+h_5)$ vanishes.
On this locus, there is a map from ${\mathcal F}$
to ${\mathcal F}^*(h_0)$ and $D\subset Z$ is where that degenerates.
That is, $D$ is the zero of a section of
$\det ({\mathcal F}^*(h_0))\otimes (\det{\mathcal F})^{-1} 
=(\det{\mathcal F})^{-2}(4h_0)={\mathcal O}(2h_0+2h_3-2h_5)$ on $Z$,
which means
\beq
[D]=(2h_0+2h_3-2h_5)|_Z.
\eeq

Now, we are ready to compute the Euler numbers of
$Z,D,C$ and hence of $\wt{Y}$.
Since $Z\subset \PP({\mathcal E})$ is the zero of $S(p)x$,
which is a section of the vector bundle
${\mathcal G}={\mathcal O}(h_0+h_5)^{\oplus 4}\oplus{\mathcal O}(h_3+h_5)$,
its Euler number is
\beq
\chi(Z)=\int_{\PP({\mathcal E})}c_{\rm top}({\mathcal G})\cdot
{c(\PP({\mathcal E}))\over c({\mathcal G})}=-88.
\eeq
Since $D\subset \PP({\mathcal E})$ is the zero of a section of
${\mathcal G}\oplus {\mathcal O}(2h_0+2h_3-2h_5)$,
its Euler number is
\beq
\chi(D)=\int_{\PP({\mathcal E})}c_{\rm top}({\mathcal G})(2h_0+2h_3-2h_5)
\cdot{c(\PP({\mathcal E}))\over c({\mathcal G})(1+2h_0+2h_3-2h_5)}=-88.
\eeq
Thus, $\chi(C)=\chi(D)/2=-44$ (the curve $C$ has genus $23$).
Applying (\ref{chitilY}), we find
\beq
\chi(\wt{Y})=2(-88)-3(-44)=-44,
\eeq
which means $h^{2,1}(\wt{Y})-h^{1,1}(\wt{Y})=22$.
This is consistent with $h^{2,1}(\wt{Y})=24$ and $h^{1,1}(\wt{Y})=2$,
that follows physically from the result in phase I${}_+$ by
the deformation invariance of the Hodge diamond of RR ground states.

Let us try to compute the intersection numbers.
Put $M_a:=\pi^*(H_a|_Y)$ for $a=0,3$.
Using $\int_{\wt{Y}}\pi^*\eta=2\int_Y\eta$ for $\eta\in {\rm H}^6(Y)$
and (\ref{relPIV})-(\ref{volPIV}), we find
\beq
M_0^3=4,\quad
M_0^2M_3=M_0M_3^2=M_3^3=10.
\eeq
Using the Riemann-Roch formula
$\chi(M_a)={1\over 12}c_2(\wt{Y})\cdot M_a+{1\over 3!}M_a^3$
and \underline{assuming} $\chi(M_a)=h^0(\PP,H_a)$, which is $4$ for $a=0$
and $5$ for $a=3$,
we find
\beq
c_2(\wt{Y})\cdot M_0=40,\qquad
c_2(\wt{Y})\cdot M_3=40.
\eeq

$\wt{Y}$ is simply connected.
To show this, we first note that
$Y\subset \PP$ is the zero of $\det S(p)$, which is a section
of the line bundle ${\mathcal O}(3H_0+2H_3)$ on $\PP$. And this
line bundle is very ample, that is,
the map $\PP\to \PP({\rm H}^0({\mathcal O}(3H_0+2H_3))^*)\cong\PP^{110}$
is a smooth embedding, as one can explicitly see.
By the Lefschetz hyperplane theorem, we find
\beq
\pi_1(Y)\cong\pi_1(\PP)=\{1\}.
\label{Ysimp}
\eeq
Next, we apply the van Kampen theorem for $\wt{Y}=(\wt{Y}-C)\cup \wt{U}_C$
and $Y=(Y-C)\cup U_C$, where $\wt{U}_C$ and $U_C$ are tubular neighborhoods
of $C$ in $\wt{Y}$ and $Y$. Writing
$\wt{S}_C=(\wt{Y}-C)\cap \wt{U}_C$ and $S_C=(Y-C)\cap U_C$, the theorem reads
\beq
\pi_1(\wt{Y})\cong \pi_1(\wt{Y}-C)\mathop{\ast}_{\pi_1(\wt{S}_C)}\pi_1(C),
\qquad
\pi_1(Y)\cong \pi_1(Y-C)\mathop{\ast}_{\pi_1(S_C)}\pi_1(C).
\label{vanKampen}
\eeq
Note that $\wt{S}_C$ and $S_C$ are homotopy equivalent to
 $S^3$ and $S^3/\Z_2$ bundles on $C$, and hence we have exact sequences
\beq
1\to\pi_1(\wt{S}_C)\to\pi_1(C)\to 1,\qquad
\Z_2\to \pi_1(S_C)\to\pi_1(C)\to 1.
\label{hES}
\eeq
From (\ref{vanKampen}) and (\ref{hES}), we find
\beq
\pi_1(\wt{Y})\cong \pi_1(\wt{Y}-C),\qquad
\pi_1(Y)\cong {\pi_1(Y-C)\over {\rm Im}({\rm Ker}\,(\pi_1(S_C)\to \pi_1(C)))}.
\label{Isms}
\eeq
Since $\wt{Y}-C$ is a smooth double cover of $Y-C$, we have
$|\pi_1(Y-C)|=2|\pi_1(\wt{Y}-C)|$, and note also that $|\pi_1(Y)|=1$ by
(\ref{Ysimp}). By the second isomorphism of (\ref{Isms}),
we find
\beq
1={2|\pi_1(\wt{Y}-C)|\over |{\rm Im}({\rm Ker}\,(\pi_1(S_C)\to \pi_1(C)))|}.
\eeq
By the second exact sequence of (\ref{hES}), the denominator is at most $2$.
The only possibility is that the denominator is indeed $2$ and that
$|\pi_1(\wt{Y}-C)|=1$, that is, $\pi_1(\wt{Y}-C)=\{1\}$.
By the first isomorphism of (\ref{Isms}), we obtain
\beq
\pi_1(\wt{Y})=\{1\}.
\label{wtYsimp}
\eeq
This is what we wanted to show. As a corollary of this, we also obtain
$\mathrm{H}_1(\wt{Y},\Z)=\{0\}$ and $h^{1,0}(\wt{Y})=h^{2,0}(\wt{Y})=0$.

\subsubsection{Proof of $p\not\in F_{\rm IV}$ in the dual model}\label{subsub:detail}

Here
we show that the vacuum equations of the dual model also require
$p\not\in F_{\rm IV}$ in phase IV, that is,
\beq
p^{5,6}\ne 0\quad\mbox{and}\quad p^{1\ldots 4}\ne 0.
\eeq
We recall that the D-term equations are
\beqa
&&-2|p^{1\ldots 4}|^2-|p^5|^2+|p^6|^2-|\wt{x}^{1\ldots 4}|^2
+2|s_{ij}|^2+|s_{i5}|^2=\wt{\zeta}_1.
\label{U11D}\\
&&-|p^5|^2-|p^6|^2-|\wt{x}^5|^2+|s_{i5}|^2+2|s_{55}|^2=\wt{\zeta}_2,
\label{U12D}\\
&&\sum_{I=1}^5\left(\overline{x}_a^Ix_b^I-\overline{x}_b^Ix_a^I\right)=0,
\quad a,b=1,\ldots, 4,
\label{O4D}
\eeqa
and the F-term equations are
\beqa
&&S^{ij}_ks_{ij}+S^{5j}_kp^6s_{5j}+S^{55}_k(p^6)^2s_{55}=0,
\quad k=1,\ldots, 4,\label{FF1}\\
&&S^{5j}_5s_{5j}+S^{55}_5p^6s_{55}=0,\label{FF2}\\
&&S^{5j}_kp^ks_{5j}+(S^{55}_5p^5+2S^{55}_kp^kp^6)s_{55}=0,\label{FF3}\\
&&S^{IJ}(p)+(\wt{x}^I\wt{x}^J)=0,\quad I,J=1,\ldots, 5,\label{FF4}\\
&&s_{IJ}\wt{x}^J=0,\quad I=1,\ldots, 5.\label{FF5}
\eeqa
In phase IV where $\wt{\zeta}_1<\wt{\zeta}_2<0$,
the equations (\ref{U11D})
and (\ref{U12D}) require
\beq
(p^{5,6},\wt{x}^5)\ne 0,\qquad
(p^{1\ldots 4},\wt{x}^{1\ldots 4},s_{55})\ne 0,\qquad
(p^{1\ldots 5},\wt{x}^{1\ldots 4})\ne 0.
\label{firstineq}
\eeq

Suppose $p^{5,6}=0$. Then, (\ref{firstineq})
implies $\wt{x}^5\ne 0$, and the F-term equations imply
that $(\wt{x}^5\wt{x}^I)=0$ for all $I$.
Since $(\wt{x}^5\wt{x}^5)=0$, we can find a real orthogonal frame
with respect to which
$\wt{x}^5=(c_5,c_5i,0,0)^T$ for some $c_5\ne 0$,
and the other orthogonality means that
$\wt{x}^j=(c_j,c_ji,\ast,\ast)$ for some $c_j\in \C$ for
$j=1,\ldots, 4$. Then, the $SO(4)$ D-term equation with $(a,b)=(1,2)$
reads
\beq
0=\sum_{I=1}^5\overline{c}_I\cdot (c_Ii)-\sum_{I=1}^5
\overline{c_Ii}\cdot c_I=2i\sum_{I=1}^5|c_I|^2
\eeq
which is impossible since $c_5\ne 0$. This proves that $p^{5,6}\ne 0$.

Suppose $p^{1\ldots 4}=0$. Then the F-term equations would
imply $(\wt{x}^i\wt{x}^j)=0$ for $i,j=1,\ldots, 4$.
We would also have $p^5\ne 0$. To show this,
let us suppose otherwise, i.e. $p^5=0$ in addition to $p^{1\ldots 4}$.
Then, the last of (\ref{firstineq}) would mean that $\wt{x}^{1\ldots 4}\ne 0$.
At the same time, (\ref{FF4}) would mean
$(\wt{x}^I\wt{x}^J)=0$ but then there is no other solution
to the $SO(4)$ D-term equations than $\wt{x}^I=0$, in contradiction to
$\wt{x}^{1\ldots 4}\ne 0$. Recall that $S^{5i}_5\ne 0$ follows from (C).
We may assume that $S^{5i}_5=c\delta^i_1$ for $c\ne 0$ by a change of coordinates
if necessary, so that (\ref{FF4}) reads
\beq
(\wt{x}^i\wt{x}^j)=0,\qquad
(\wt{x}^i\wt{x}^5)=-c\,\delta^i_1p^5\,\,(\,\ne\, 0).
\eeq
Then, we can find a real orthonormal frame
with respect to which
\beqa
&&\wt{x}^1=(c_1,c_1i,0,0)^T,\nn\\
&&\wt{x}^j=(c_j,c_ji,d_j,d_ji)^T~~\mbox{for}~~j=2,3,4,\nn\\
&&\wt{x}^5=(a_5,b_5,c_5,d_5)^T,
\eeqa
with
\beqa
&&c_1(a_5+ib_5)\ne 0,\label{nonzz}\\
&&c_j(a_5+ib_5)+d_j(c_5+id_5)=0\quad\mbox{for $j=2,3,4$.}
\eeqa
The F-term equations (\ref{FF5}) then imply
\beq
\sum_{j=1}^4s_{Ij}c_j+s_{I5}a_5=i\sum_{j=1}^4s_{Ij}c_j+s_{I5}b_5=0
\eeq
from which it follows that
$s_{I5}(a_5+ib_5)=0$.
Since we know that $a_5+ib_5\ne 0$ from (\ref{nonzz}), this means that
$s_{I5}=0$. In particular, $s_{55}=0$. Then, the difference of
(\ref{U11D}) and (\ref{U12D}) reads
\beq
2|p^6|^2-|\wt{x}^{1\ldots 4}|^2+|\wt{x}^5|^2+2|s_{ij}|^2
=\wt{\zeta}_1-\wt{\zeta}_2<0,
\eeq
which implies
\beq
|\wt{x}^{1\ldots 4}|^2>|\wt{x}^5|^2.
\label{1big5}
\eeq
On the other hand, the $SO(4)$ D-term equations (\ref{O4D})
for $(a,b)=(1,2)$ and $(3,4)$ read
\beq
2i\sum_{j=1}^4|c_j|^2+\overline{a}_5b_5-\overline{b}_5a_5=0,
\qquad2i\sum_{j=2}^4|d_j|^2+\overline{c}_5d_5-\overline{d}_5c_5=0,
\eeq
which imply
\beq
|\wt{x}^{1\ldots 4}|^2\leq |\wt{x}^5|^2.
\label{1leq5}
\eeq
(\ref{1big5}) and (\ref{1leq5}) contradict.
This completes the proof that $p^{1\ldots 4}\ne 0$.

\subsection{Phase I${}_-$}

In phase I${}_-$ where $\zeta_2>\zeta_1>0$, the D-term equations forbid
\beqa
F_{\rm I_-}&=&\{u_{1\ldots 4}=v_{1\ldots 4}=p^6=0\}\cup\{u_5=p^{1\ldots 4}=0\}
\cup\{v_5=p^{1\ldots 4}=0\}\nn\\
&&~~~~~\cup\{u_5=v_5=0\}\cup
\{u_{1\ldots 5}=0\}\cup \{v_{1\ldots 5}=0\}.
\label{FI-}
\eeqa
The identity component $G_0$ of the gauge group acts freely
on the space of solutions to the D-term equations,
and the quotient defines a smooth non-compact
toric variety $V_{\rm I_-}=(V-F_{\rm I_-})/G_{0\C}$.
Under Condition (C), we may impose the constraint
\beq
p^5=S_5(u,v,p^6)=0.\label{pS5}
\eeq
Let $X_{\rm I_-}$ be the $\Z_2=G/G_0$ quotient of the locus (\ref{pS5})
in $V_{\rm I_-}$.
The theory reduces at low energies to the hybrid LG/sigma model
with target $X_{\rm I_-}$ and the superpotential $W_{\rm I_-}$ induced from $W$.

The critical locus of $W_{\rm I_-}$
is the union of components $Z$, $C_1,\ldots, C_{10}$
where $Z$ is a Calabi-Yau threefold (quotient of the
intersection of four quadrics in $\PP^7$ by a free involution)
and the $C_i$ are rational curves.
$Z$ has ten conifold points at which the $C_i$'s intersect.
The $U(1)_V$ R-symmetry acts trivially on $Z$ but non-trivially on
the $C_i$'s. This suggests that the model is a bad hybrid (or pseudo hybrid)
in the sense of \cite{Aspinwall:2015zia}.

\subsection{Phase II}

In phase II where $\zeta_2<0,\zeta_1+\zeta_2>0$, the D-term equations forbid
\beq
F_{\rm II}=\{p^{5,6}=0\}\cup\{p^6=u_{1\ldots 4}=0\}\cup
\{p^6=v_{1\ldots 4}=0\}\cup\{u_{1\ldots 5}=0\}\cup\{v_{1\ldots 5}=0\}.
\label{FII}
\eeq
The group $G_0$ acts freely
on the space of solutions to the D-term equations,
and the quotient defines a smooth non-compact
toric variety $V_{\rm II}=(V-F_{\rm II})/G_{0\C}$.
Under Condition (C), we may impose the constraint
\beq
p^{1\ldots 4}=S_{1\ldots 4}(u,v,p^6)=0.\label{pS1234}
\eeq
Let $X_{\rm II}$ be the $\Z_2=G/G_0$ quotient of the locus
(\ref{pS1234}) in $V_{\rm II}$.
The theory reduces at low energies to the
hybrid model with target $X_{\rm II}$ and the superpotential $W_{\rm II}$
induced from $W$.

The critical locus of $W_{\rm II}$
is the union of components $Z'$, $C'_1,\ldots, C'_6$
where $Z'$ is a Calabi-Yau threefold ($\Z_2$-quotient of the
intersection of five symmetric bilinears in $\PP^4\times \PP^4$)
and the $C'_i$ are rational curves.
$Z'$ has six conifold points at which the $C'_i$'s intersect.
The $U(1)_V$ R-symmetry acts trivially on $Z'$ but non-trivially on
the $C'_i$'s. This suggests that the model is a bad hybrid.

\subsection{Phase III}

In phase III where $\zeta_1+\zeta_2<0, \zeta_1-\zeta_2>0$,
the D-term equations forbid
\beq
F_{\rm III}=\{p^{5,6}=0\}\cup\{p^{1\ldots 5}=0\}\cup\{u_{1\ldots 4}=p^6=0\}
\cup\{v_{1\ldots 4}=p^6=0\}.
\label{FIII}
\eeq
It is possible to have $u=v=0$ at which the F-term equations are all
satisfied. At this locus, the $O(2)$ subgroup of the gauge group
is unbroken.
There is also a locus, such as
$p^{1\ldots 4}=u_5=v_5=0$, $p^5\ne 0$, $u_{1\ldots 4}\ne 0$,
$v_{1\ldots 4}\ne 0$, which satisfies the F-term equations if
$S^{ij}_ku_iv_j=0$, $S^{5j}_5u_j=S^{5j}_5v_j=0$.
There the gauge group is completely broken.
Therefore, there is a mixture of strongly coupled vacua and
weakly coupled vacua, and it is not easy to tell what the low energy theory
is.

The dual theory cannot be of help.
The weakly coupled vacua in the original theory correspond to strongly coupled
vacua in the dual. Indeed, for $p^{1\ldots 4}=0$ and
$p^5\ne 0$, the matrix $S(p)$ is of the form
\beq
S(p)=\left(\begin{array}{cc}
0_{4\times 4}&S^{i5}_5p^5\\
S^{5j}_5p^5&S^{55}_5p^5p^6
\end{array}\right)
\eeq
and has rank 2. It is possible to find vacua with such $p$'s,
and in such a vacuum $(\wt{x}^I_a)$ has rank 2. They are strongly coupled vacua
with an unbroken $SO(2)$ subgroup of $SO(4)\subset \wt{G}$.

To summarize, we are unable to find the nature of the low energy theory in
phase III.

\subsection{Phase V}

In phase V where $\zeta_1<0,\zeta_2>0$, the D-term equations forbid
\beqa
F_{\rm V}&=&\{p^{1\ldots 5}=0\}\cup\{u_5=v_5=0\}\cup\{u_{1\ldots 5}=0\}
\cup\{v_{1\ldots 5}=0\}\nn\\
&&~~~~\cup\{p^{1\ldots 4}=u_5=0\}\cup\{p^{1\ldots 4}=v_5=0\}.
\label{FV}
\eeqa
The group $G_0$ acts freely
on the space of solutions to the D-term equations
except at the locus
$p^5=p^6=u_{1\ldots 4}=v_{1\ldots 4}=0$ with the stabilizer
$\{(1,\pm 1,\pm 1)\}\cong \Z_2$, and
the quotient $V_{\rm V}=(V-F_{\rm V})/G_{0\C}$ is
a non-compact toric orbifold.
Under Condition (C), we may impose the constraint (\ref{pS5}).
Let $X_{\rm V}$ be the $\Z_2=G/G_0$ quotient of the locus (\ref{pS5})
in $V_{\rm V}$.
The theory reduces at low energies to
the hybrid model with target $X_{\rm V}$ and the superpotential $W_{\rm V}$
induced from $W$.

The critical locus of $W_{\rm V}$
is the union of components $Z''$, $C''_1,\ldots, C''_{10}$
where $Z''$ is the $\Z_2=G/G_0$ quotient of
a weighted projective space $\PP^3_{[2222]}$ and the $C''_i$ are
teardrops $\PP^1_{[21]}$.
$Z''$ intersects with each $C''_i$ at the $\Z_2$ point of the latter.
The $U(1)_V$ R-symmetry acts trivially on $Z''$ but non-trivially on
the $C''_i$'s. This suggests that the model is a bad hybrid.

%%%%%%%%%%%%%%%%%%%%%%%%%%%%%%%%%%%%%%%%%%%%%%%%%%%%%%%%%%%%%%%%%%%%%%%%%%%
\section{K\"ahler Moduli}
\label{sec-kahler}
In this section and the next, we determine the regularity conditions 
of the parameters of the system. The criterion is that there is no non-compact
flat direction in the effective target space \cite{Witten:1993yc}.
In this section, we focus on the condition on the FI-theta parameters,
$t_1$ and $t_2$. We shall identify the discriminant locus where there is
a non-compact Coulomb branch or mixed Coulomb-Higgs branch
\cite{Witten:1993yc,Morrison:1994fr}. 
The complement is the ``K\"ahler moduli space'' $\mathfrak{M}_K$,
that is, the space of exactly marginal twisted chiral parameters
of the infra-red SCFTs.
In Section~\ref{sec-pf}, we shall revisit the same problem
by examining the Picard-Fuchs operator.

%%%%%%%%%%%%%%%%%%%%%%%%%%%%%%%%%%%%%%%%%%%%%%%%%%%%%%%%%%%%%%%%%%%%%%%%%
\subsection{The original model}
\label{sec-coulomb}
It is convenient to work with the scalar components
$(\sigma_0,\sigma_3,\sigma_4)$ of the vector multiplet for
$U(1)_0\times U(1)_3\times U(1)_4$ since the corresponding
theta angle has simple periodicity, i.e., $(2\pi \Z)^{\oplus 3}$.
They are related to the ones $(\sigma_1,\sigma_2,\sigma_h)$ for
$U(1)_1\times U(1)_2\times SO(2)$ via (\ref{o2toruspar}) as
$\sigma_0=2\sigma_1$, $\sigma_3=\sigma_1+\sigma_2$ and
$\sigma_4=\sigma_1+\sigma_h$.
The $O(2)$ Weyl reflection 
$(\sigma_1,\sigma_2,\sigma_h)\mapsto (\sigma_1,\sigma_2,-\sigma_h)$
is given by
$(\sigma_0,\sigma_3,\sigma_4)\mapsto(\sigma_0,\sigma_3,\sigma_0-\sigma_4)$. If
we write $t_1$ and $t_2$ for the FI-theta parameter of $U(1)_1\times
U(1)_2$, the tree level twisted superpotential is
\begin{align}
\widetilde{W}_{\it tree}=-t_1\sigma_1-t_2\sigma_2
=-\frac{t_1-t_2}{2}\sigma_0-t_2\sigma_3,
\end{align}
where for this choice of gauge group
$\widetilde{W}_{\Theta}(\sigma)=0$. The periodicity of these
parameters is $({\frac{t_1-t_2}{2}},t_2)\equiv
(\frac{t_1-t_2}{2},t_2)+2\pi i (n,m)$ for $n$ and $m$ integers. It
would be more appropriate to use $t_0$ and $t_3$ defined by
$\widetilde{W}_{tree}=-t_0\sigma_0-t_3\sigma_3$, but we informally use
$t_1$ and $t_2$ as they are convenient to compare with the phase
diagram \ref{fig-twoparphases}.

Since $O(2)$ does not have any roots, the quantum corrections
given in (\ref{coulombweff}) come only from integrating out the massive
matter fields. With that, the vacuum equations
$\partial_{\sigma_a}\widetilde{W}_{\it eff} \equiv 0$ (mod $2\pi
i\mathbb{Z}$) yield
\begin{eqnarray}
&\displaystyle
\e^{-\frac{t_1-t_2}{2}}=\frac{(\sigma_0-\sigma_4)^4(\sigma_0-\sigma_3)}
{(-\sigma_0)^4(-\sigma_0+\sigma_3+\sigma_4)},\quad
\e^{-t_2}=\frac{(-\sigma_0+\sigma_3+\sigma_4)(\sigma_3-\sigma_4)}
{(-\sigma_3)(\sigma_0-\sigma_3)},\nonumber\\
&\displaystyle
\frac{\sigma_4^4(-\sigma_0+\sigma_3+\sigma_4)}
{(\sigma_0-\sigma_4)^4(\sigma_3-\sigma_4)}=1.
\label{C2paraO2}
\end{eqnarray}
For $y:=\sigma_3/\sigma_0$ and $z:=\sigma_4/\sigma_0$, %the $O(2)$ Weyl
%group action is $(y,z)\mapsto (y,1-z)$, 
and the equations (\ref{C2paraO2}) read
\begin{equation}
\e^{-\frac{t_1-t_2}{2}}=\frac{(1-z)^4(1-y)}{-1+y+z},\quad
\e^{-t_2}=\frac{(-1+y+z)(y-z)}{-y(1-y)},\quad
\frac{z^4(-1+y+z)}{(1-z)^4(y-z)}=1.
\label{deff}
\end{equation}
The last equation factorizes as
\begin{equation}
\label{fzdef}
(2z-1)(y-f(z))=0,\qquad
f(z):=\frac{z(1-z)(z^2-z+1)}{2z^2-2z+1},
\end{equation}
and there are two solutions: (i) $2z-1=0$
and (ii) $y=f(z)$. For these we have
\beqa
\mbox{(i)}&&
\e^{-\frac{t_1-t_2}{2}}=2^{-3}\frac{1-y}{2y-1},\qquad
\e^{-\frac{t_1+t_2}{2}}=-2^{-5}\frac{2y-1}{y},\label{B1}\\
%\\&&\e^{-t_1}=-2^{-8}{1-y\over y},\qquad
%\e^{-t_2}=-{\left(y-{1\over 2}\right)^2\over y(1-y)}.\nn
\mbox{(ii)}&&
\e^{-\frac{t_1-t_2}{2}}=-\upsilon^2+3\upsilon-1,\qquad
\e^{-\frac{t_1+t_2}{2}}={\upsilon^3\over 1-\upsilon},\label{B2}
%\\&&\e^{-t_1}=-{(1-z)^3u^3p_4(z)\over z^2-z+1},\qquad
%\e^{-t_2}={(1-z)^3z^3\over (z^2-z+1)p_4(z)}\nn
\eeqa
where we used the Weyl invariant $\upsilon=z(1-z)$ in the latter.
Equations (\ref{B1}) and (\ref{B2}) show the location of the
singularity. This encodes an amoeba whose spines reproduce part of the
classical phase boundaries. Let us see how this works explicitly. We
start with the component (i). At $y=1$ we have
$e^{-\frac{t_1-t_2}{2}}=0,e^{-\frac{t_1+t_2}{2}}=const.$ which implies
$t_1=-t_2>0$. This is the phase boundary separating phase II and
III. For $y=\frac{1}{2}$,
$e^{-\frac{t_1-t_2}{2}}=\infty,e^{-\frac{t_1+t_2}{2}}=0$ and therefore
$t_1=0, t_2>0$. This is the phase boundary between phase V and phase
I${}_-$. Finally, $y=0$ yields the phase boundary $t_1=t_2<0$
separating phases III and IV. On the component (ii) the roots of 
$\upsilon^2-3\upsilon+1$
correspond to the phase boundary separating phases II and III. $\upsilon=0$
corresponds to the phase boundary between phase I${}_-$ and
I${}_+$. Another end of the second component is
at $\upsilon=\infty$ which supplies the phase boundary $t_1<0,t_2=0$ between
phase IV and V. Finally, at $\upsilon=1$ we get
$e^{-\frac{t_1-t_2}{2}}=const.,e^{-\frac{t_1+t_2}{2}}=\infty$ and
therefore $t_1=t_2<0$. This is the boundary between phases III and IV.

The boundary between phases I${}_+$ and II is missing. 
In fact, there is an additional discriminant locus 
associated to the mixed Coulomb-Higgs branch where $U(1)_3$ is unbroken
and $\sigma_3$ is arbitrarily large.
On this branch, $u_5,v_5,p^5,p^6$ are heavy and
should be integrated out.
This yields the following effective FI-theta parameters
\begin{align}
t^0_{\it eff}(\sigma)=&\frac{t_1-t_2}{2}+\log (\sigma_0-\sigma_3)
-\log(-\sigma_0+\sigma_3+\sigma_4)\nonumber\\
=&\frac{t_1-t_2}{2}+\pi i 
+\log\left[\left(1-\frac{\sigma_0}{\sigma_3}\right)
\left(1-\frac{\sigma_0-\sigma_4}{\sigma_3}\right)^{-1}\right],
\label{effFImixed0}\\
t^3_{\it eff}(\sigma)=&t_2
+\log\left[\left(1-\frac{\sigma_0-\sigma_4}{\sigma_3}\right)
\left(1-\frac{\sigma_4}{\sigma_3}\right)
\left(1-\frac{\sigma_0}{\sigma_3}\right)^{-1}\right],\label{effFImixed3}\\
t^4_{\it eff}(\sigma)=&\log\left[
\left(1-\frac{\sigma_0-\sigma_4}{\sigma_3}\right)
\left(1-\frac{\sigma_4}{\sigma_3}\right)^{-1}\right].
\label{effFImixed4}
\end{align}
The scalar potential of the effective theory is
\begin{align}
U_{\it eff}=&\left|\sigma_4 u_{1...4}\right|^2
+\left|(\sigma_0-\sigma_4)v_{1...4}\right|^2
+\left|-\sigma_0 p^{1...4}\right|^2\nonumber\\
&+\frac{1}{2}\sum_{a,b}
(e_{\it eff})^2_{ab}
\bigl(\mu^a_{\it eff}-\zeta^a_{\it eff}(\sigma)\bigr)
\bigl(\mu^b_{\it eff}-\zeta^b_{\it eff}(\sigma)\bigr)
\nonumber\\
&+\sum_k\left|\sum_{i,j}S^{ij}_ku_iv_j\right|^2
+\sum_j\left|\sum_{k,i}S^{ij}_kp^ku_i\right|^2
+\sum_i\left|\sum_{k,j}S^{ij}_kp^kv_j\right|^2.
\end{align}
Here $\zeta^a_{\it eff}(\sigma):={\rm Re}\,t^a_{\it eff}(\sigma)$ and
\begin{align}
\mu^0_{\it eff}=|v_{1...4}|^2-|p^{1...4}|^2,\qquad
\mu^3_{\it eff}=0,\qquad
\mu^4_{\it eff}=|u_{1...4}|^2-|v_{1...4}|^2.
\end{align}
There are also theta angles $\theta^a_{\it eff}(\sigma):=-{\rm
  Im}\,t^a_{\it eff}(\sigma)$. 
When 
\begin{align}
\mbox{(iii)${}_+$}\qquad
\e^{-t_2}=1,\quad\zeta_1\gg 0,\qquad\qquad
\label{extcomp}
\end{align}
the effective theory at arbitrarily large $\sigma_3$ has supersymmetric
vacua in which $u_{1\ldots 4}$ and $v_{1\ldots 4}$ are both non-zero,
breaking the gauge symmetry to $U(1)_3$ and forcing $\sigma_0=\sigma_4=0$.
That is, there is a non-compact mixed Coulomb-Higgs branch.
Thus, we need to include
(\ref{extcomp}) as a part of
the discriminant locus, which accounts for the missing phase boundary.
Since the discriminant locus must be an analytic subspace, we expect that
the condition $\zeta_1\gg 0$ can be removed.
There are indeed supersymmetric vacua in the opposite regime $\zeta_1\ll 0$
of the same line $\e^{-t_2}=1$, but all or most of them
have $u_{1\ldots 4}=v_{1\ldots 4}=0$, leaving
the $O(2)$ subgroup also unbroken. The theory is strongly coupled and
the classical analysis is not reliable.
To be sure, for now we count only (\ref{extcomp}) as a part of
the discriminant locus. We shall reconsider the other region $\zeta_1\ll 0$
in the dual model in Sections~\ref{subsec:dualC} and \ref{subsec:detailonmixed}
and directly in Section~\ref{subsec:detailonmixed}.

Apart from this, there are no further mixed branches. To confirm this,
one has to systematically analyze all field configurations where a
continuous subgroup of the gauge group is unbroken.  To
illustrate how a situation where there is no mixed branch manifests
itself, we consider the situation where $U(1)_0$ is unbroken and
$\sigma_0$ can become arbitrarily large. This happens when
$p^{1...4}$, $p^6$, $u^5$ and $v_{1...4}$ are heavy and can
be integrated out. In this case the scalar potential of the effective
theory is
\begin{align}
\label{eff-potential}
U_{\it eff}=&\left|-\sigma_3 p^5\right|^2
+\left|\sigma_4u_{1...4}\right|^2
+\left|(\sigma_3-\sigma_4) v_5\right|^2\nonumber\\
&+\frac{1}{2}\sum_{a,b}
(e_{\it eff})^{2,{ab}}
\left(\mu^a_{\it eff}-\zeta^a_{\it eff}(\sigma)\right)
\left(\mu^b_{\it eff}-\zeta^b_{\it eff}(\sigma)\right)
\nonumber\\
&+\left|\sum_{j}S^{j5}_5u_jv_5\right|^2+\sum_j\left|S^{j5}_5p^5v_5\right|^2
+\left|\sum_i S^{i5}_5p^5u_i \right|^2,
\end{align}
with
\begin{align}
\mu^0_{\it eff}=0,\qquad
\mu^3_{\it eff}=-|p^5|^2+|v_5|^2,\qquad
\mu^4_{\it eff}=|u_{1...4}|^2-|v_{5}|^2
\end{align}
and
\begin{align}
t^0_{\it eff}(\sigma)=&\frac{t_1-t_2}{2}+\pi i 
+\log\left[\left(1-\frac{\sigma_4}{\sigma_0}\right)^4
\left(1-\frac{\sigma_3}{\sigma_0}\right)
\left(1-\frac{\sigma_3+\sigma_4}{\sigma_0}\right)^{-1}\right],\nonumber\\
t^3_{\it eff}(\sigma)=&t_2
+\pi i+\log\left[\left(1-\frac{\sigma_3+\sigma_4}{\sigma_0}\right)
\left(1-\frac{\sigma_3}{\sigma_0}\right)^{-1}\right],\nonumber\\
t^4_{\it eff}(\sigma)=&\pi i+\log\left[
\sigma_0^{-3}\left(1-\frac{\sigma_3+\sigma_4}{\sigma_0}\right)
\left(1-\frac{\sigma_4}{\sigma_0}\right)^{-4}\right].
\end{align}
We are looking for a mixed branch where only
 $U(1)_0$ is unbroken,
so that $\sigma_3=\sigma_4=0$ while
$\sigma_0$ can have arbitrary values. There the effective D-term equations
reduce to
\begin{align}
\label{eff-dterms}
0=&\frac{\zeta_1-\zeta_2}{2}\nonumber\\
-|p^5|^2+|v_5|^2=&\zeta_2\nonumber\\
|u_{1...4}|^2-|v_5|^2=&\log\frac{1}{|\sigma_0|^3}.
\end{align} 
It is straightforward to show that it is not possible to get
$U_{eff}=0$ for arbitrary $\sigma_0$. By the first D-term equation, we
have $\zeta_1=\zeta_2$ and we distinguish between the two cases
$\zeta_2>0$ and $\zeta_2<0$. If $\zeta_2>0$, by the second equation of
(\ref{eff-dterms}), we have $v_5\neq 0$, but then the F-term equations
in the last line of (\ref{eff-potential}) enforce $p^5=0$. Then the
second D-term equation reduces to $|v_5|^2=\zeta_2$ and the third one
becomes
\begin{align}
|u_{1...4}|^2-\zeta_2=\log\frac{1}{|\sigma_0|^3}.
\end{align}
For a fixed $\zeta_2$ it is impossible to satisfy this
equation for an arbitrarily large $\sigma_0$. For $\zeta_2<0$ the
second equation in (\ref{eff-dterms}) yields $p^5\neq 0$. From the
F-term equations it then follows that $v_5=0$. Then the last equation
in (\ref{eff-dterms}) reduces to
\begin{align}
|u_{1...4}|^2=\log\frac{1}{|\sigma_0|^3},
\end{align}
which cannot be satisfied for large $\sigma_0$. Hence we conclude that
there is no mixed branch with unbroken $U(1)_0$.

A systematic analysis of all possibilities of unbroken $U(1)$s shows
that there are no further mixed branches.
Note that no work is necessary to show that there is no mixed
branch with unbroken $U(1)_4$.
When $\sigma_4$ is large, one integrates out the fields
that are charged under $U(1)_4$. These are the $u_I$ and $ v_I$. The
effective potential is the same as for an $O(2)$-theory with five
fundamentals. In \cite{Hori:2011pd} it was shown that $O(k)$-theories
with $N$ fundamentals with trivial theta angle
have no Coulomb branch
if and only if $N-k$ is odd. This is indeed the case here, and
therefore there is no mixed branch with unbroken $U(1)_4$.

\subsection{The dual model}\label{subsec:dualC}

\newcommand{\wtz}{\wt{z}}
\newcommand{\wtu}{\wt{\upsilon}}

Let us next identify the discriminant locus in the dual model.
As a maximal torus of the gauge group $\wt{G}$, we take
\beqa
{U(1)_1\times U(1)_2\times SO(2)\times SO(2)\over
\{(\pm 1,\pm 1,\pm {\bf 1}_2,\pm {\bf 1}_2)\}}\!\!&\cong&\!\!
U(1)_0\!\times \! U(1)_3\!\times \!U(1)_4\!\times\! U(1)_5
\label{dualMT}\\
{[}(z_1,z_2,h_1,h_2){]}~~~&\longmapsto&~~~
(z_1^2,z_1z_2,z_1h_1,z_1h_2)
\nn
\eeqa
We shall work with the fields $(\sigma_0,\sigma_1,\sigma_4,\sigma_5)$
corresponding to the group on the right hand side of (\ref{dualMT}).
The $SO(4)$ Weyl group acts on them as
\beq
(\sigma_0,\sigma_3,\sigma_4,\sigma_5)\to
(\sigma_0,\sigma_3,\sigma_5,\sigma_4),~
(\sigma_0,\sigma_3,\sigma_0-\sigma_4,\sigma_0-\sigma_5).
\eeq
In the Coulomb branch analysis, we disregard possible solutions to
the vacuum equation at the fixed points:\footnote{This is an empirical rule
which can sometimes be supported by detailed argument \cite{Hori:2006dk}.
See \cite{Aharony:2016jki} for a recent proposal on this point.}
\beq
\sigma_4=\sigma_5,\qquad
\sigma_4+\sigma_5=\sigma_0.
\eeq
The tree level twisted superpotential is
\beq
\widetilde{W}_{\it tree}=-\wt{t}_1\sigma_1-\wt{t}_2\sigma_2
=-{\wt{t}_1-\wt{t}_2\over 2}\sigma_0-\wt{t}_2\sigma_3.
 \eeq
The $U(1)_0\!\times \! U(1)_3\!\times \!U(1)_4\!\times\! U(1)_5$ charges
of the fields are
\beq
\begin{array}{|cc|cc|cc|cc|}
\hline
p^{1...4}&(-1,0,0,0)&
\wt{u}_{\uparrow}^{1...4}&(-1,0,1,0)&\wt{u}_{\uparrow}^5&(0,-1,1,0)&
s_{ij}&(1,0,0,0)\\
p^5&(0,-1,0,0)&
\wt{v}_{\uparrow}^{1...4}&(0,0,-1,0)&\wt{v}_{\uparrow}^5&(1,-1,-1,0)&
s_{i5}&(0,1,0,0)\\
p^6&(1,-1,0,0)&
\wt{u}_{\downarrow}^{1...4}&(-1,0,0,1)&\wt{u}_{\downarrow}^5&(0,-1,0,1)&
s_{55}&(-1,2,0,0)\\
&&\wt{v}_{\downarrow}^{1...4}&(0,0,0,-1)&\wt{v}_{\downarrow}^5&(1,-1,0,-1)
&&
\\
\hline
\end{array}
\eeq
The equations that determine the full Coulomb branch are
\beqa
&&\e^{-{\wt{t}_1-\wt{t}_2\over 2}}={\sigma_0^{10}(\sigma_0-\sigma_3-\sigma_4)
(\sigma_0-\sigma_3-\sigma_5)(\sigma_0-\sigma_3)\over
(-\sigma_0+\sigma_4)^4(-\sigma_0+\sigma_5)^4(-\sigma_0)^4
(-\sigma_0+2\sigma_3)}\nn\\
&&\e^{-\wt{t}_2}={\sigma_3^4(-\sigma_0+2\sigma_3)^2\over
(-\sigma_3+\sigma_4)(\sigma_0-\sigma_3-\sigma_4)
(-\sigma_3+\sigma_5)(\sigma_0-\sigma_3-\sigma_5)
(-\sigma_3)(\sigma_0-\sigma_3)}\nn\\
&&1={(-\sigma_0+\sigma_4)^4(-\sigma_3+\sigma_4)\over
(-\sigma_4)^4(\sigma_0-\sigma_3-\sigma_4)},\quad
1={(-\sigma_0+\sigma_5)^4(-\sigma_3+\sigma_5)\over
(-\sigma_5)^4(\sigma_0-\sigma_3-\sigma_5)}
\label{dCeqn}
\eeqa
For $y:=\sigma_3/\sigma_0$, $z:=\sigma_4/\sigma_0$,
$w:=\sigma_5/\sigma_0$, the $SO(4)$ Weyl group action
is
\beq
(y,z,w)\to (y,w,z),~(y,1-z,1-w),
\eeq
and the disregarded locus is
\beq
z=w,\quad z+w=1.
\eeq
The last two equations of (\ref{dCeqn}) read
\beq
(2z-1)\left(y-f(z)\right)
=(2w-1)\left(y-f(w)\right)=0,
\eeq
where $f(z)$ is as in (\ref{fzdef}).
Note that
\beq
f(z)-f(w)={1\over 2}(w-z)(z+w-1)\left\{{1\over
(2z^2-2z+1)(2w^2-2w+1)}+1\right\}.
\eeq
Since $z=w$ and $z+w=1$ are disregarded, 
we have either 
(i) $y=f(z)$ and $(2z^2-2z+1)(2w^2-2w+1)=-1$ 
or (ii) $y=f(z)$ and $2w=1$, up to the Weyl group action.
In these cases, we have (with $\wtu:=z(1-z)$)
\beqa
\mbox{(i)}&&
\e^{-{\wt{t}_1-\wt{t}_2\over 2}}={\wtu^2-3\wtu+1\over 2\wtu^2-4\wtu+1},
\quad
\e^{-{\wt{t}_1+\wt{t}_2\over 2}}=-{2\wtu^2-4\wtu+1\over \wtu(1-\wtu)},
\qquad\label{Bd1}\\
\mbox{(ii)}&&
\e^{-{\wt{t}_1-\wt{t}_2\over 2}}=-2^3 {\wtu^2-3\wtu+1\over (1-2\wtu)^2},\quad
\e^{-{\wt{t}_1+\wt{t}_2\over 2}}=
2^5  {(1-\wtu)^3\over \wtu(1-2\wtu)^2}.
\qquad
\label{Bd2}
\eeqa
There is an additional discriminant locus
associated with a mixed Coulomb-Higgs branch with $U(1)_3$ unbroken.
We look at the regime where $\sigma_3$ is large and 
all the matter fields other than $p^{1\ldots 4}$, $\wt{x}^{1\ldots 4}$ and $s_{ij}$
are integrated out. 
When
\beq
\mbox{(iii)${}_-$}\qquad\e^{-\wt{t}_2}=4,\quad\wt{\zeta}_1\ll 0,\qquad\qquad
\label{extcompd}
\eeq
the effective theory at arbitrarily large $\sigma_3$ has supersymmetric
vacua. At most of them, non-zero values of the matter fields
break the gauge symmetry to $U(1)_3$ and force
$\sigma_0=\sigma_{SO(4)}=0$.
That is, there is a non-compact mixed Coulomb-Higgs branch.
Thus, we need to include (\ref{extcompd}) as a part of the discriminant.
Again, we expect that the condition $\wt{\zeta}_1\ll 0$
can be removed. There are indeed supersymmetric vacua in the other regime
of the line $\e^{-\wt{t}_2}=4$, but all of them are strongly coupled.
Therefore, we take only (\ref{extcompd}) for now.

\subsection{Summary}

\begin{figure}
\begin{center}
\input{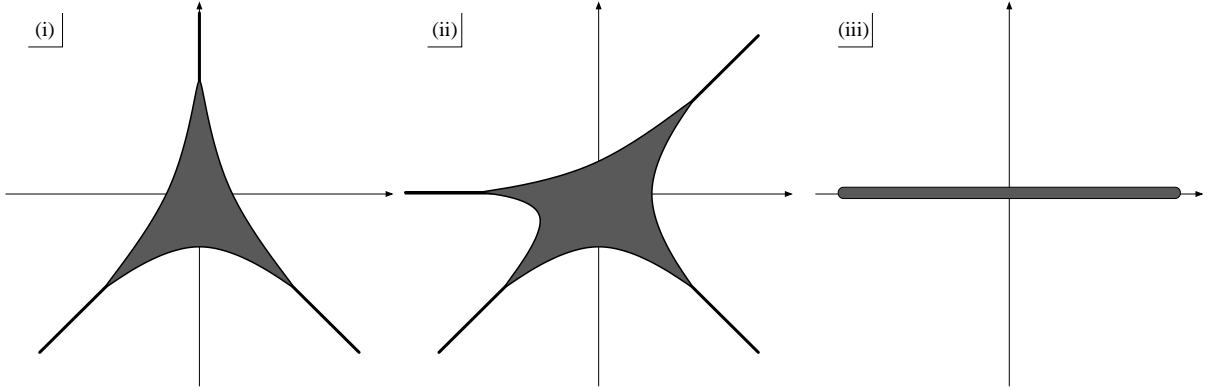}
\end{center}\caption{The three components of the discriminant locus}
\label{fig:discriminant}
\end{figure}

To summarize, as the discriminant locus,
we identified two complete components, (i) at (\ref{B1}) and (ii) at (\ref{B2}),
plus one half-line (iii)${}_+$ at (\ref{extcomp})  in the original model,
and
two complete components, (i) at (\ref{Bd1})
and (ii) at (\ref{Bd2}), plus one half-line (iii)${}_-$ at (\ref{extcompd})
 in the dual model.
Under
\beq
\e^{-{t_1-t_2\over 2}}=-2^{-3}\e^{-{\wt{t}_1-\wt{t}_2\over 2}},\qquad
\e^{-{t_1+t_2\over 2}}=-2^{-5}\e^{-{\wt{t}_1+\wt{t}_2\over 2}}.
\label{dualitymap}
\eeq
the complete components for the original and the dual
are mapped to each other by
\beqa
\mbox{(i)}&&y={\wtu(1-\wtu)\over 1-2\wtu},\\
\mbox{(ii)}&&\upsilon={1-\wtu\over 1-2\wtu},
\eeqa
while the two half-lines (iii)${}_+$ and (iii)${}_-$ are opposite
regimes of a complete line
\beq
\mbox{(iii)}\quad
\e^{-t_2}=1\Longleftrightarrow \e^{-\wt{t}_2}=4.
\qquad
\label{lineiii}
\eeq
This suggests that (\ref{dualitymap}) is the map of the parameters under
the duality, and that the complete line
(\ref{lineiii}) is indeed one component of the discriminant locus.
Altogether, the discriminant consists of the three complete
components, (i), (ii) and (iii), as shown in
Figure~\ref{fig:discriminant}.

\subsection{Some detail on the mixed branch}
\label{subsec:detailonmixed}

\newcommand{\Cmixed}{C${}_{\rm mixed}$}

Let us describe the detail of the mixed branch
supported at the component (iii) of the discriminant.
We assume the following genericity condition on $S^{ij}_k$:
\begin{description}
\item[Condition (\Cmixed):]
If $u_{1\ldots 4}\ne 0$ and $v_{1\ldots 4}\ne 0$ satisfy $S^{ij}_ku_iv_j=0$,
then the $4\times 8$ matrix $(Su,Sv)$ has rank $4$. 
\end{description}
Here $Su$ stands for
the $4\times 4$ matrix whose $(i,j)^{\rm th}$ entry is
$S^{jk}_iu_k$ (and similarly for $Sv$).
This is different form Condition (C) for the regularity of
the full theory.
We assume (\Cmixed) just for simplicity of the following discussion.
Nothing is wrong
even if it is violated --- it is just that the mixed branch 
would be non-compact also in the Higgs direction.

For $\zeta_1\gg 0$, it is best to work in the original model where the 
effective theory at large $\sigma_3$ has matter fields
$p^{1\ldots 4}$, $u_{1\ldots 4}$ and $v_{1\ldots 4}$.
The D-term equations require that both
$u_{1\ldots 4}$ and $v_{1\ldots 4}$ have non-zero values,
breaking the gauge group to $U(1)_3$.
The F-term equations read
\beq
S^{ij}_ku_iv_j=0,\quad
S^{ij}_kp^kv_j=0,\quad
S^{ij}_kp^ku_i=0.
\label{Fmixed}
\eeq
Under (\Cmixed), these require $p^{1\ldots 4}=0$.
Thus, the vacuum manifold at a fixed $\sigma_3$ is the quotient of 
\beq
\wt{X}_{\rm mixed}=\{S^{ij}_ku_iv_j=0\}\subset \PP^3\times\PP^3
\eeq 
by the exchange
$u\leftrightarrow v$. Under (\Cmixed), $\wt{X}_{\rm mixed}$ is a smooth
K3 surface on which the exchange acts freely.
Thus, the mixed branch is the product of the Coulomb branch for
$U(1)_3$ and the Higgs branch $X_{\rm mixed}=\wt{X}_{\rm mixed}/\Z_2$ 
which is an Enriques surface.

For $\zeta_1\ll 0$ (i.e. $\wt{\zeta}_1\ll 0$),
it is better to work in the dual model 
where the effective theory at large $\sigma_3$ has matter fields
$p^{1\ldots 4}$, $\wt{x}^{1\ldots 4}$ and $s_{ij}$.
The D-term equations require $p^{1\ldots 4}\ne 0$, breaking
$U(1)_0$, and
the F-term equations read
\beq
S^{ij}_ks_{ij}=0, \quad
S^{ij}(p)+(\wt{x}^i\wt{x}^j)=0,\quad
s_{ij}\wt{x}^j=0.
\label{Fmixedd}
\eeq
Under (\Cmixed), the $4\times 4$ matrix
$S^{ij}(p)$ for $p^{1\ldots 4}\ne 0$ has rank at least $2$.
Then, $S^{ij}(p)s_{jk}=0$ that follows from the latter two of (\ref{Fmixedd})
implies that $s_{jk}$ has rank at most $2$, so that one may write
$s_{jk}=u^*_jv^*_k+u^*_kv^*_j$ for some $u^*_{1\ldots 4}$ and $v^*_{1\ldots 4}$.
Then under (\Cmixed), the equations (\ref{Fmixedd}) enforce
either $u^*_{1\ldots 4}=0$ or $v^*_{1\ldots 4}=0$, hence
$s_{jk}=0$. Thus, the vacuum manifold is
\beq
\wt{Y}_{\rm mixed}=\left\{\,\, (p^{1\ldots 4}\ne 0, \wt{x}^{1\ldots 4})\,\,
\Bigl|
~\mbox{$SO(4)$ stability},~
S^{ij}(p)+(\wt{x}^i\wt{x}^j)=0~\right\}\mbox{\Large $/$}
\wt{G}_{-\C}.
\eeq
where $\wt{G}_-=(U(1)_1\times SO(4))/\{(\pm 1,\pm {\bf 1}_4)\}$.
For the part where $S^{ij}(p)$ has rank $4$ and $3$ the gauge group
$\wt{G}_{-}$ acts freely, but at the points where the rank is exactly $2$,
there is a stabilizer isomorphic to $SO(2)$. Thus, unfortunately,
there are bad points in the quotient.
Note that there is a map $\wt{Y}_{\rm mixed}\to \PP^3$ that forgets
$\wt{x}^{1\ldots 4}$. It is a double cover that is ramified over the
locus of rank $\leq 3$, i.e. $\{\det S^{ij}(p)=0\}$ which is a
singular K3 surface.

A part of this can be seen also in the original model.  The $U(1)_0$
D-term equation requires $p^{1\ldots 4}\ne 0$, breaking $U(1)_0$ and
forcing $\sigma_0=0$.  Then, under (\Cmixed), the F-term equations and
the $U(1)_4$ D-term equation force $u_{1\ldots 4}=v_{1\ldots
  4}=0$. Thus, $\zeta_1\ll 0$ is a strongly coupled phase where the
$O(2)$ is unbroken\footnote{If (\Cmixed) is violated, we may have
  special weakly coupled vacua where $u_{1\ldots 4}$ and $v_{1\ldots
    4}$ are non-zero.}.  On the generic points of $[p]\in \PP^3$ where
$S^{ij}(p)$ has rank $4$, all $x_i$'s are massive. But there is a
non-trivial twisted superpotential for the vector multiplet of $O(2)$,
as shown in (\ref{effFImixed4}), and there is an isolated critical
point at $\sigma_4=0$. Alternatively, the theory has a single $O(2)$
doublet $x_5$ with a large twisted mass $\sigma_3$. As shown in
\cite{Hori:2011pd}, such a theory has two massive vacua.  (Recall that
our $O(2)$ is $O_+(2)$, and see Eqn (3.15) in \cite{Hori:2011pd}.)
This matches with the picture obtained in the dual theory that the
effective target space is a double cover of $\PP^3$. Note that
$U(1)_0$ is Higgsed, $U(1)_3$ is in Coulomb phase and $O(2)$ is
confined.  Thus, it is a mixed Higgs-Coulomb-confining branch in the
original theory.  Similarly, in the regime $\wt{\zeta}_1\gg 0$ on
(iii), we have a mixed Higgs-Coulomb-confining branch in the dual
theory.

Note that the nature of the mixed branch is very different between the
two opposite regimes, $\zeta_1\gg 0$ and $\zeta_1\ll 0$.
Even the dimension of the effective target spaces are different ---
$X_{\rm mixed}$ is an Enriques surface
 while $\wt{Y}_{\rm mixed}$ is a (singular) Calabi-Yau threefold.
This is not an immediate problem since it is not that we have a family of
Higgs/confining branch theories in isolation
--- they appear only in the large $\sigma_3$ regimes.
But still, it is interesting to see that we can have such very
different types of
mixed branch theories on the same discriminant component.

%%%%%%%%%%%%%%%%%%%%%%%%%%%%%%%%%%%%%%%%%%%%%%%%%%%%
%%%%%%%%%%%%%%%%%%%%%%%%%

\section{Complex Moduli}
\label{sec:genericity}

In this section, we determine the regularity condition of
the superpotential so that the Higgs branch is compact
\cite{Aspinwall:2015zia,AspPleTalk}.
This is relevant to find the ``complex moduli space''
$\mathfrak{M}_C$, that is, the space of exactly marginal
chiral parameters of the infra-red SCFTs.
Note that the conditions must be the
same for all phases since the moduli space of SCFTs must be the
direct product $\mathfrak{M}_K\times \mathfrak{M}_C$.
Indeed, we shall find one condition, Condition (C), that works in all phases.
The condition is found in phase I${}_+$ straightforwardly, but
it is very non-trivial to confirm that it also works in other phases
(except in phase I${}_-$).
We shall also derive some consequences of Condition (C)
which are used in earlier sections.

\subsection{Phase I${}_+$}

Recall that the superpotential can be written
as $W=\sum_{I=1}^5p^IS_I(u,v,p^6)$, with
\beqa
S_i(u,v,p^6)&=&S_i^{jk}u_jv_k+S^{5j}_ip^6(u_5v_j+v_5u_j)+
S^{55}_i(p^6)^2u_5v_5,\qquad i=1,\dots, 4,\nn\\
S_5(u,v,p^6)&=&S^{5j}_5(u_5v_j+v_5u_j)+S^{55}_5p^6u_5v_5.
\eeqa
In phase I${}_+$, the range of the fields $u,v,p^6$ is bounded by the D-term
equations --- they form the homogeneous coordinates of a compact space
--- provided that $p^1,\ldots, p^5$ are bounded.
Hence, the only source of non-compactness comes from the fields
$p^1,\ldots, p^5$. They enter into the F-term potential quadratically.
Thus, the condition in phase I${}_+$ is that the mass matrix
of the fields $p^1,\ldots, p^5$ has full rank.
That is,
\begin{description}
\item[Condition (C):] If $(u,v,p^6)\not\in F_{\rm I_+}$ solves the equations
\beq
S_1(u,v,p^6)=\cdots=S_5(u,v,p^6)=0,
\label{FtilX}
\eeq
then the $5\times (5+5+1)=5\times 11$ matrix
\beq
M:=\left(\left({\partial S_I\over\partial u_J}\right),
\left({\partial S_I\over\partial v_J}\right),
\left({\partial S_I\over\partial p^6}\right)\right)
\label{defM}
\eeq
has rank $5$.
\end{description}
This is an open condition: First, note that
the matrix $M$ annihilates $(u,0,0)^T$ and $(0,v,0)^T$ if $(u,v,p^6)$
solves (\ref{FtilX}). Thus the number of
conditions that $M$ has rank $4$ or less is $(5-4)\times (11-2-4)=5$.
This is generically impossible to satisfy since $\wt{X}$ only
has dimension $3$.

Since (\ref{FtilX}) is the defining equation for $\wt{X}$
and (\ref{defM}) is its first order differential, (C) is
equivalent to the condition for smoothness of $\wt{X}$. (This is always the
case in the usual geometric phases.)
It turns out that this condition also implies that the $\Z_2$ action on
$\wt{X}$ that exchanges $u$ and $v$ is free,
so that the quotient $\wt{X}/\Z_2=X$ is also smooth.
Let us prove this.

Suppose there is $(u,u,p^6)\not\in F_{\rm I_+}$ that solves the equations
(\ref{FtilX}). (We would like to show that
$M$ would have rank 4 or less, in contradiction to Condition (C).)
Note that, at this point, the first two $5\times 5$ matrix factors of $M$
are identical and that they are of rank 4 or less
since
${\partial S_I\over \partial u_J}\cdot u^J=S_I=0$ for a non-zero $u$.
We are not done yet since there are also the last $5\times 1$ entries
${\partial S_I\over \partial p^6}$ in $M$. Here we note that
\beqa
\left.\left({\partial S_I\over\partial u_J}\right)\right|_{u=v}
\!\!\!\!\!\cdot
\left(
\begin{array}{c}
-u_j
\\
u_5
\end{array}\right)
&=&\left(\begin{array}{c}
-S^{jk}_iu_ku_j-S^{5j}_ip^6u_5u_j+S^{5k}_ip^6u_ku_5+S^{55}_i(p^6)^2u_5^2
\\
-S^{5j}_5u_5u_j+S^{5k}_5u_ku_5+S^{55}_5p^6u_5^2
\end{array}\right)\nn\\
&=&\left(\begin{array}{c}
2S^{5k}_ip^6u_ku_5+2S^{55}_i(p^6)^2u_5^2
\\
S^{55}_5p^6u_5^2
\end{array}\right)
=p^6\left.\left({\partial S_I\over \partial p^6}\right)\right|_{u=v}
\!\!\!\!\!,
\eeqa
where we used $S_i(u,u,p^6)=0$ in the second equality.
This means that, as long as $p^6\ne 0$, the
last $5\times 1$ entries of $M$ is a linear combination of the first
$5\times 5$ entries of $M$, so that $M$ has rank $4$ or less, in contradiction
to Condition (C). When $p^6=0$, a separate discussion is needed.
In this case, the equations are $S^{jk}_iu_ju_k=0$ ($k=1,\ldots, 4$)
and $S_5^{5j}u_5u_j=0$. Note that $u_5=0$ is not allowed since
$u_5=v_5=0$ would be in the deleted set $F_{\rm I_+}$. Thus, the latter equation
is equivalent to $S^{5j}_5u_j=0$. In this case,
\beq
\left({\partial S_I\over \partial u_J}\right)
=\left(\begin{array}{cc}
S^{jk}_iu_k&0\\
S^{5j}_5u_5&0
\end{array}\right),
\eeq
is of rank 3 or less since it annihilates the column vector
$(u_j,0)^T$. Since the last $5\times 1$ entries contribute at most
rank one, $M$ is of rank $4$ or less, again in contradiction
to Condition (C). This completes the proof of the claim that
$\wt{X}$ misses the diagonal $u=v$.

\subsection{Phase IV}\label{subsec:Cprime}

In phase IV, the range of the fields $p^1,\ldots, p^6$ is bounded
by the D-term equations --- they form homogeneous coordinates of
a compact space --- provided that the fields
$u$ and $v$ are bounded. We shall show that Condition (C)
ensures that the vacuum equations in phase IV
force $u$ and $v$ to vanish, removing the danger of non-compactness.
In the dual model, phase IV is the usual geometric phase where
the target space $\wt{Y}$ is defined by the equations $\partial_{s_{IJ}}W=0$.
Hence the condition is that the mass matrix for the fields
$s_{IJ}$ is of full rank on $\wt{Y}$, or equivalently,
smoothness of the variety $\wt{Y}$. We shall also see that
this is ensured by Condition (C).

As a preparation, we derive two consequences of Condition (C).

\subsubsection{Consequence 1:~ $S^{5i}_5\ne 0$}

First consequence of (C) is that $S^{5i}_5\ne 0$ for some $i=1,\ldots, 4$.
Indeed, if $S^{5i}_5$ were all zero, $(u,v,p^6)$ with $v_5=p^6=0$,
$S^{ij}_ku_iv_j=0$, $u_5\ne 0$, $u_{1\ldots 4}\ne 0$ and $v_{1\ldots 4}\ne 0$
would solve the equations (\ref{FtilX}) and
\beq
M=\left(\left(\begin{array}{cc}
S^{jk}_iv_k&0\\
0&0\end{array}\right),
\left(\begin{array}{cc}
S^{jk}_iu_k&0\\
0&0\end{array}\right),
\left(\begin{array}{c}
S^{5k}_iu_5v_k\\
0
\end{array}\right)\right).
\eeq
This matrix has rank 4 or less,
in contradiction to Condition (C).

\subsubsection{Consequence 2:~ ${\rm rank}\,S(p)\geq 3$ for $p\not\in F_{\rm IV}$}

The second consequence of (C) is that
the matrix $S(p)$ has at least rank 3
if $p$ represents a point of $\PP_{\rm IV}$, i.e., if $p\not\in F_{\rm IV}$.

Suppose there is $p_*\not\in F_{\rm IV}$ such that $S(p_*)$ has rank $2$ or less.
We first assume that $p_*^6$ is non-zero.
Then, on dimensional grounds, it is possible to find $[(u,v,p_*^6)]\in \wt{X}$
such that $S(p_*)u=S(p_*)v=0$. There, we can show that
 $(p_*^1,\ldots, p_*^5)\cdot M(u,v,p_*^6)=0$.
Indeed,
\beq
(p_*^1,\ldots, p_*^5)\cdot M
=\Bigl(\underbrace{S^{JK}(p_*)v_K}_{0}, \underbrace{S^{JK}(p_*)u_K}_{0},
\partial_{p^6}S^{KL}(p_*)u_Kv_L
\Bigr),
\eeq
and the last entry,
which is
\beq
\partial_{p^6}S^{KL}(p_*)u_Kv_L
=p^jS^{5k}_j(u_5v_k+v_5u_k)+2p^jS^{55}_jp^6u_5v_5+p^5S^{55}_5u_5v_5,
\eeq
also vanishes provided $p_*^6\ne 0$ since
\beqa
0&=&u_5S^{5I}(p_*)v_I+v_5S^{5I}(p_*)u_I\nn\\
&=&S^{5k}(p_*)u_5v_k+S^{55}(p_*)u_5v_5+S^{5k}(p_*)v_5u_k+S^{55}(p_*)u_5v_5\nn\\
&=&p_*^6\partial_{p^6}S^{KL}(p_*)u_Kv_L
+p_*^5\Bigl(
\underbrace{S^{5k}_5(u_5v_k+v_5u_k)+S^{55}_5p_*^6u_5v_5}_{S_5(u,v,p_*^6)=0}
\Bigr).
\eeqa
As $(p_*^1,\ldots, p_*^5)\ne 0$, this means that $M(u,v,p_*^6)$ has
rank $4$ or less, in contradiction to Condition (C).

We need a separate discussion for the case $p_*^6=0$.
Note that $p_*^5\ne 0$ since $p_*\not\in F_{\rm IV}$.
In this case, $S^{ij}(p_*)=S^{ij}_kp_*^k$, $S^{5j}(p_*)=S^{5j}_5p_*^5$
and $S^{55}(p_*)=0$. Recall that $S^{5j}_5\ne 0$ and hence $S^{5j}(p_*)\ne 0$.
Then, we see that
$S(p_*)$ has rank $2$ or less (actually rank exactly $2$) if and only if
\beq
S^{ij}_kp_*^k=x_*^iS^{5j}_5+x_*^jS^{5i}_5
\label{SPst}
\eeq
for some $x_*^i$. But we can show that presence of such $p_*^k$
contradicts Condition (C).
That is, we can find $p_!^5, u, v$ such that
$[(u,v,0)]\in \wt{X}$ and that $(p_*^1,\ldots, p_*^4,p_!^5)\cdot M(u,v,0)=0$.
Indeed, $[(u,v,0)]\in \wt{X}$ means $(u,v,0)\not\in F_{\rm I_+}$ and
$$
\mbox{(a)}~~ S^{ij}_ku_iv_j=0,\qquad 
\mbox{(b)}~~ S^{5j}_5(u_5v_j+v_5u_j)=0, 
$$
and $(p_*^1,\ldots, p_*^4,p_!^5)\cdot M(u,v,0)=0$ reads
\beqa
&&\mbox{(1)}~~ x_*^jv_j+p_!^5v_5=0,\qquad
\mbox{(2)}~~ S^{5j}_5v_j=0,\nn\\
&&\mbox{(3)}~~ x_*^ju_j+p_!^5u_5=0,\qquad
\mbox{(4)}~~ S^{5j}_5u_j=0,\nn\\
&&\mbox{(5)}~~ p_*^jS^{5k}_j(u_5v_k+v_5u_k)+p_!^5S^{55}_5u_5v_5=0.\nn
\eeqa
Note that one of the four equations in (a)
follows from (2) and (4) provided (\ref{SPst}) holds.
Note also that (b) follows from (2) and (4) as well.
The number of equations is therefore $3+5=8$.
The number of variables is $11-3=8$ where $11$ comes from
$u_{1\ldots 5}, v_{1\ldots 5}, p_!^5$ and $-3$ comes from the gauge group action.
Thus, there is a solution, and the contradiction against (C)
is confirmed.

This completes the proof under (C)
that $S(p)$ has rank at least $3$ for $p\not\in F_{\rm IV}$.

\subsubsection{The proof --- the original model}\label{subsub:iiandiii}

Suppose $(u,v,p)$ solves the vacuum equations in phase IV. In particular,
$p\not\in F_{\rm IV}$ and $(u,v,p)$ solves the F-term equations.
In view of $W=\sum_{I=1}^5p^IS_I(u,v,p^6)$, we see that the F-term equations
$\partial_{u_I}W=\partial_{v_I}W=\partial_{p^6}W=0$ read
$(p^1,\ldots, p^5)\cdot M=0$,
where $M$ is the $5\times 11$ matrix defined by (\ref{defM}).
Since $(p^1,\ldots, p^5)\ne 0$ by $p\not\in F_{\rm IV}$,
this means that $M$ has rank 4 or less.
Since $(u,v,p^6)$ solves the equations (\ref{FtilX}), this means
by Condition (C) that $(u,v,p^6)$ has to land in $F_{\rm I_+}$.
We now show that $u=v=0$ under (C), in each of the five
components of $F_{\rm I_+}$:\\
(i) \underline{$u_{1\ldots 5}=0$:}\,
Then $v_{1\ldots 5}=0$ by the $O(2)$ D-term equation.
\\
(ii) \underline{$v_{1\ldots 5}=0$:}\, Then $u_{1\ldots 5}=0$ for the same reason.\\
(iii) \underline{$u_{1\ldots 4}=p^6=0$:}\,
Then, $p^5\ne 0$ by $p\not\in F_{\rm IV}$.
By the F-term equation,
\beq
0=S(p)u=\left(\begin{array}{cc}
S^{i5}_5p^5u_5\\
0
\end{array}\right).
\eeq
Since $S^{i5}_5\ne 0$ (a consequence of (C)), this means
$u_5=0$. Therefore, $u=0$, and hence $v=0$ by the $O(2)$ D-term equation.
\\
(iv) \underline{$v_{1\ldots 4}=p^6=0$:}\, Then $u=v=0$ for the same reason.\\
(v) \underline{$u_5=v_5=0$:}\, Then, the equations are
$S^{ij}_ku_iv_j=0$,
$S^{Ij}(p)u_j=S^{Ij}(p)v_j=0$, i.e.,
\beqa
&&S^{ij}_ku_iv_j=0,\quad k=1,\ldots, 4,\label{Eq12}\\
&&S^{ij}_kp^ku_j=S^{ij}_kp^kv_j=0,\quad i=1,\ldots, 4,\label{Eq13}\\
&&S^{5j}_5p^5u_j+S^{5j}_kp^kp^6u_j=S^{5j}_5p^5v_j+S^{5j}_kp^kp^6v_j=0.
\label{Eq14}
\eeqa
Suppose $u=v=0$ fails. By the $O(2)$
D-term equation, this means that $u_i\ne 0$ \underline{and} $v_j\ne 0$ for some
$i,j\in\{1,\ldots, 4\}$. Let us now put
$U=(u_1,\ldots, u_4,U_5)$ and $V=(v_1,\ldots, v_4,V_5)$ and ask
if there is $(U,V,p^6_!)\not\in F_{{\rm I}_+}$ solving (\ref{FtilX})
but $M$ has rank 4 or less. We may put $p^6_!=0$ and still have
$(U,V,0)\not\in F_{{\rm I}_+}$ provided $(U_5,V_5)\ne (0,0)$.
Then, the first four equations in (\ref{FtilX}) are equivalent to
(\ref{Eq12}) and the last one reads
\beq
S^{5j}_5(U_5v_j+V_5u_j)=0,\label{EqA}
\eeq
Also, by (\ref{Eq13}) the matrix $M$ at $(U,V,0)$ satisfies
$(p^1,\ldots, p^4,0)\cdot M=0$ provided
\beq
p^iS^{5j}_i(U_5v_j+V_5u_j)=0.\label{EqB}
\eeq
The equations (\ref{EqA}) and (\ref{EqB}) have a solution with
$(U_5,V_5)\ne (0,0)$ provided
\beq
\det\left(\begin{array}{cc}
S^{5j}_5v_j&S^{5j}_5u_j\\
S^{5j}_ip^iv_j&S^{5j}_ip^iu_j
\end{array}\right)=0.
\eeq
This is indeed the case since (\ref{Eq14}) has a solution
with $(p^5,p^6)\ne (0,0)$. Therefore, we are able to draw
a contradiction to (C).
This proves that $u=v=0$.

\subsubsection{The proof --- the dual model}

In phase IV, the range of the fields
$(p,\wt{x})$ is bounded by the D-term equations and
a part of the F-term equations:
\beq
S^{IJ}(p)+(\wt{x}^I\wt{x}^J)=0.\label{StilXtilX}
\eeq
Therefore, the only source of non-compactness comes from
the fields $s_{IJ}=s_{JI}$ with $I,J=1,\ldots 5$.
Since they enter into the F-term potential quadratically,
the condition in phase IV is that the mass matrix has the full rank,
or equivalently, the F-term equations force $s_{IJ}=0$.
The F-term equations imply, if we use (\ref{StilXtilX}),
\beqa
&&
S(p)^{IJ}s_{JK}=0,\qquad I,K=1,\ldots 5,\label{Sseq}\\
&&
\partial_{p^{\alpha}}S^{IJ}(p)s_{IJ}=0,\qquad \alpha=1,\ldots 6.
\label{pSs}
\eeqa
Since $p\not\in F_{\rm IV}$, the matrix $S(p)$ has rank $3$ or higher
(a consequence of (C)).
Then (\ref{Sseq}) requires that $s_{JK}$ is of rank $2$ or less and can
be written as $s_{IJ}=u_Iv_J+v_Iu_J$ for some $u_I$'s and $v_J$'s satisfying
\beq
S^{IJ}(p)u_J=S^{IJ}(p)v_J=0.
\label{SuSv}
\eeq
The first five equations of (\ref{pSs}) are nothing but
the equations (\ref{FtilX}), and (\ref{SuSv}) together with
the last of (\ref{pSs}) is equivalent to $(p^1,\ldots, p^5)\cdot M=0$.
Since $(p^1,\ldots, p^5)\ne 0$ by $p\not\in F_{\rm IV}$,
this means by Condition (C) that $(u,v,p^6)$ must land in $F_{\rm I_+}$.
Then, we can reuse most of the argument in the original model
(Section~\ref{subsub:iiandiii}), and show that $s_{IJ}=0$ is enforced
under Condition (C).
The $O(2)$ D-term equation was important in the original model but cannot
be used here. However, that is not necessary since we only need
\underline{either}
$u=0$ \underline{or} $v=0$ to conclude $s_{IJ}=0$:
In components (i) and (ii), $s_{IJ}=0$  from the outset.
In component (iii) ({\it resp}. (iv)), $u=0$ ({\it resp}. $v=0$)
is derived without the $O(2)$ D-term equation.
In component (v), $u_i\ne 0$ \underline{and} $v_j\ne 0$ for some
$i,j\in \{1,\ldots, 4\}$
is the only non-trivial possibility to exclude. So again, no need for
the $O(2)$ D-term equation.

This must be equivalent to the condition for smoothness of $\wt{Y}$.
Indeed, this can be seen explicitly.
Extending the analysis in
\cite{Hori:2011pd} for a similar problem, we see that the smoothness
condition goes as follows:
\begin{description}
\item
{\it Take a point $[p]\in C$ so that $S(p)$ has rank $3$.
Then, the linear map ${\rm Sym}^2\C^5\to \C^6$ represented by 
the $6\times 15$ matrix $N_{\alpha}^{\,{}^{(IJ)}}(p)=\partial_{p^{\alpha}}S^{IJ}(p)$
has maximal rank ($=3$)
when restricted to the subspace ${\rm Sym}^2{\rm Ker}\,S(p)$.}
\end{description}
The conclusion part
is obviously equivalent to ``{\it (\ref{Sseq}) and (\ref{pSs}) require  
 $s=0$.}''

\subsection{Phase I${}_-$}

Under Condition (C),
the vacuum equations in phase I${}_-$ require
\beq
p^5=0.
\eeq
To show this, suppose $p^5\ne 0$. Then, by Condition (C),
$(u,v,p^6)$ must land in $F_{\rm I_+}$.
In view of the deleted set (\ref{FI-}), the only possibility is
$p^6=v_{1\ldots 4}=0$, $u_{1\ldots 4}\ne 0$, $v_5\ne 0$ (or $u\leftrightarrow v$
exchanged case). But then the F-term equations include
$S^{i5}_5p^5v_5=0$ which is impossible since $S^{i5}_5\ne 0$ by (C).
This proves that $p^5=0$. Since $p^5$ enters into the superpotential
at most linearly, this means that
we may impose $p^5=S_5(u,v,p^6)=0$ without loosing massless degrees of
freedom.

Now let us show that the Higgs branch is compact.
We consider the cases $p^{1\ldots 4}=0$ and $p^{1\ldots 4}\ne 0$
separately.\\
\underline{$p^{1\ldots 4}=0$:}~
Possibly non-zero fields are $u, v, p^6$, but their
charges under $U(1)_1\times U(1)_2$ lie strictly inside a half space
of the charge lattice. Therefore, their values are
bounded by the $U(1)_1\times U(1)_2$ D-term equations.\\
\underline{$p^{1\ldots 4}\ne 0$:}~
By (C), we must have
$p^6=v_{1\ldots 4}=0$, $u_{1\ldots 4}\ne 0$, $v_5\ne 0$ (or $u\leftrightarrow v$).
Possibly non-zero fields are then
$p^{1\ldots 4}$, $u_{1\ldots 4}$, $u_5$, $v_5$, and their
charges under the $U(1)$ subgroup
$\{(z^{-1},z^2,z)\}\subset U(1)_0\times U(1)_3\times U(1)_4$
are all positive ($1$, $1$, $4$, $1$ respectively).
Thus, their values are bounded by the corresponding D-term equation.

With a little more work, we can also find what the Higgs branch is.
For \underline{$p^{1\ldots 4}=0$},
we must have $u_5\ne 0$ and $v_5\ne 0$ which breaks
$G_0$ to the subgroup $\{(z^2,z,z)\}\cong U(1)$. Under this $U(1)$,
the possibly non-zero fields $p^6,u_{1\ldots 4}, v_{1\ldots 4}$ all
 have charge $1$, defining a $\PP^8$, or a $\PP^7$ if we take into
account $S_5(u,v,p^6)=0$ which is linear in these variables.
The non-trivial F-term equations
$S_1(u,v,p^6)=\cdots =S_4(u,v,p^6)=0$ are quadratic in these variables.
Thus, we have the intersection of
four quadrics in $\PP^7$ (a Calabi-Yau threefold).
$\Z_2=G/G_0$ acts freely on it. Let $Z$ be the quotient.
$Z$ has a conifold singularity at $p^6=v_{1\ldots 4}=0$,
$u_{1\ldots 4}=u^*_{1\ldots 4}$ where $u^*_{1\ldots 4}\ne 0$ satisfy the following
equations for some $p_*^{1\ldots 4}\ne 0$:
\beqa
&&S^{5j}_5u^*_j=0,\nn\\
&&p_*^kS^{ij}_ku^*_j=0,\qquad i=1,\ldots, 4,\label{auxFeq}\\
&&p_*^kS^{5j}_ku^*_j=0.\nn
\eeqa
Up to scaling, there are ten such $(p_*^{1\ldots 4}, u_{1\ldots 4}^*)$.
For \underline{$p^{1\ldots 4}\ne 0$}, we have
$p^6=v_{1\ldots 4}=0$, $u_{1\ldots 4}\ne 0$, $v_5\ne 0$, and
the non-trivial F-term equations are nothing but (\ref{auxFeq}).
Thus, there are ten isolated solutions for $(p^{1\ldots 4},u_{1\ldots 4})$
up to scale.
The non-zero values of $v_5$ and $u_{1\ldots 4}$ break $G_0$
to $\{(z^{-1},1,1)\}\cong U(1)$ under which the
remaining fields $p^{1\ldots 4}$ and $u_5$ both have charge $1$.
Thus, we have $\PP^1$ minus one point with $p^{1\ldots 4}=0$.
The deleted point is nothing but one of the ten singular points of $Z$.
In conclusion, the Higgs branch is a singular Calabi-Yau threefold
$Z$ and ten rational curves rooted at ten conifold points of $Z$.
The behaviour of the superpotential near the roots is
\beq
W\sim p(x_1x_2+x_3x_4).\label{Wroot}
\eeq
Indeed, the equation $\dd W=0$ reads
\beq
x_1x_2+x_3x_4=0,\quad
px_1=px_2=px_3=px_4=0,
\eeq
and Crit$(W)$ is the union of $\{p=0, x_1x_2+x_3x_4=0\}$ (a conifold)
and $\{p\,{\rm free},\,x_1=x_2=x_3=x_4=0\}$ (a line) which touch each other at
the origin.

\subsection{Phase II}

Under Condition (C),
the vacuum equations in phase II require
\beq
p^{1\ldots 4}=0.
\eeq
To show this, suppose $p^{1\ldots 4}\ne 0$.
Since $p^{5,6}\ne 0$ (\ref{FII}),
we have $p^{1\ldots 6}\not\in F_{\rm IV}$.
Then, by the analysis of phase IV, this implies under Condition (C) that
$u=v=0$, but that is forbidden (\ref{FII}).
This proves that $p^{1\ldots 4}=0$.
Since $p^{1\ldots 4}$ enters into the superpotential
at most linearly, this means that
we may impose $p^{1\ldots 4}=S_{1\ldots 4}(u,v,p^6)=0$
without loosing massless degrees of
freedom.

Let us show that the Higgs branch is compact.
We consider the cases $p^5=0$ and $p^5\ne 0$
separately.\\
\underline{$p^5=0$:}~
Possibly non-zero fields are $u,v,p^6$, and their values are bounded
by the D-term equations.\\
\underline{$p^5\ne 0$:}~
Then, by Condition (C), $(u,v,p^6)$ must land in $F_{\rm I_+}$.
In view of the deleted set (\ref{FII}), we must have $u_5=v_5=0$.
Then, possibly non-zero fields are $p^5, p^6, u_{1\ldots 4}, v_{1\ldots 4}$,
and their values are bounded by the D-term equations.

With a little more work, we can also find what the Higgs branch is.
For \underline{$p^5=0$}, we must have $p^6\ne 0$ which breaks $G_0$
to the subgroup $\{(zw,zw,w)\}\cong U(1)\times U(1)$. Under this,
the possibly non-zero fields $u_{1\ldots 5}$ and $v_{1\ldots 5}$ have
charge $(0,1)$ and $(1,0)$, defining $\PP^4\times \PP^4$.
The non-trivial F-term equations $S_1(u,v,p^6)=\cdots =S_5(u,v,p^6)=0$
are of degree $(1,1)$ and define a Calabi-Yau threefold.
Let $Z'$ be its quotient by $\Z_2=G/G_0$.
It has a conifold singularity at $u_5=v_5=0$ and $u_{1\ldots4}\ne 0$
$v_{1\ldots 4}\ne 0$ such that
\beq
\begin{array}{l}
S^{ij}_ku_iv_j=0,\quad k=1,\ldots, 4,\\
S^{5j}_5u_j=S^{5j}_5v_j=0.
\end{array}
\label{auxFeq2}
\eeq
Up to scaling, there are six such $(u,v)$.
For \underline{$p^5\ne 0$}, we have $u_5=v_5=0$, $u_{1\ldots4}\ne 0$
$v_{1\ldots 4}\ne 0$, and the non-trivial F-term equations are nothing but
(\ref{auxFeq2}). Thus, there are six isolated solutions up to scale.
The non-zero values of $u_{1\ldots4}$ and $v_{1\ldots 4}$ break
$G_0$ to $U(1)_3$ under which the remaining fields
$p^5$ and $p^6$ both have charge $-1$. Thus, we have $\PP^1$ minus one point
with $p^5=0$. The deleted point is nothing but one of the six
singular points of $Z'$. Thus, the Higgs branch is a singular
Calabi-Yau threefold $Z'$ and six rational curves rooted at
six singular conifold points of $Z'$. The behaviour of the superpotential near
the roots is as in (\ref{Wroot}).

\subsection{Phase III}

Let us show that the Higgs branch in phase III is compact under Condition (C).
We consider the cases $p^{1\ldots 4}=0$ and $p^{1\ldots 4}\ne 0$ separately.
\\
\underline{$p^{1\ldots 4}=0$:}~
In view of the deleted set (\ref{FIII}), we must have $p^5\ne 0$.
On the other hand, the F-term equations require
$S^{i5}_5p^5u_5=S^{i5}_5p^5v_5=0$. Since (C) implies $S^{i5}_5\ne 0$, we
must have $u_5=v_5=0$. Possibly non-zero fields are thus
$p^5,p^6,u_{1\ldots 4},v_{1\ldots 4}$, and their values are bounded by
the D-term equations.\\
\underline{$p^{1\ldots 4}\ne 0$:}~
Since $p^{5,6}\ne 0$ (\ref{FIII}), we have $p^{1\ldots 6}\not\in F_{\rm IV}$.
Then, by the analysis in phase IV, this implies under Condition (C) that
$u=v=0$. Possibly non-zero fields are thus
$p^{1\ldots 6}$ and their values are bounded by the D-term equations.

\subsection{Phase V}

Under Condition (C), the vacuum equations in phase V
require
$$
\mbox{(i)~} p^{1\ldots 4}\ne 0,\quad\,\,
\mbox{(ii)~} p^5=p^6=0,\quad\,\,
\mbox{(iii) $u_{1\ldots 4}=0$ or $v_{1\ldots 4}=0$.}
$$
Indeed, in view of the deleted set (\ref{FV}),
$p^{1\ldots 4}=0$ would imply $p^5\ne 0$, $u_5\ne 0$ and $v_5\ne 0$
but that is inconsistent with the F-term equations
$S^{i5}_5p^5u_5=S^{i5}_5p^5v_5=0$ as we know $S^{i5}_5\ne 0$ from (C).
This establishes (i). Then, if we assume $p^{5,6}\ne 0$,
we have $p^{1\ldots 6}\not\in F_{\rm IV}$. By the analysis in phase IV,
this implies under (C) that $u=v=0$, but that is forbidden (\ref{FV}).
This establishes (ii).
To show (iii), suppose both $u_{1\ldots 4}$ and $v_{1\ldots 4}$
are non-zero. Then, since $u_5=v_5=0$ is a part of the
deleted set (\ref{FV}), this means that
$(u,v,p^6=0)\not\in F_{\rm I_+}$. Then, by (C), we must have
$p^{1\ldots 5}=0$, but that is inconsistent with $p^{1\ldots 4}\ne 0$.
This proves (iii).
Since $p^5$ enters into the superpotential
at most linearly, (ii) means that
we may impose $p^5=S_5(u,v,p^6)=0$ without loosing massless degrees of
freedom.

Let us show that the Higgs branch is compact.
In view of (iii) and the symmetry, we may assume $v_{1\ldots 4}=0$.
Possibly non-zero fields are then
$p^{1\ldots 4},u_{1\ldots 4}, u_5, v_5$.
Their values are bounded by the D-term equation for
$U(1)\cong \{(z^{-1}, z^2, z)\}\subset G_0$ (see
 the analysis in phase I${}_-$).

We can also determine what the Higgs branch is.
For $u_{1\ldots 4}=v_{1\ldots 4}=0$, the F-term equations are all satisfied
(recall $p^5=p^6=0$). In view of the deleted set (\ref{FV}), we must have
$u_5\ne 0$ and $v_5\ne 0$ which break $G_0$ to the
$U(1)$ subgroup $\{(z^2,z,z)\}$. Under this, the remaining fields
$p^{1\ldots 4}$ have charge $-2$, defining the weighted projective space
$\PP^3_{[2222]}$.
For $u_{1\ldots 4}\ne 0$ and $v_{1\ldots 4}=0$, we have
$v_5\ne 0$, and non-trivial F-term equations are the same as
(\ref{auxFeq}). Thus, we have ten isolated pairs for
$(p^{1\ldots 4},u_{1\ldots 4})$ up to scale. The non-zero values of $p^{1\ldots 4}$
and $v_5$ break $G_0$ to $\{(1,z,z)\}\cong U(1)$ under which
the remaining two fields $u_{1\ldots 4}$ and $u_5$ have charge $1$ and $2$
respectively. Thus we have the weighted projective line (or the teardrop)
$\PP^1_{[12]}$ minus the $\Z_2$ point $u_{1\ldots 4}=0$.
The deleted point is nothing but a point
of $\PP^3_{[2222]}/\Z_2$.
Thus, the Higgs branch is the union of a $\PP^3_{[2222]}/\Z_2$
and ten teardrops.

\subsection{The Moduli Space}\label{subsec:countcplx}

Transformations of variables that commute with the gauge symmetry are
\beqa
&&x_i\to a_i^{\,j}x_j+c_ix_5p^6,\quad x_5\to a_5^{\,5}x_5,\nn\\
&&p^i\to b^i_{\,j}p^j,\quad p^5\to b^5_{\,5}p^5+b_kp^kp^6,\quad
p^6\to b^6_{\,6}p^6,
\eeqa
where $(a_i^{\, j}), a_5^{\,5}, (b^i_{\, j}), b^5_{\, 5}, b^6_{\, 6}$ 
are invertible.
They induce the following transformations of the space ${\mathcal S}$
of coefficients $(S^{ij}_k,\ldots)$
\beqa
&&S^{ij}_k\to S^{mn}_la_m^{\,i}a_n^{\, j}b^l_{\,k},\nn\\
&&S^{5j}_5\to S^{5m}_5a_5^{\,5}a_m^{\,j}b^5_{\,5},\nn\\
&&S^{5j}_k\to S^{5m}_la_5^{\,5}a_m^{\,j}b^l_{\,k}b^6_{\,6}
+S^{5m}_5a_5^{\,5}a_m^{\,j}b_k+S^{im}_la_m^{\,j}b^l_{\,k}c_i,\\
&&S^{55}_5\to S^{55}_5(a_5^{\,5})^2b^5_{\,5}b^6_{\,6}
+2S^{5j}_5a_5^{\,5}b^5_{\,5}c_j,\nn\\
&&S^{55}_k\to S^{55}_l(a_5^{\,5})^2b^l_{\,k}(b^6_{\,6})^2
+S^{55}_5(a_5^{\,5})^2b_kb^6_{\,6}
+S^{ij}_lb^l_{\,k}c_ic_j
+2S^{5j}_5a_5^{\,5}b_kc_j+2S^{5j}_la_5^{\,5}b^l_{\,k}b^6_{\,6}c_j.\nn
\eeqa
The transformations from the complexified gauge group,
\beqa
&&a_i^{\,j}=\lambda_1\delta_i^j,\quad
a_5^{\,5}=\lambda_2,\nn\\
&&b^i_{\,j}=\lambda_1^{-2}\delta^i_j,\quad
b^5_{\,5}=\lambda_1^{-1}\lambda_2^{-1},\quad
b_j=0,\quad
b^6_{\,6}=\lambda_1\lambda_2^{-1},\quad
c_j=0,\nn
\eeqa
form the subgroup that acts trivially on ${\mathcal S}$.
Let ${\mathcal G}$ be the effective group of transformations, and
let ${\mathcal S}_{(\rm C)}\subset {\mathcal S}$ be the subset consisting of
$(S^{ij}_k,\ldots)$ that satisfies Condition (C).
Then, the complex moduli space is the quotient
\beq
\mathfrak{M}_C={\mathcal S}_{(\rm C)}/{\mathcal G}.
\label{MC}
\eeq
Note that the dimension of ${\mathcal S}$ is $40+4+16+1+4=65$
while ${\mathcal G}$ has dimension $16+1+16+1+4+1+4-2=43-2=41$.
Thus, $\mathfrak{M}_C$ has dimension $65-41=24$,
in agreement with the number of complex moduli of $X$ and $\wt{Y}$.

We expect that ${\mathcal G}$ acts on ${\mathcal S}_{(\rm C)}$
without continuous stabilizer. Indeed, a point with continuous stabilizer
would correspond to a continuous symmetry of either
$X$ or $\wt{Y}$, but that would be impossible since both
$X$ and $\wt{Y}$ have $h^{2,0}=0$.
Therefore, the quotient (\ref{MC}) should be a good one.
It would be nicer to show this mathematically.

%%%%%%%%%%%%%%%%%%%%%%%%%%%%%%%%%%%%%%%%%%%%%%%%%%%%%%%%%%%%%%%%%%%%%%%%%%%%%
\section{Mirror Symmetry}
\label{sec-mirror}
Mirror symmetry for Calabi-Yaus which are not complete
  intersections in toric ambient spaces is still mostly an open
  problem. However, since phase I${}_+$ is a free
  $\mathbb{Z}_2$-quotient of a complete intersection in a toric
  variety it is possible to work with well-established methods of
  toric geometry and mirror symmetry. In the following we will
recompute the topological data of this Calabi-Yau, determine its mirror and
the Picard-Fuchs operator and compute the Yukawa couplings and
Gromov-Witten invariants. Once we have the Picard-Fuchs operator we
are also able to compute the Gromov-Witten invariants in phase IV, up
to normalization.

\subsection{Phase I${}_+$}
\label{sec-phase1}
%%%%%
\subsubsection{Toric analysis and topological data}
\label{sec-toric}
The present example is of particular interest not only because it has
two geometric phases but also because we are able to compute the
mirror in phase I${}_+$ using toric geometry. 
%This is possible because
%phase I${}_+$ is a free $\mathbb{Z}_2$-quotient of a three-parameter
%complete intersection of codimension five in a toric ambient
%space. 
Recall that we denote the two-parameter model by $X$ and the
three-parameter model by $\widetilde{X}$. Their respective mirrors are
denoted by $X^{\vee}$ and $\widetilde{X}^{\vee}$. The analysis relies
heavily on the machinery of toric geometry and the toric mirror
construction by Batyrev and Borisov
\cite{Batyrev:1994pg,Batyrev:1995ca}. The main references for toric
mirror symmetry of complete intersection Calabi-Yaus are
\cite{Hosono:1993qy,Hosono:1994ax} and the book by Cox and Katz
\cite{coxkatz}. A nice exposition focusing on complete intersections
can also be found in \cite{Klemm:2004km}, mirror symmetry for free
quotients has been discussed for instance in \cite{Braun:2007vy}. The
complexity of the calculations requires the use of specialized
computer programs, most importantly the toric geometry package PALP
\cite{Kreuzer:2002uu,Braun:2012vh}.

As can be seen already from (\ref{basis1}) and (\ref{basis2}) the
relation to a toric three-parameter model stems from the fact that the
maximal torus of $O(2)$ is $SO(2)\simeq U(1)$. In order to reveal some more 
properties of the toric ambient space, we make a change of
basis. Starting from (\ref{basis1}), we choose the following linear
combinations of the charge vectors:
$\{Q_{U(1)_1},Q_{U(1)_2},Q_{SO(2)}\}\rightarrow\{Q_{U(1)_1}+Q_{U(1)_2}
+Q_{SO(2)},Q_{U(1)_2},Q_{U(1)_2}+Q_{SO(2)}\}$. The
first of the new charge vectors shows that there is a $\mathbb{Z}_2$
fixed point since all charges are either $0$ or $2$. Modding out the
$\mathbb{Z}_2$ simply means dividing the charge vector by two. Further
subtracting this new vector from the second and the third vector we
arrive at the following:
\begin{align}
\label{basis3}
\begin{array}{rrrrrrr||c}
p^{1...4}&p^5&p^6&u_{1...4}&u_5&v_{1...4}&v_5&FI\\
\hline
-1&-1&0&1&1&0&0&(\zeta_1+\zeta_2)/2\\
-2&-1&1&1&0&1&0&\zeta_1\\
-1&-1&0&0&0&1&1&(\zeta_1+\zeta_2)/2
\end{array}
\end{align}
Let us focus on phase I${}_+$ where $p^1=\ldots=p^5=0$. The charges of
the remaining fields are all positive and define a toric ambient space
given by $\mathbb{P}^8$ with two $\mathbb{P}^4$s blown up in
orthogonal directions. This geometry is smooth. A codimension $5$
complete intersection as given by the F-terms of the GLSM in this
phase is a three-parameter Calabi-Yau with the three parameters
corresponding to the volumes of $\mathbb{P}^8$ and the two
$\mathbb{P}^4$s. As one can read off from the FI-parameters, the
volumes of the two $\mathbb{P}^4$s get identified. There is a
$\mathbb{Z}_2$ that exchanges the two $\mathbb{P}^4$s. This confirms
again that $h^{1,1}(X)=2$ and that phase I${}_+$ is free
$\mathbb{Z}_2$-quotient of a complete intersection of codimension five
of in the ambient space defined by (\ref{basis3}).

\begin{figure} 
\begin{center}
\input{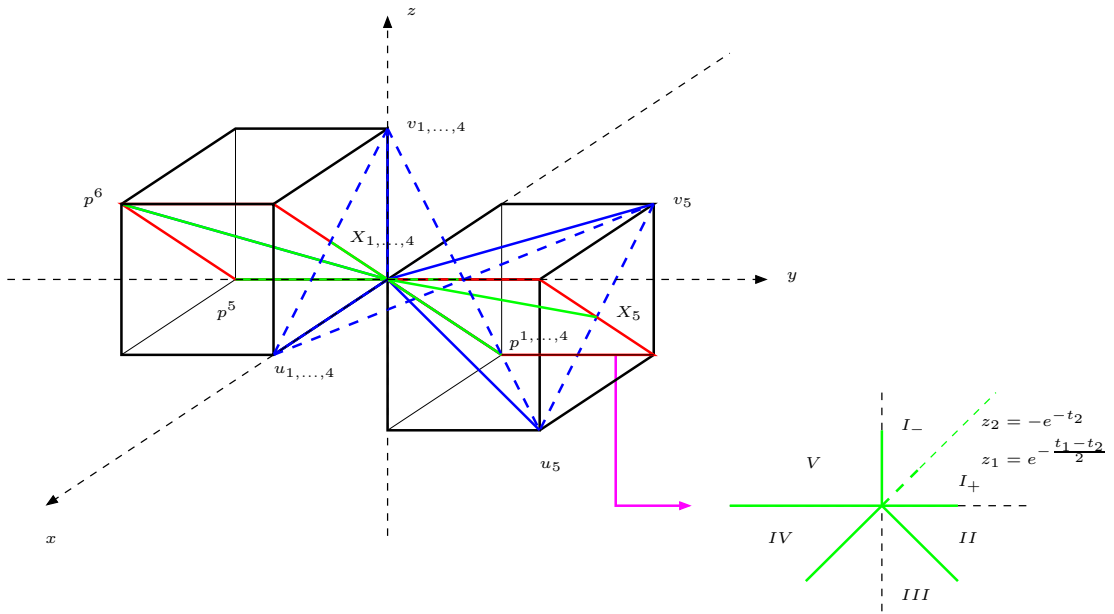}
\end{center}
\caption{Phases of the two-parameter model from a three-parameter
  complete intersection.}\label{fig-twoparproj}
\end{figure}

In the following we will also require the K\"ahler cone, and its
dual, the Mori cone, which is given by a particular basis. We give
this for later reference:
\begin{align}
\label{basis4}
\begin{array}{rrrrrrr||c}
p^{1...4}&p^5&p^6&u_{1...4}&u_5&v_{1...4}&v_5&FI\\
\hline
-1&1&0&1&0&0&-1&(\zeta_1-\zeta_2)/2\\
0&-1&-1&0&1&0&1&\zeta_2\\
-1&1&0&0&-1&1&0&(\zeta_1-\zeta_2)/2
\end{array}
\end{align}
The phase diagram of the three-parameter model (with arbitrary FI
parameters), which coincides with the secondary fan of the associated
toric variety, is depicted on the left side of figure
\ref{fig-twoparproj}. In \cite{Halverson:2013eua} this toric model
associated with a non-Abelian GLSM was called Cartan model.
Identifying the FI parameters as in the charge table amounts to
projecting into the $x-z$-plane. While the $p$-fields are already in
the plane the $u$- and $v$-fields combine into the fundamentals $x$
along the dashed blue lines. From this picture one can also clearly
see the appearance of the extra phase boundary that separates phases
I${}_-$ and I${}_+$: the dashed blue lines connecting the charge
vectors of $u_{1...4}$ with $v_5$ (resp. $v_{1...4}$ with
$u_5$) project onto the extra phase boundary. This combines into the
information encoded in the non-Abelian D-term.

Given the data of the three-parameter model $\widetilde{X}$ it is
possible to determine the mirror $\widetilde{X}^{\vee}$. Mirror
symmetry for complete intersections in toric varieties is connected to
the following data associated to lattice polytopes in dual integer
lattices, called the M- and the N-lattice:
\begin{eqnarray}                                        \label{cicy-nef}
        \Delta=\Delta_1+\ldots+\Delta_r&& \Delta^{\circ}
        =\langle \nabla_1,\ldots,\nabla_r\rangle_{\mathrm{conv}}\nonumber\\
                &(\nabla_n,\Delta_m)\geq-\delta_{nm}&\\
        \nabla^{\circ}=\langle \Delta_1,\ldots,\Delta_r\rangle_{\mathrm{conv}}
        &&\nabla=\nabla_1+\ldots+\nabla_r\nonumber
\end{eqnarray}
Here $r$ is the codimension of the Calabi-Yau and the defining equations
$f_i=0$ are sections of $\mathcal{O}(\Delta_i)$. The decomposition of
the M-lattice polytope $\Delta\subset M_{\mathbb{R}}$ into a
Minkowski sum $\Delta=\Delta_1+\ldots+\Delta_r$ is dual to a nef
(numerically effective) partition of the vertices of a reflexive
polytope $\nabla\subset N_{\mathbb{R}}$ such that the convex hulls
$\langle\nabla_i\rangle_{\mathrm{conv}}$ of the respective vertices
and $0\in N$ only intersect at the origin. The hypersurface equations
are then given by
\begin{align}
\label{toriccicy}
f_m=\sum_{w_k\in\Delta_m}c_k^m\prod_{n=1}^r\prod_{\nu_i\in\nabla_n}
x_i^{\langle \nu_i,w_k\rangle+\delta_{mn}}.
\end{align}
Mirror symmetry is realized by exchanging the M- and N-lattices. The
software package PALP provides the routine {\tt nef.x} which computes
the polytopes and the nef partitions for a complete intersection
Calabi-Yau from the weight matrix given by the three $U(1)$-charges of
the GLSM. The resulting M-lattice polytope which describes the ambient
space has $41$ vertices and $12740$ points, the N-lattice polytope has
$11$ vertices plus the interior point at the origin. Explicitly, the
vertices of $\nabla \in N$ are:
\begin{align}
\begin{array}{rrrrrrrrrrr}
 \nu_1&\nu_2&\nu_3&\nu_4&\nu_5&\nu_6&\nu_7&\nu_8&\nu_9&\nu_{10}&\nu_{11}\\
  -1& 0& 0& 0& 0& 0& 0& 0& 1& 0& 0\\
   -1& 0& 1& 0& 0& 0& 0& 0& 0& 0& 0\\
   -1& 0& 0& 1& 0& 0& 0& 0& 0& 0& 0\\
    0& 0& 0& 0& 1& 0& 0& 0& 0& 0&-1\\
    0& 0& 0& 0& 0& 1& 0& 0& 0& 0&-1\\
    0& 0& 0& 0& 0& 0& 1& 0& 0& 0&-1\\
    1& 0& 0& 0& 0& 0& 0&-1& 0& 1&-1\\
   -1& 1& 0& 0& 0& 0& 0& 1& 0& 0& 0\\
\hline
    1& 1& 1& 1& 1& 1& 1& 0& 1& 0& 1\\
    1& 0& 1& 1& 0& 0& 0& 1& 1& 0& 0\\
    0& 0& 0& 0& 1& 1& 1& 0& 0& 1& 1
\end{array}
\end{align}
The columns above the horizontal line are the vertices
$\nu_i\in\mathbb{Z}^8$ of the polytope in the eight-dimensional
ambient space. The three lines below the horizontal line denote the
linear relations $\sum_ia_i\nu_i=0$ between the vertices. This
coincides with the basis (\ref{basis3}) of the $U(1)$s in the
GLSM. Each of the eight-dimensional vertices is associated to a toric
divisor $D_i$. The information about the complete intersection is
encoded in the nef partition. For this example there are $241$, $16$
of which cannot be related through symmetries of the polytope. The one
we are looking for is
\begin{align}
E=\{(D_1,D_5),(D_3,D_6),(D_4,D_7),(D_9,D_{11}),(D_2,D_8,D_{10})\}. 
\end{align}
The corresponding hypersurfaces have degrees
$\{(2,1,1),(2,1,1),(2,1,1),(2,1,1),(1,1,1)\}$ which is exactly what we
have for the complete intersection in phase I${}_+$.

We can read off the Laurent polynomials defining the mirror from the
vertices of the N-lattice polytope via
$f_r=\sum_{E_r}a_{i,r}x^{\nu_{i,r}}$, where $E_r$ are the elements of
the nef-partition, each of which also contains the origin. Therefore
we get for our example:
\begin{align}
\label{twoparmirror}
f_1=&1+\alpha_1\frac{x_7}{x_1x_2x_3x_8}+\alpha_5x_4\nonumber\\
f_2=&1+\alpha_3x_2+\alpha_6x_5\nonumber\\
f_3=&1+\alpha_4x_3+\alpha_7x_6\nonumber\\
f_4=&1+\alpha_9x_1+\alpha_{11}\frac{1}{x_4x_5x_6x_7}\nonumber\\
f_5=&1+\alpha_2x_8+\alpha_{8}\frac{x_8}{x_7}+\alpha_{10}x_7,
\end{align}
where the constants $\alpha_i$ redundantly encode the complex
structure parameters on the mirror. For the mirror symmetry
calculations it is not necessary to match the variables $x_i$ with
$u_I,v_I,p^6$ but it is useful to confirm that one can recover the
equations for the complete intersection that was obtained from the
GLSM. This can be done by calculating the dual of the Gorenstein cone
associated to the nef-partition above.  In the M-lattice, four
elements of the nef partition have $25$ points and one has nine
points. This corresponds to the number of monomials of the defining
equation of the complete intersection in phase I${}_+$. Computing the
hypersurface equations using the polytope data as in (\ref{toriccicy})
one can indeed recover the defining equations of the complete
intersection.

Now that we have the toric data of the three-parameter model, we can
determine its topological characteristics and the intersection ring
and compare with the results of section \ref{sec-i+}. The Mori
generators encode the linear relations in the intersection ring and
can be determined by an algorithm which requires a maximal star
triangulation of the N-lattice polytope. In our example the only
non-vertex of the polytope is the origin. Therefore there is only one
such triangulation. The Mori generators of a complete intersection of
codimension $r$ are of the form
$l^{(a)}=(l_{0,1}^{(a)},\ldots,l_{0,r}^{(a)};l_1^{(a)},\ldots,l_n^{(a)})$
with $a=1,\ldots,h^{1,1}({\tilde{X}})$ and
$\sum_{m=1}^rl_{0,m}^{(a)}+\sum_{i=1}^nl_i^{(a)}=0$. Using the {\tt
  mori.x} routine of PALP and the information about the degrees of the
complete intersections one gets the following result for the
generators of the Mori cone:
\begin{align}
\label{moricone}
\begin{array}{ccrrrrrrrrrrrrrrrrc}
l^{(1)}&=(&-1,&-1,&-1,&-1,&0;
0,&  1,&  0,&  0,&  1,&  1,&  1,& -1,&  0,&  0,&  1&)\\
l^{(2)}&=(&0,&0,&0,&0,&-1;
0,& -1,&  0,&  0,&  0,&  0,&  0,&  1,&  0,&  1,&  0&)\\
l^{(3)}&=(&-1,&-1,&-1,&-1,&0; 
1,&  1,&  1,&  1,&  0,&  0,&  0,&  0,&  1,& -1,&  0&)
\end{array}.
\end{align}
The entries to the right of the semicolon coincide with the basis
(\ref{basis4}). The entries to the left of the semicolon encode the
hypersurface degrees of the complete intersection in the given basis.

The intersection ring of the ambient variety has the form
$\mathbb{Z}[D_1,\ldots,D_n]/\langle I_{lin}+I_{non-lin}\rangle$, where
the linear relations $I_{lin}$ can be read off from the Mori
generators and the non-linear relations $I_{non-lin}$ are encoded in
the Stanley-Reisner ideal. By adjunction, one gets the intersection
ring of the complete intersection by modding out by the hypersurface
ideals encoded in the nef partition which we denote by $I_{CY}$. For
the present example the linear relations are
\begin{align}
I_{lin}=&\{D_1-D_9,D_2-D_{11}+D_{10},D_3-D_9,D_4-D_9,D_5-D_{11},D_6-D_{11},
\nonumber \\
&D_7-D_{11},D_8+D_{11}-D_9-D_{10}\}.
\end{align}
The Stanley-Reisner ideal can be obtained by using the PALP program
{\tt mori.x}:
\begin{align}
I_{SR}=\{D_8 D_{10},D_5 D_6 D_7 D_{10} D_{11},D_2 D_5 D_6 D_7 D_{11},
D_1 D_3 D_4 D_8 D_9,D_1 D_2 D_3 D_4 D_9\}.
\end{align}
The ideal of the complete intersection is
\begin{align}
I_{CY}=\prod_{r=1}^5\sum_{i\in E_r}D_{i,r}=
(D_1+D_5)(D_3+D_6)(D_4+D_7)(D_9+D_{11})(D_2+D_8+D_{10}).
\end{align}
We choose the basis
\begin{align}
J_1=D_9\quad J_2=D_{10}-D_9\quad J_3=D_{11}.
\end{align}
Given this data, we obtain the following triple intersection
numbers of the three-parameter Calabi-Yau $\widetilde{X}$:
\begin{equation}
\{J_1^3=J_3^3=5,J_1^3J_2=J_2J_3^2=11,J_1^2J_3=J_1J_3^2=10,J_1J_2^2=J_2^2J_3=15,
J_1J_2J_3=14,J_2^3=16\}.
\end{equation}
This coincides with (\ref{3partriple}) computed in section \ref{sec-i+} under the identification $x=J_1,y=J_3,z=J_2$.  The Chern class of the complete intersection can be computed using the formula
\begin{align}
c(\widetilde{X})=\prod_{i=1}^n(1+D_i)/\prod_{m=1}^r(1+\sum_{j\in E_r}D_{j,m}).
\end{align} 
This indeed yields the results for second Chern class and the Euler number of $\widetilde{X}$ we have already computed. The calculation of the topological characteristics of $X$ proceeds exactly like in section \ref{sec-i+}, and we will not repeat it here. In the following we will only slightly alter our notation and write  $H_0\equiv\tilde{J}_1= J_1-J_3,H_2\equiv J_2$ so that we have
\begin{align}
\tilde{J}_1^3=35,\qquad \tilde{J}_1^2J_2=25,\qquad \tilde{J}_1J_2^2=15,\qquad 
J_2^3=8,
\end{align}
and
\begin{align}
c_2\cdot \tilde{J}_1=50\qquad c_2\cdot J_2=32.
\end{align}
%%%%%%%
\subsubsection{Picard-Fuchs equations and discriminant}
\label{sec-pf}
For the mirror calculation we have to go to the large complex
structure limit. This information is encoded in the Mori cone. The
complex structure moduli on the mirror at the large complex structure
limit are $z_a=(-1)^{\prod_m l_{0,m}}\prod_i\alpha_i^{l_i^{(a)}}$,
where we insert (\ref{moricone}). Therefore we identify
\begin{align}
\label{o2cs}
z_1=\frac{\alpha_2\alpha_5\alpha_6\alpha_7\alpha_{11}}{\alpha_{8}}\quad 
z_2=-\frac{\alpha_{8}\alpha_{10}}{\alpha_2}\quad 
z_3=\frac{\alpha_1\alpha_2\alpha_3\alpha_4\alpha_9}{\alpha_{10}}.
\end{align}
For this choice of moduli, modding out by the freely acting
$\mathbb{Z}_2$ amounts to setting $z_1=z_3$.

Using the toric information we can explicitly compute the fundamental
period $\varpi_0(z_1,z_2,z_3)$ of $\widetilde{X}^{\vee}$. The
Picard-Fuchs operators of $X^{\vee}$ are then determined by making an
ansatz for a differential operator and requiring that the fundamental
period is annihilated. Once the Picard-Fuchs system is identified one
can employ the mirror symmetry machinery to determine the B-model
Yukawa couplings and the mirror map. With help of the intersection
data, we then compute the normalized Yukawa couplings in phase
I${}_+$.

Given the Laurent polynomials (\ref{twoparmirror}) of the complete
intersection of the mirror and the moduli at the large complex
structure limit (\ref{o2cs}), the fundamental period of
$\widetilde{X}^{\vee}$ is given by the residue integral
\begin{align}
\varpi_0(\widetilde{X}^{\vee})=\frac{1}{(2\pi i)^{11}}
\int_{\gamma}\prod_i\frac{dx_i}{x_i}\frac{1}{f_1f_2f_3f_4f_5},
\end{align}
where $\gamma$ is a suitably chosen cycle. The result is
\begin{align}
\label{w0ytilde}
\varpi_0(\widetilde{X}^{\vee})=\sum_{a,b,c}\binom{a+c}{c}^4\binom{b}{a-b+c}
\binom{2b-a-c}{-a+b}z_1^az_2^bz_3^c.
\end{align}
The sum is such that $a-b+c\geq0$, $-a+b\geq0$ and $b-c\geq0$. The
calculation is very similar to the mirror symmetry calculation of
\cite{Hosono:2011np} and we give details in appendix
\ref{app-w0calc}. The expression is symmetric under the exchange of
$z_1$ and $z_3$. The fundamental period of the two-parameter model
$X^{\vee}$ is then obtained by setting $z_3=z_1$. The first few terms
in the expansion are
\begin{align}
\varpi_0(X^{\vee})=1+2z_1z_2+16z_1^2z_2+34z_1^2z_2^2+\ldots.
\end{align}
Next we make an ansatz for the Picard-Fuchs system and impose the
condition that the Picard-Fuchs operators annihilate
$\varpi_0(X^{\vee})$. As expected for a two-parameter model we find
one degree two and one degree three operator
\begin{align}
\label{pfsystem}
\mathcal{L}_1=&5 \theta_1^2-19 \theta_2 \theta_1+20 \theta_2^2+z_1z_2^2
\left(-12\theta_2^2+12 \theta_1 \theta_2-12 \theta_2+12 \theta_1\right)
   \nonumber\\
&+z_1\left(5 \theta_1^2-4 \theta_2 \theta_1+5 \theta_1-12
   \theta_2^2-10 \theta_2\right)+z_2\left(-\theta_1^2-19 \theta_2
   \theta_1-20 \theta_1+20 \theta_2^2+20 \theta_2\right)
   \nonumber\\
&+z_1z_2\left(-11 \theta_1^2+8 \theta_2 \theta_1-11 \theta_1-24
   \theta_2^2-22 \theta_2-12\right) \\
\mathcal{L}_2=&3 \theta_1^3-13 \theta_2 \theta_1^2+20 \theta_2^2 \theta_1-10
   \theta_2^3\nonumber\\
&+z_1\left(3 \theta_1^3-4 \theta_2 \theta_1^2+6
   \theta_1^2-4 \theta_2^2 \theta_1-10 \theta_2 \theta_1+3
   \theta_1-4 \theta_2^2-6 \theta_2\right) \nonumber\\
&+z_2\left(-10
   \theta_2^3+20 \theta_1 \theta_2^2-20 \theta_2^2-10 \theta_1^2
   \theta_2+30 \theta_1 \theta_2-10 \theta_2-10 \theta_1^2+10
   \theta_1\right) \nonumber\\
&+z_1z_2\left(4 \theta_2 \theta_1^2+4 \theta_1^2-4
   \theta_2^2 \theta_1+4 \theta_1-4 \theta_2^2-4 \theta_2\right),
\end{align}
where $\theta_i=z_i\frac{\partial}{\partial z_i}$. The power series
expansions of the remaining periods can be easily determined from the
Picard-Fuchs operators. There are two linearly independent
$\log$-solutions, two $\log^2$ solutions and one $\log^3$
solution. The first few terms of the expansions are given in appendix
\ref{app-periods}.

Given the two Picard-Fuchs operators we can also calculate the
Gauss-Manin system and the monodromy matrix. For a suitable basis
$\Pi$ of $H^3(X^{\vee})$ the Gauss-Manin system reads
\begin{align}
\theta_i\Pi=M_i\Pi\qquad i=1,2.
\end{align} 
We choose
$\Pi=(\int\Omega,\theta_1\int\Omega,\theta_2\int\Omega,\theta_1^2\int\Omega,
\theta_1\theta_2\int\Omega,\theta_1^3\int\Omega)$, where $\Omega$ is
the holomorphic threeform of $X^{\vee}$.  The matrices $M_{1,2}$ can
be obtained from the Picard-Fuchs equations and their
derivatives. Evaluated at $z_1=0$ and $z_2=0$, respectively, and
transformed into their Jordan normal forms, one obtains ($1/(2\pi i)$
times) the logarithms of the monodromy matrices $\mathcal{T}=e^{2\pi i
  M}$, where $M=a_iM_i$ for $a_i>0$ \cite{Candelas:1993dm}. Note that
the relation between the connection matrix $M$ and the monodromy
matrix $\mathcal{T}$ only works if the eigenvalues of $M_i(0)$ are
zero. In phase I${}_+$ we get for the Jordan normal forms:
\begin{align}
M_1|_{z_1=0}\sim M_2|_{z_2=0}\sim\left(
\begin{array}{llllll}
 0 & 1 & 0 & 0 & 0 & 0 \\
 0 & 0 & 0 & 0 & 0 & 0 \\
 0 & 0 & 0 & 1 & 0 & 0 \\
 0 & 0 & 0 & 0 & 1 & 0 \\
 0 & 0 & 0 & 0 & 0 & 1 \\
 0 & 0 & 0 & 0 & 0 & 0
\end{array}
\right)
\end{align}
This shows that the monodromy in phase I${}_+$ is maximally unipotent. 
%%%%%%%%%%
\subsubsection{Yukawa couplings and Gromov-Witten invariants}
Now we have all the ingredients to compute the Gromov-Witten
invariants of $X$. There are several methods available. We will choose
the following, as described for instance in section 5.6 of
\cite{coxkatz}. We first compute the B-model Yukawa couplings
$\kappa_{z_1z_jz_k}=\int_{X^{\vee}}\Omega\wedge
\nabla_{\theta_i}\nabla_{\theta_j}\nabla_{\theta_k}\Omega$ up to
normalization. Here $\nabla$ denotes the covariant derivative with
respect to the Gauss-Manin connection. A further ingredient is the
mirror map. With the two $\log$-solutions $\varpi_{1,1}$ and
$\varpi_{1,2}$ (cf. appendix \ref{app-periods}), it is
\begin{align}
t_1(z_1,z_2)=\frac{\varpi_{1,1}}{\varpi_0}\qquad t_2(z_1,z_2)=
\frac{\varpi_{1,2}}{\varpi_0}.
\end{align}
Using the inverse mirror map $q_i=e^{2\pi i t_i(z)}=z_i+\ldots$ we can
extract from that the normalized Yukawa couplings in the A-model as
follows
\begin{align}
\kappa_{t_it_jt_k}=\frac{(2\pi i)^3}{\varpi_0^2}
\left(\frac{t_i}{z_l}\frac{\partial z_l}{\partial t_i}\right)
\left(\frac{t_j}{z_m}\frac{\partial z_m}{\partial t_j}\right)
\left(\frac{t_k}{z_n}\frac{\partial z_n}{\partial t_k}\right)\kappa_{z_kz_lz_m}.
\end{align}
From this, we can read off the A-model Yukawa couplings which have the
following expansion in terms of the Gromov-Witten invariants $n_{\beta}$:
\begin{align}
\kappa_{t_it_jt_k}=\int_X J_i\cdot J_j\cdot J_k+
\sum_{\beta\neq 0}n_{\beta}\frac{q^{\beta}}{1-q^{\beta}}
\int_{\beta}J_i\int_{\beta}J_j\int_{\beta}J_k,
\end{align}
where $J_i,J_j,J_k\in H^2(X,\mathbb{Z})$ and $\beta$ is the homology
class of a rational curve in $X$. For a particular Yukawa coupling in
the two-parameter case, say $\kappa_{t_1t_1t_2}$, this looks as
follows. Choosing $\beta=a J_1+bJ_2$ one gets
\begin{align}
\kappa_{t_1t_1t_2}=\int_X J_1\cdot J_1\cdot J_2+\sum_{(a,b)\neq(0,0)}n_{a,b}
\frac{a^2 b q_1^aq_2^b}{1-q_1^aq_2^b}.
\end{align}
The first term is the triple intersection number of divisors
Poincar\'e dual to the $J_i$, which, by abuse of notation, we also
call $J_i$. Fixing one of the intersection numbers corresponds to
choosing the normalization of (one of) the Yukawa couplings. This
information is not contained in the solutions of the Picard-Fuchs
equations but can be fixed by the topological data of $X$.

The first step is to use the Picard-Fuchs equations and their
derivatives to express the four Yukawa couplings in the B-model in
terms of a single one. One finds for example
{\footnotesize\begin{align}{ \kappa_{z_1z_1z_2}=\frac{ \left(32
      \left(5 z_2^2+6 z_2+1\right) z_1^3-4 \left(35 z_2^2+172
      z_2-23\right) z_1^2+\left(-20 z_2^2+185 z_2+85\right) z_1+5 (7
      z_2+5)\right)}{(z_2+1) \left(64 (z_2+1) z_1^3-20 (7 z_2-9)
      z_1^2-20 (z_2-8) z_1+35\right)}\kappa_{z_1z_1z_1}.}
\end{align}}
Using the fact that the Yukawa couplings and their derivatives satisfy
linear differential equations \cite{Hosono:1993qy,coxkatz}, we obtain
the remaining Yukawa coupling by using
\begin{align}
\int_{X^{\vee}}\Omega\wedge \nabla_{z_1}^4\Omega =2\theta_1 \kappa_{z_1z_1z_1}.
\end{align} 
Once again using the Picard-Fuchs equations we get the Yukawa coupling
up to an integration constant $c$
\begin{align}
\kappa_{z_1z_1z_1}=\frac{c (-4 z_1 (z_1 (-35 z_2+16 z_1 (z_2+1)+45)
-5 (z_2-8))-35)}{(32 z_1 (8 z_1+1) z_2-1)
   \left(z_1 \left(11 z_2+z_1 \left(z_1 (z_2+1)^2-z_2 (z_2+14)+3\right)
+3\right)+1\right)}.
\end{align}
 The other Yukawa couplings are
{\scriptsize \begin{align}
\kappa_{z_1z_1z_2}=&\frac{c\left(-32 (z_2+1) (5 z_2+1) z_1^3+
4 (z_2 (35 z_2+172)-23) z_1^2+5 (z_2 (4 z_2-37)-17) z_1-5
   (7 z_2+5)\right)}{(z_2+1) (32 z_1 (8 z_1+1) z_2-1) 
\left(z_1 \left(11 z_2+z_1 \left(z_1
   (z_2+1)^2-z_2 (z_2+14)+3\right)+3\right)+1\right)} \nonumber\\
\kappa_{z_1z_2z_2}=&\frac{c\left(-(16 z_1+15) (z_1+1)^2
+20 z_1 ((7-8 z_1) z_1+1) z_2^2+5 (z_1 (z_1 (16 z_1+73)-46)-7)
   z_2\right)}{(z_2+1) (32 z_1 (8 z_1+1) z_2-1) 
\left(z_1 \left(11 z_2+z_1 \left(z_1 (z_2+1)^2-z_2
   (z_2+14)+3\right)+3\right)+1\right)} \nonumber\\
\kappa_{z_2z_2z_2}=& \frac{c \left(-8 (z_1+1)^3+20 z_1 ((7-8 z_1) z_1+1) z_2^3+
5 \left(z_1 \left(56 z_1^2+38
   z_1-51\right)-7\right) z_2^2-5 (z_1 (2 z_1 (8 z_1+7)+51)+8) z_2\right)}
{(z_2+1)^2 (32 z_1 (8 z_1+1)
   z_2-1) \left(z_1 \left(11 z_2+z_1 \left(z_1 (z_2+1)^2-z_2 (z_2+14)+3\right)
+3\right)+1\right)}.\nonumber\\
\end{align}}
The Yukawa couplings are of the form $\frac{p(z)}{q(z)\Delta}$, where
$p,q$ are polynomials in $z_{1,2}$ and $\Delta$ is the discriminant
\begin{align}
\label{discriminant}
\Delta=(z_2+1) (32 z_1 (8 z_1+1) z_2-1) \left(z_1 \left(11 z_2+z_1 
\left(z_1 (z_2+1)^2-z_2
   (z_2+14)+3\right)+3\right)+1\right).
\end{align}
As expected from the analysis of the GLSM it has three
components. They match with the Coulomb branch analysis. The first
factor $-z_2=1$ obviously corresponds to the mixed branch with
$t_2=0$. Using the parametric expressions (\ref{deff}) from the
Coulomb branch analysis and the identification
$z_1=e^{-\frac{t_1-t_2}{2}},z_2=-e^{-t_2}$ we can identify the second
factor of (\ref{discriminant}) with branch (i) and the third factor
with branch (ii) as discussed in section \ref{sec-coulomb}.
 
The mirror map is easily computed from the periods. The calculation of
its inverse requires inverting series in two variables. This is
efficiently implemented in the Mathematica package INSTANTON
\cite{instanton}.  Computing the A-model Yukawa couplings and matching
the classical term with the triple intersection numbers we find that
the integration constant $c$ has to be set to $c=1$. Since we only
need one triple intersection number for the normalization, the
appearance of the other three is a non-trivial check. The instanton
numbers for low degrees are
\begin{align}
\begin{array}{c|rrrrrr}
d_1\backslash d_2& 0&1&2&3&4&5\\
\hline
0&- & 3 & 0 & 0 & 0 & 0 \\
1& 10 & 40 & 0 & 0 & 0 & 0 \\
2& 0 & 185 & 140 & 0 & 0 & 0 \\
3& 0 & 45 & 1150 & 280 & 0 & 0 \\
4& 0 & -20 & 3210 & 10875 & 1260 & 0 \\
5& 0 & 3 & 640 & 62428 & 80912 & 4592 \\
\end{array}.
\end{align}
This completes our discussion of phase I${}_+$. 
%%%%%%%%%%%%%%%%%%%%%%%%%%%%%%%%%%%%%%%%%%%%%%%%%%%%%%%%%%%%%%%%%%%%%%%%%%
\subsection{Phase IV}
\label{sec-phase4}
Phase IV is harder to analyze since the powerful machinery of toric
geometry is no longer at our disposal for this determinantal
variety. Since we have the Picard-Fuchs operator we can still obtain a
lot of information about this phase.

At first, we can confirm that phase IV is indeed geometric in the
sense that the limiting point in the moduli space is a point of
maximally unipotent monodromy. In phase I${}_+$ we have chosen the
coordinates $z_1,z_2$. The coordinates in the other phases can be read
off from the the phase diagram in figure \ref{fig-twoparphases} by
simply transforming from one coordinate patch to the other.  Up to a
numerical factors $\rho,\sigma$, we have to make the following change
of coordinates to get to phase IV:
\begin{align}
\label{phase4coord}
z_{1,\mathrm{IV}}=\frac{\sigma}{z_1}\qquad z_{2,\mathrm{IV}}=\frac{\rho}{z_2}.
\end{align}
 In order to compute the Jordan forms of the Gauss-Manin connections
 in the other phases, one can either transform the Picard-Fuchs
 operators or the connection matrices themselves. After rescaling the
 holomorphic threeform $\Omega\rightarrow z_1z_2\Omega$, one finds
\begin{equation}
M_{1}|_{z_{1,\mathrm{IV}}=0}\sim\left(
\begin{array}{llllll}
 0 & 1 & 0 & 0 & 0 & 0 \\
 0 & 0 & 0 & 0 & 0 & 0 \\
 0 & 0 & 0 & 1 & 0 & 0 \\
 0 & 0 & 0 & 0 & 1 & 0 \\
 0 & 0 & 0 & 0 & 0 & 1 \\
 0 & 0 & 0 & 0 & 0 & 0
\end{array}
\right)
\quad
M_2|_{z_{2,\mathrm{IV}}=0}\sim\left(
\begin{array}{llllll}
 0 & 0 & 0 & 0 & 0 & 0 \\
 0 & 0 & 0 & 0 & 0 & 0 \\
 0 & 0 & 0 & 1 & 0 & 0 \\
 0 & 0 & 0 & 0 & 1 & 0 \\
 0 & 0 & 0 & 0 & 0 & 1 \\
 0 & 0 & 0 & 0 & 0 & 0
\end{array}
\right).
\end{equation}
This confirms that the monodromy is maximally unipotent in phase IV.

In the other phases relating the Gauss-Manin connection to the
monodromy is not as simple, because in some cases the difference
between eigenvalues is an integer. While a transformation to zero
eigenvalues always exists \cite{deligne,Candelas:1993dm} it is hard to
find in practice. In these cases one can also check that the solutions
of the Picard-Fuchs equations are not as expected for a large complex
structure limit. Therefore we conclude that only $z_1=z_2=0$ and
$z_{1,\mathrm{IV}}=z_{2,\mathrm{IV}}=0$ correspond to large complex
structure points and all the other phases are of hybrid type. For
completeness, we list the Jordan normal forms of the connection
matrices of the other phases in appendix \ref{app-gaussmanin}.

Using (\ref{phase4coord}) the Picard-Fuchs operators in phase IV are
\begin{align}
\mathcal{L}^{\mathrm{IV}}_1=&-12 \theta_2^2 \sigma  \rho ^2+
12 \theta_1 \theta_2 \sigma  \rho ^2+z_2^2\left(5 \sigma  \theta_1^2-
4\theta_2 \sigma  \theta_1+\sigma  \theta_1-12 \theta_2^2 \sigma 
-18 \theta_2 \sigma -6 \sigma
   \right)\nonumber\\
&+z_1z_2^2\left(5 \theta_1^2-19 \theta_2 \theta_1-9 \theta_1+20 \theta_2^2+21
   \theta_2+6\right)\nonumber\\
&+z_2\left(-11 \rho  \sigma  \theta_1^2+8 \theta_2 \rho  \sigma  \theta_1-3
   \rho  \sigma  \theta_1-24 \theta_2^2 \rho  \sigma -18 \theta_2 \rho  \sigma
 -6 \rho  \sigma \right)\nonumber\\
&   +z_1z_2\left(-\rho  \theta_1^2-19 \theta_2 \rho  \theta_1-\rho  \theta_1
+20 \theta_2^2 \rho
   +\theta_2 \rho \right) \nonumber\\
\mathcal{L}^{\mathrm{IV}}_2=&-4 \theta_2 \rho  \sigma  \theta_1^2+
4 \theta_2^2 \rho  \sigma  \theta_1+z_1\left(10 \rho  \theta_2^3-20
   \theta_1 \rho  \theta_2^2-10 \rho  \theta_2^2+10 \theta_1^2 \rho  \theta_2+
10 \theta_1 \rho 
   \theta_2\right)\nonumber\\
&+z_2\left(-3 \sigma  \theta_1^3+4 \theta_2 \sigma  \theta_1^2+\sigma 
   \theta_1^2+4 \theta_2^2 \sigma  \theta_1+6 \theta_2 \sigma  \theta_1+
2 \sigma  \theta_1\right)\nonumber\\
&   +z_1z_2\left(-3 \theta_1^3+13 \theta_2 \theta_1^2+4 \theta_1^2-
20 \theta_2^2 \theta_1-14
   \theta_2 \theta_1-3 \theta_1+10 \theta_2^3+10 \theta_2^2+3 \theta_2\right).
\end{align}
For the sake of readability we have we written
$z_{i,\mathrm{IV}}\equiv z_i,
\theta_{i,\mathrm{IV}}\equiv\theta_i$. Up to order $3$, the expansion
of the fundamental period is
\begin{align}
\varpi_0(\widetilde{Y}^{\vee})=1-\frac{z_2}{2 \rho }+\frac{z_1  z_2}{2 \rho  \sigma }+
\frac{3 z_2^2}{8 \rho ^2}-\frac{21 z_1 z_2^2}{16 \rho ^2 \sigma }-
\frac{5 z_2^3}{16 \rho ^3}+\ldots
\end{align}
We can also compute the Yukawa couplings, depending on the parameters
$\rho,\sigma$, and including an integration constant $c$. The result
is
{\scriptsize\begin{align} \kappa^{\mathrm{IV}}_{z_1z_1z_1}=&
  \frac{c \left(4 \rho \sigma \left(-16 \sigma ^2+35 \sigma z_1+5
    z_1^2\right)-z_2 \left(64\sigma ^3+180 \sigma ^2 z_1+160 \sigma
    z_1^2+35 z_1^3\right)\right)}{\left(32 \rho \sigma (8\sigma
    +z_1)-z_1^2 z_2\right) \left(\rho ^2 \sigma ^2 (\sigma -z_1)+\rho
    \sigma z_2 \left(2\sigma ^2-14 \sigma z_1+11 z_1^2\right)+z_2^2
    (\sigma +z_1)^3\right)}\nonumber
  \\ \kappa^{\mathrm{IV}}_{z_1z_1z_2}=&-\frac{c \left(20 \rho ^2
    \sigma (\sigma -z_1) (8 \sigma +z_1)+\rho z_2 \left(192
    \sigma^3-688 \sigma ^2 z_1+185 \sigma z_1^2+35 z_1^3\right)+z_2^2
    (\sigma +z_1) (4 \sigma +5 z_1)(8 \sigma +5 z_1)\right)}{(\rho
    +z_2) \left(32 \rho \sigma (8 \sigma +z_1)-z_1^2
    z_2\right)\left(\rho ^2 \sigma ^2 (\sigma -z_1)+\rho \sigma z_2
    \left(2 \sigma ^2-14 \sigma z_1+11 z_1^2\right)+z_2^2 (\sigma
    +z_1)^3\right)}\nonumber\\ \kappa^{\mathrm{IV}}_{z_1z_2z_2}=&-\frac{c
    \left(20 \rho ^2 \sigma (\sigma -z_1) (8 \sigma +z_1)+\rho z_2
    \left(192 \sigma^3-688 \sigma ^2 z_1+185 \sigma z_1^2+35
    z_1^3\right)+z_2^2 (\sigma +z_1) (4 \sigma +5 z_1)(8 \sigma +5
    z_1)\right)}{(\rho +z_2) \left(32 \rho \sigma (8 \sigma
    +z_1)-z_1^2 z_2\right)\left(\rho ^2 \sigma ^2 (\sigma -z_1)+\rho
    \sigma z_2 \left(2 \sigma ^2-14 \sigma z_1+11z_1^2\right)+z_2^2
    (\sigma
    +z_1)^3\right)}\nonumber\\ \kappa^{\mathrm{IV}}_{z_2z_2z_2}=&-\frac{c
    \left(20 \rho ^3 \sigma (\sigma -z_1) (8 \sigma +z_1)-5 \rho ^2
    z_2 \left(56 \sigma^3+38 \sigma ^2 z_1-51 \sigma z_1^2-7
    z_1^3\right)+5 \rho z_2^2 \left(16 \sigma ^3+14 \sigma^2 z_1+51
    \sigma z_1^2+8 z_1^3\right)+8 z_2^3 (\sigma +z_1)^3\right)}{(\rho
    +z_2)^2 \left(32\rho \sigma (8 \sigma +z_1)-z_1^2 z_2\right)
    \left(\rho ^2 \sigma ^2 (\sigma -z_1)+\rho \sigma z_2 \left(2
    \sigma ^2-14 \sigma z_1+11 z_1^2\right)+z_2^2 (\sigma
    +z_1)^3\right)}
\end{align}}

From the classical limit $z_1=z_2=0$ we can extract the triple intersection numbers
\begin{align}
J_{1,\mathrm{IV}}^3=-\frac{1}{4}\frac{c}{\rho^2\sigma^2}\qquad 
J_{1,\mathrm{IV}}^2J_{2,\mathrm{IV}}=J_{1,\mathrm{IV}}J_{2,\mathrm{IV}}^2=J_{2,\mathrm{IV}}^3
=-\frac{5}{8}\frac{c}{\rho^2\sigma^2}.
\end{align}
We note that in each of the intersection numbers the same combination
of the unknown constants appears. Therefore they are determined up to
an overall factor. Comparing with the topological analysis of phase IV
in section \ref{sec-iv} we find agreement if we set
$\frac{c}{\rho^2\sigma^2}=-16$. Finally, we can extract the
Gromov-Witten invariants up to the three constants. The result for low
degrees is
\begin{align}
\begin{array}{c|ccccc}
d_1\backslash d_2& 0&1&2&3&4\\
\hline
0&-&\frac{5 c}{8 \rho ^3 \sigma ^2}
&-\frac{5 c (4 \rho +3)}{256 \rho ^4 \sigma ^2}
&\frac{c\left(5-80 \rho ^2\right)}{3456 \rho ^5 \sigma ^2}
&\frac{15 c \left(16 \rho ^2-1\right)}{32768 \rho^6 \sigma ^2}\\
1&\frac{3 c}{64 \rho ^2 \sigma ^3}&-\frac{235 c}{256 \rho ^3 \sigma ^3}
&\frac{345c}{512 \rho ^4\sigma ^3}&-\frac{345 c}{2048 \rho ^5 \sigma ^3}
&\frac{235 c}{16384 \rho ^6 \sigma ^3}\\
2&\frac{3 c (7-8 \sigma )}{4096 \rho ^2 \sigma ^4}
&\frac{15 c}{512 \rho ^3 \sigma ^4}
&\frac{5 c(1504 \rho  \sigma -12551)}{65536 \rho ^4 \sigma ^4}
&\frac{2885 c}{2048 \rho ^5 \sigma ^4}
&-\frac{5c \left(8832 \rho ^2 \sigma +92021\right)}{524288 \rho ^6 \sigma ^4}\\
3&\frac{c \left(7-16 \sigma ^2\right)}{9216 \rho ^2 \sigma ^5}
&\frac{5 c}{1024 \rho ^3 \sigma^5}&\frac{1405 c}{32768 \rho ^4 \sigma ^5}
&\frac{5 c \left(12032 \rho ^2 \sigma^2-626729\right)}
{1769472 \rho ^5 \sigma ^5}&\frac{617225 c}{131072 \rho ^6 \sigma ^5}\\
4&\frac{3 c \left(135-448 \sigma ^2\right)}{2097152 \rho ^2 \sigma ^6}
&\frac{45 c}{32768 \rho ^3\sigma ^6}
&\frac{15 c \left(661-256 \rho  \sigma ^2\right)}{1048576 \rho ^4 \sigma ^6}
&\frac{50145c}{524288 \rho ^5 \sigma ^6}
&\frac{5 c \left(12852224 \rho ^2 \sigma ^2-528741767\right)}
{536870912\rho ^6 \sigma ^6}
\end{array}
\end{align}
We can try to make an educated guess for the choice of unknowns by
looking for the minimal values of $\{\rho,\sigma,c\}$ such that the
coefficients in the fundamental period positive integers and the
Gromov-Witten invariants are integer. The most obvious choice
compatible with our results of the triple intersection numbers seems
to be
\begin{align}
\sigma=-\frac{1}{8}\qquad\rho=-\frac{1}{4}\qquad c=-\frac{1}{64}.
\end{align} 
Considering the relation (\ref{dualitymap}) between the FI-theta parameters of the GLSM and its dual, we observe that this choice of constants in (\ref{phase4coord}) is consistent with the identification
\begin{align}
 \frac{1}{z_{1,IV}}=e^{-\frac{(\tilde{t}_1-\tilde{t}_2)}{2}}\qquad \frac{1}{z_{2,IV}}=-e^{-\tilde{t}_2}. 
\end{align}
Fixing the constants in this way gives the following Gromov-Witten invariants:
\begin{align}
\begin{array}{c|rrrrr}
d_1\backslash d_2& 0&1&2&3&4\\
\hline
0&-&40&10&0&0\\
1&6&470&1380&1380&470 \\
2&-6&120& 15630& 92320& 229880\\
3&6&-160& 5620& 928470& 9875600\\
4&-12&360& -9930& 401160& 82613940
\end{array}
\end{align}
Since we do not know if our choice of the constants $\rho,\sigma,c$ is
the correct one, these numbers remain conjectural.
%%%%%%%%%%%%%%%%%%%%%%%%%%%%%%%%%%%%%%%%%%%%%%%%%%%%%%%%%%%%%%%%%%%%%%%%%%%%%
\section{Summary and Outlook}
\label{sec-outlook}

In this paper, we constructed and studied a two parameter non-Abelian
GLSM with six phases.
Two phases are geometric,
one weakly coupled and the other strongly coupled,
and correspond to Calabi-Yau manifolds, $X$ and $\wt{Y}$,
which are birationally inequivalent but are expected to be
derived equivalent.
Three others are hybrid phases, described by pairs
$(X_{\alpha},W_{\alpha})$ of spaces and potentials,
$\alpha=\mathrm{I}_-,\mathrm{II}, \mathrm{V}$.
These are presumably bad hybrids
because the vector $U(1)$ R-symmetry acts non-trivially on 
$\mathrm{Crit}(W_{\alpha})$.
We were unable to find the character of the remaining phase since
the original and the dual models are both strongly coupled.

Having constructed and analyzed one particular example in detail,
the next obvious task is to explore more examples
and try to systematize the analysis.
This may lead to a novel systematic construction of Calabi-Yau varieties
which parallels the systemic construction and classification
of the complete intersection Calabi-Yaus in toric varieties.
This would considerably expand our knowledge on the landscape of
Calabi-Yau varieties, or more generally, of
2d (2,2) SCFTs with charge integrality.

Of course, there are still many things that could be studied
just for our model.
Recently, techniques to compute supersymmetric partition functions
of 2d $(2,2)$ gauge theories have been developed
\cite{Benini:2012ui,Doroud:2012xw,Doroud:2013pka,Closset:2015rna,Hori:2013ika,Sugishita:2013jca,Honda:2013uca,Kim:2013ola,Benini:2013xpa,Benini:2013nda}.
In the present paper, we limited our analysis to those that can be done
with the ``classical'' methods, like undergraduate topology
and classical mirror symmetry, but
the new technology can tell us more.
For example, by studying the sphere partition function
\cite{Benini:2012ui,Doroud:2012xw,Jockers:2012dk,Gerchkovitz:2014gta,Gomis:2015yaa},
we expect to obtain more information about the hybrid phases as well as
the mysterious phase where the classical analysis gave us no clue.
In particular, we can examine whether the distances to the
limiting loci are finite, settling the question about
the badness of the hybrids.

The hemisphere partition function
\cite{Hori:2013ika,Sugishita:2013jca,Honda:2013uca} also has a good application.
From general principles, we expect that the Calabi-Yau manifolds
$X$ and $\wt{Y}$ are derived equivalent (\ref{DEXY})
and that the equivalence depends on the homotopy class of paths in
$\mathfrak{M}_K$ that connect the two phases.
But we would like to know what the equivalence is for each homotopy class.
The solution to this problem in Abelian GLSMs \cite{Herbst:2008jq}
(completed and generalized in \cite{SegalCMP,DHLGIT,BFKVGIT}) 
was found to be reproduced by analyzing the hemisphere partition function
\cite{Hori:2013ika}, and is being extended to the non-Abelian models
of \cite{Hori:2006dk,Hori:2011pd} in \cite{wip}.
When applied to our model, we expect to obtain equivalences not only
between $\mathrm{D}^b_{\it Coh}(X)$ and $\mathrm{D}^b_{\it Coh}(\wt{Y})$
but also with the categories of B-branes in the other phases, that is,
the categories of matrix factorizations of the 
pairs $(X_{\alpha},W_{\alpha})$:
\begin{center}
\begin{picture}(220,95)
\put(10,0){$\mathrm{D}^b_{\it Coh}(\wt{Y})$}
\put(0,80){$\mathrm{MF}(X_{\mathrm{V}}, W_{\mathrm{V}})$}
\put(100,80){$\mathrm{MF}(X_{\mathrm{I}_-},W_{\mathrm{I}_-})$}
\put(180,42){$\mathrm{D}^b_{\it Coh}(X)$}
\put(170,0){$\mathrm{MF}(X_{\mathrm{II}}, W_{\mathrm{II}})$}
\put(108,0){?}
\put(182,54){\vector(-1,1){20}}
\put(95,83){\vector(-1,0){25}}
\put(30,72){\vector(0,-1){58}}
\put(200,35){\vector(0,-1){23}}
\put(162,2){\vector(-1,0){40}}
\put(99,2){\vector(-1,0){40}}
\put(176,62){$\cong$}
\put(80,86){$\cong$}
\put(17,40){$\cong$}
\put(204,22){$\cong$}
\put(80,6){?}
\put(142,6){?}
\end{picture}
\end{center}

For many of the Calabi-Yau pairs that appear in the
one parameter models, including those found by R\o dland, Hosono-Takagi
and Miura, it is known that the derived equivalences fit into the framework of 
``Homological Projective Duality'' by A. Kuznetsov \cite{HPD}.
It would be interesting to see whether the derived equivalence
of our $X$ and $\wt{Y}$ in the two parameter model
also fits into this framework.
A preliminary discussion shows the appearance of some of the 
structures in the hybrid models.\footnote{We thank A.~Kuznetsov
for showing his picture on our example.}
The study on this point may shed new light on
the relation between the Homological Projective Duality and
the gauge theory understanding of
the equivalences that involves 2d Seiberg duality.

Non-birational but derived equivalent pairs of varieties
with Picard number $\geq 2$ had been known for a long time.
A ``trivial'' example is $B\times S$ and $B\times S'$ for some
variety $B$ where $S$ and $S'$ are birationally inequivalent but
derived equivalent K3 or Abelian surfaces,
related by a Fourier-Mukai functor.
We may obtain non-trivial Calabi-Yau examples if we consider manifolds
with a structure of K3 or Abelian fibration and applying
Fourier-Mukai transforms on the fibers.\footnote{We thank Y. Toda for
informing us of such a construction and
the example \cite{Christian}.}
Indeed, such a pair $(X_1,X_2)$ was found in \cite{Christian}.
They are Abelian surface fibrations over $\PP^1$
which are related by T-duality along the fibers.
(They have $(h^{1,1},h^{2,1})=(2,2)$. $X_1$ is simply connected
and $X_2$ tuns out to be the quotient
of $X_1$ by a freely acting symmetry group $\Z_8\times\Z_8$.
Hence they cannot be birationally equivalent.)
It would be interesting to see if our pair $(X,\wt{Y})$ 
is or is not of this type.

We have not completed the analysis of the topology of $X$ and $\wt{Y}$,
due to lack of our ability to do so. Most importantly, we have not proved that
the Hodge numbers of $\wt{Y}$ are $(h^{1,1},h^{2,1})=(2,24)$, although that
must certainly be the case. Also, to compute the intersection
numbers on $\wt{Y}$, we needed to make some assumptions.
These are obvious gaps in our analysis.
Another important topological information is the topological K-theory.
Our model is expected to flow to a family of SCFTs with
$\wh{c}=3$ and charge integrailty that can be used as supersymmetric
backgrounds for Type II string theory.
When we consider Type II string theory on a spacetime ${\bf X}$,
the D-brane (or Ramond-Ramond) charge is classified by the K-theory
of ${\bf X}$, of various relative types depending on the dimension of
the objects \cite{Witten:1998cd}.
For this reason, the K-theory of $X$ and $\wt{Y}$ is important.
Since the charge lattice must be stable under continuous
deformation, we expect that the 
K-theory of $X$ and that of $\wt{Y}$ are isomorphic.
Indeed that seems to be the case under (\ref{DEXY}):
according to \cite{AddingtonThomas,AddingtonBrauer},
the (topological) K-theory is derived invariant.
It was shown in \cite{Brunner:2001eg,Batyrev:2005jc,DoranMorgan}
that the torsion parts of the K-groups of a Calabi-Yau threefold
$M$ (which are the information not obtained by the Hodge numbers)
are given by
$\mathrm{Tors}(\mathrm{K}^0(M))=A(M)\oplus B(M)^*$
and $\mathrm{Tors}(\mathrm{K}^{-1}(M))=A(M)^*\oplus B(M)$
where $A(M)=\mathrm{Hom}(\pi_1(M),\Q/\Z)$ and
$B(M)=\mathrm{Tors}(\mathrm{H}^3(M,\Z))$
(the latter is called the Brauer group of $M$).
Here for a finite Abelian group $\gamma$, we write
$\gamma^*=\mathrm{Hom}(\gamma,\Q/\Z)$ for its dual.
Thus, we need to compute the fundamental group
and the Brauer group of $X$ and $\wt{Y}$, of which we only know
$\pi_1(\wt{Y})=\{1\}$ at this stage. 
A related problem is to determine the K-theory of non-geometric phases
\cite{Brunner:2001eg}. Recently, topological K-theory
of dg-categories, such as the categories of B-branes for
Landau-Ginzburg and hybrid models, has been defined
\cite{BlancK,DyckerhoffK}. It would be interesting to compute it
in the hybrid phases of our model
and check that they match with the results in the
geometric phases. And it would be interesting to see if 
the construction \cite{BlancK,DyckerhoffK} works directly in
GLSMs.

It may also be interesting to further study mirror symmetry for our
model. We have drawn all our conclusions on the mirror starting from the
Calabi-Yau $X$ in phase I${}_+$. One could also attempt to construct
the mirror of the determinantal Calabi-Yau $\widetilde{Y}$ in phase
IV. Mirrors of Pfaffian Calabi-Yaus have been proposed in
\cite{rodland98,Kanazawa:2012xya} partly based on 
\cite{boehmthesis} (see also \cite{Shimizu:2010us}). It
would be interesting to see if any of these constructions also work
for $\wt{Y}$, and also the determinantal
Calabi-Yau constructed in \cite{Hori:2013gga}.
Of course, understanding
mirror symmetry for 2d (2,2) non-Abelian gauge theories more generally
is an important problem.

\section*{Acknowledgments}
We would like to thank Paul Aspinwall, Alexey Bondal, Yalong Cao, Will Donovan,
Richard Eager, Shinobu Hosono, Alexander Kuznetsov, 
Dave Morrison, Ronen Plesser, Mauricio Romo, Emanuel Scheidegger,
Eric Sharpe, Harald Skarke, Hiromichi Takagi, Yukinobu Toda and
Johannes Walcher for discussions, instructions and encouragement.
Part of our work was supported by the
World Premier International Research Center Initiative, MEXT, Japan.
%%%%%%%%%%%%%%%%%%%%%%%%%%%%%%%%%%%%%%%%%%%%%%%%%%%%%%%%%%%%%%%%%%%%%%%%%%%%
\appendix
\section{Additional details on mirror symmetry}
\subsection{Evaluation of the fundamental period of $\widetilde{X}^{\vee}$}
\label{app-w0calc}
Here we give further details on the evaluation of the fundamental
period (\ref{w0ytilde}) of $\widetilde{X}^{\vee}$. We introduce a
short-hand notation for the Laurent polynomials (\ref{twoparmirror}):
\begin{align} 
f_1=&1+\alpha_1\frac{x_7}{x_1x_2x_3x_8}+\alpha_5x_4:=1+\beta_1+
\beta_5\nonumber\\
f_2=&1+\alpha_3x_2+\alpha_6x_5:=1+\beta_3+\beta_6\nonumber\\
f_3=&1+\alpha_4x_3+\alpha_7x_6:=1+\beta_4+\beta_7\nonumber\\
f_4=&1+\alpha_9x_1+\alpha_{11}\frac{1}{x_4x_5x_6x_7}:=1+\beta_9+
\beta_{11}\nonumber\\
f_5=&1+\alpha_2x_8+\alpha_{8}\frac{x_8}{x_7}+\alpha_{10}x_7:=
1+\beta_2+\beta_{8}+\beta_{10},
\end{align}
where the $\beta_1=\alpha_1\frac{x_7}{x_1x_2x_3x_8}$, etc. This we
insert into the residue formula for the fundamental period given by
\begin{align}
\varpi_0(\tilde{Y})=\frac{1}{(2\pi i)^{11}}
\int_{\gamma}\prod_i\frac{dx_i}{x_i}\frac{1}{f_1f_2f_3f_4f_5}.
\end{align}
We rewrite the integrand as follows
\begin{align}
K=&\frac{1}{f_1f_2f_3f_4f_5}\nonumber\\
=&\frac{1}{1+\beta_1+\beta_5}\frac{1}{1+\beta_3+\beta_6}
\frac{1}{1+\beta_4+\beta_7}
\frac{1}{1+\beta_9+\beta_{11}}
\frac{1}{1+\beta_2+\beta_{8}+\beta_{10}}\nonumber\\
=&\sum_{n_1=0}^{\infty}(-1)^{n_1}(\beta_1+\beta_5)^{n_1}
\sum_{n_2=0}^{\infty}(-1)^{n_2}(\beta_3+\beta_6)^{n_2}
\sum_{n_3=0}^{\infty}(-1)^{n_3}(\beta_4+\beta_7)^{n_3}\times\nonumber\\
&\times\sum_{n_4=0}^{\infty}(-1)^{n_4}(\beta_9+\beta_{11})^{n_4}
\sum_{n_5=0}^{\infty}(-1)^{n_5}(\beta_2+\beta_{8}+\beta_{10})^{n_5}\nonumber\\
=&\sum_{n_1=0}^{\infty}(-1)^{n_1}\sum_{k_1=1}^{n_1}
\left(\begin{array}{c}n_1\\k_1\end{array}\right)
\beta_1^{k_1}\beta_5^{n_1-k_1}\sum_{n_2=0}^{\infty}(-1)^{n_2}
\sum_{k_2=1}^{n_2}\left(\begin{array}{c}n_2\\k_2\end{array}\right)
\beta_3^{k_2}\beta_6^{n_2-k_2}\times\nonumber\\
&\times\sum_{n_3=0}^{\infty}(-1)^{n_3}\sum_{k_3=1}^{n_3}
\left(\begin{array}{c}n_3\\k_3\end{array}\right)
\beta_4^{k_3}\beta_7^{n_3-k_3}\times
\sum_{n_4=0}^{\infty}(-1)^{n_4}
\sum_{k_4=1}^{n_4}\left(\begin{array}{c}n_4\\k_4\end{array}\right)
\beta_9^{k_4}\beta_{11}^{n_4-k_4}\times\nonumber\\
&\times\sum_{n_5=0}^{\infty}(-1)^{n_5}\sum_{k_5=0}^{n_5}
\sum_{l_5=0}^{n_5-k_5}\left(\begin{array}{c}n_5\\k_5\end{array}\right)
\left(\begin{array}{c}n_5-k_5\\l_5\end{array}\right)
\beta_2^{k_5}\beta_{8}^{l_5}\beta_{10}^{n_5-k_5-l_5}.
\end{align}
Only those products of the $\beta_i$ which are independent of $x_i$
contribute to the residue integral. Which monomials these are is
encoded in the Mori generators (\ref{moricone}):
\begin{align}
\frac{\beta_2\beta_5\beta_6\beta_7}{\beta_8}=
\frac{\alpha_2\alpha_5\alpha_6\alpha_7}{\alpha_8}=
z_1\quad \frac{\beta_8\beta_{10}}{\beta_2}=
-z_2\quad \frac{\beta_1\beta_2\beta_3\beta_4\beta_9}{\beta_{10}}=z_3
\end{align}
The fundamental period is therefore a power series of the form
\begin{align}
\sum_{a,b,c\geq 0}c_{abc}z_1^az_2^bz_3^c=
\sum_{a,b,c}c_{abc}(-1)^b\beta_1^c\beta_2^{a-b+c}\beta_3^c\beta_4^c\beta_5^a
\beta_6^a\beta_7^a\beta_8^{-a+b}\beta_9^c\beta_{10}^{b-c}\beta_{11}^a
\end{align}
Comparing coefficients with the integrand $K$ above we find
\begin{align}
&n_1=n_2=n_3=n_4=a+c \qquad k_1=k_2=k_3=k_4=c\nonumber\\
&n_5=b\qquad k_5=a-b+c\qquad l_5=-a+b\nonumber\\
&a-b+c\geq0\qquad -a+b\geq 0\qquad b-c\geq0
\end{align}
Inserting this into the residue integral, we arrive at the result
(\ref{w0ytilde}) for the fundamental period.
%%%%%
\subsection{Periods of $X^{\vee}$}
\label{app-periods}
Here we give the first few terms of the power series expansion of the
solutions to the Picard-Fuchs system (\ref{pfsystem}). The first few
terms of the series expansion of the two $\log$-solutions are:
\begin{align}
\varpi_{1,1}=&-\frac{1}{3} z_1^3+16 \log (z_1) z_2 z_1^2+15 z_2
   z_1^2+\frac{z_1^2}{2}+2 z_2^2 z_1+2 \log (z_1) z_2 z_1+5 z_2
   z_1-z_1\nonumber\\
&+\frac{z_2^3}{3}-\frac{z_2^2}{2}+\log (z_1)+z_2+\ldots\nonumber\\
\varpi_{1,2}=&\frac{2 z_1^3}{3}+
16 \log (z_2) z_2 z_1^2+34 z_2 z_1^2-z_1^2-2 z_2^2
   z_1+2 \log (z_2) z_2 z_1+2 z_1\nonumber\\
&-\frac{z_2^3}{3}+\frac{z_2^2}{2}+\log
   (z_2)-z_2+\ldots
\end{align}
The $\log^2$ solutions have the following form:
\begin{align}
\varpi_{2,1}=&\log ^2(z_1)+\frac{10}{19} \log (z_2) \log (z_1)+\left(\frac{28
   z_2}{19}-\frac{18 z_1}{19}\right) \log (z_1)\nonumber\\
&-\frac{32 z_1}{19}+\log
   (z_2) \left(\frac{10 z_2}{19}-\frac{10 z_1}{19}\right)+\frac{18
   z_2}{19}+\ldots\nonumber\\
\varpi_{2,2}=&\log ^2(z_2)+\frac{40}{19} \log (z_1) \log (z_2)+\left(\frac{36
   z_1}{19}+\frac{2 z_2}{19}\right) \log (z_2)+\frac{100 z_1}{19}\nonumber\\
&+\log
   (z_1) \left(\frac{80 z_1}{19}-\frac{40 z_2}{19}\right)-\frac{42
   z_2}{19}+\ldots
\end{align}
The $\log^3$ solution has the following expansion:
\begin{align}
\varpi_3=&\log ^3(z_1)+\frac{15}{7} \log (z_2) \log ^2(z_1)+\left(\frac{9
   z_1}{7}+\frac{6 z_2}{7}\right) \log ^2(z_1)+\frac{9}{7} \log ^2(z_2)
   \log (z_1)\nonumber\\
&+\frac{12}{7} z_1 \log (z_1)+\log (z_2) \left(\frac{6
   z_1}{7}+\frac{12 z_2}{7}\right) \log (z_1)+\frac{8}{35} \log
   ^3(z_2)-\frac{24 z_1}{7}\nonumber\\
&+\log ^2(z_2) \left(\frac{3 z_1}{35}+\frac{3
   z_2}{5}\right)+\frac{18}{35} \log (z_2) z_2-\frac{36 z_2}{35}+\ldots
\end{align}
%%%%%
\subsection{Gauss-Manin connection matrices}
\label{app-gaussmanin}
Up to numerical scaling factors the local coordinates in the various
phases are expressed in terms of the coordinates of $z_{1,2}$ of phase
I${}_+$ as follows
\begin{align}
z_{1,\mathrm{II}}=\frac{1}{z_1}&\quad z_{2,\mathrm{II}}=z_1z_2^2\nonumber\\
z_{1,\mathrm{III}}=\frac{1}{\sqrt{z_1}z_2}&\quad z_{2,\mathrm{III}}
=\sqrt{z_1}\nonumber\\
z_{1,\mathrm{IV}}=\frac{1}{z_1}&\quad z_{2,\mathrm{IV}}=\frac{1}{z_2}\nonumber\\
z_{1,\mathrm{V}}=z_2&\quad z_{2,\mathrm{V}}=\frac{1}{z_1z_2}\nonumber\\
z_{1,\mathrm{I}_-}=z_1z_2&\quad z_{2,\mathrm{I}_-}=\frac{1}{z_1}.
\end{align}
We compute Jordan normal forms of the the Gauss-Manin connection
matrices evaluated at $z_{i,\ast}=0$ by transforming the Picard-Fuchs
system of phase I${}_+$ (without rescaling the holomorphic
three-form). The results for the phases other then I${}_+$ and IV
already given in the main text are
\begin{align}
M_{z_{1,\mathrm{II}}}=\left(
\begin{array}{llllll}
 0 & 1 & 0 & 0 & 0 & 0 \\
 0 & 0 & 1 & 0 & 0 & 0 \\
 0 & 0 & 0 & 1 & 0 & 0 \\
 0 & 0 & 0 & 0 & 0 & 0 \\
 0 & 0 & 0 & 0 & 1 & 1 \\
 0 & 0 & 0 & 0 & 0 & 1
\end{array}
\right)
\quad
M_{z_{2,\mathrm{II}}}=\left(
\begin{array}{llllll}
 0 & 1 & 0 & 0 & 0 & 0 \\
 0 & 0 & 0 & 0 & 0 & 0 \\
 0 & 0 & 0 & 1 & 0 & 0 \\
 0 & 0 & 0 & 0 & 1 & 0 \\
 0 & 0 & 0 & 0 & 0 & 1 \\
 0 & 0 & 0 & 0 & 0 & 0
\end{array}
\right)
\end{align}
 \begin{align}
M_{z_{1,III}}=\left(
\begin{array}{llllll}
 1 & 0 & 0 & 0 & 0 & 0 \\
 0 & 1 & 0 & 0 & 0 & 0 \\
 0 & 0 & 1 & 1 & 0 & 0 \\
 0 & 0 & 0 & 1 & 1 & 0 \\
 0 & 0 & 0 & 0 & 1 & 1 \\
 0 & 0 & 0 & 0 & 0 & 1
\end{array}
\right)
\quad
M_{z_{2,\mathrm{III}}}=\left(
\begin{array}{llllll}
 0 & 1 & 0 & 0 & 0 & 0 \\
 0 & 0 & 1 & 0 & 0 & 0 \\
 0 & 0 & 0 & 1 & 0 & 0 \\
 0 & 0 & 0 & 0 & 0 & 0 \\
 0 & 0 & 0 & 0 & 1 & 1 \\
 0 & 0 & 0 & 0 & 0 & 1
\end{array}
\right)
\end{align}
\begin{align}
M_{z_{1,\mathrm{V}}}=\left(
\begin{array}{llllll}
 \frac{1}{2} & 1 & 0 & 0 & 0 & 0 \\
 0 & \frac{1}{2} & 1 & 0 & 0 & 0 \\
 0 & 0 & \frac{1}{2} & 1 & 0 & 0 \\
 0 & 0 & 0 & \frac{1}{2} & 0 & 0 \\
 0 & 0 & 0 & 0 & 1 & 1 \\
 0 & 0 & 0 & 0 & 0 & 1
\end{array}
\right)
\quad
M_{z_{2,\mathrm{V}}}=\left(
\begin{array}{llllll}
 1 & 1 & 0 & 0 & 0 & 0 \\
 0 & 1 & 0 & 0 & 0 & 0 \\
 0 & 0 & 1 & 1 & 0 & 0 \\
 0 & 0 & 0 & 1 & 1 & 0 \\
 0 & 0 & 0 & 0 & 1 & 1 \\
 0 & 0 & 0 & 0 & 0 & 1
\end{array}
\right)
\end{align}
\begin{align}
M_{z_{1,\mathrm{I}_-}}=\left(
\begin{array}{llllll}
 0 & 1 & 0 & 0 & 0 & 0 \\
 0 & 0 & 0 & 0 & 0 & 0 \\
 0 & 0 & 0 & 1 & 0 & 0 \\
 0 & 0 & 0 & 0 & 1 & 0 \\
 0 & 0 & 0 & 0 & 0 & 1 \\
 0 & 0 & 0 & 0 & 0 & 0
\end{array}
\right)
\quad
M_{z_{2,\mathrm{I}_-}}=\left(
\begin{array}{llllll}
 \frac{1}{2} & 1 & 0 & 0 & 0 & 0 \\
 0 & \frac{1}{2} & 1 & 0 & 0 & 0 \\
 0 & 0 & \frac{1}{2} & 1 & 0 & 0 \\
 0 & 0 & 0 & \frac{1}{2} & 0 & 0 \\
 0 & 0 & 0 & 0 & 1 & 1 \\
 0 & 0 & 0 & 0 & 0 & 1
\end{array}
\right).
\end{align}
None of these matrices have a structure compatible with maximally
unipotent monodromy. This, and explicitly solving the Picard-Fuchs
equations in these phases, gives further evidence that phases I${}_+$
and IV are the only geometric phases.

\bibliographystyle{fullsort}
\bibliography{paper}
\end{document}